\documentclass[a4paper,11pt]{article}
\usepackage{jheppub}
\usepackage{bm}

\usepackage[missing=]{gitinfo2}
\usepackage{xcolor}
\definecolor{darkred}{rgb}{0.5,0.15,0.15}
\hypersetup{colorlinks=true,urlcolor=darkred,linkcolor=darkred,citecolor=darkred}
\usepackage{soul}
\usepackage{graphicx}
\usepackage{epsfig}
\usepackage{amsmath}
\usepackage{amssymb}
\usepackage{amsthm}
\usepackage{indentfirst}
\usepackage{xspace}
\usepackage{multirow}
\usepackage{hyperref}
\usepackage{verbatim}
\usepackage{subcaption}
\usepackage{tikz-cd}
\usepackage{capt-of}

\usepackage{color}

\usepackage{tikz}
\usetikzlibrary{calc}   %
\usepackage{varioref}

\usetikzlibrary{topaths}
\usetikzlibrary{decorations}
\usetikzlibrary{decorations.pathmorphing}

\numberwithin{equation}{section}

\newcommand{\fad}{\operatorname{\Phi}_{\mathsf{b}}}

\newcommand{\balpha}{\boldsymbol{\alpha}}

\newcommand{\be}{\begin{equation}}
\newcommand{\ee}{\end{equation}}
\newcommand{\ba}{\begin{aligned}}
\newcommand{\ea}{\end{aligned}}

\def\IF{{\mathbb F}}

\def\IR{{\mathbb R}}
\def\IZ{{\mathbb Z}}

\newcommand{\ri}{{\rm i}}
\newcommand{\rd}{{\rm d}}

\newcommand{\E}{{\mathrm e}}

\DeclareMathOperator{\re}{{\rm e}}
\DeclareMathOperator{\Tr}{Tr}

\usepackage{color}

\newcommand{\bs}{\boldsymbol}

\usepackage{scalerel,stackengine}
\stackMath
\newcommand\reallywidehat[1]{%
\savestack{\tmpbox}{\stretchto{%
  \scaleto{%
    \scalerel*[\widthof{\ensuremath{#1}}]{\kern-.6pt\bigwedge\kern-.6pt}%
    {\rule[-\textheight/2]{1ex}{\textheight}}
  }{\textheight}%
}{0.5ex}}%
\stackon[1pt]{#1}{\tmpbox}%
}
\begin{document}

\preprint{{\flushright
CERN-TH-2023-060\\
}}

\title{Connecting topological strings and spectral theory via non-autonomous Toda equations}
\author[1,2,3]{Pavlo Gavrylenko,}
\author[1,4]{Alba Grassi,}
\author[1]{Qianyu Hao}
\affiliation[1]{Section de Math\'{e}matiques, Universit\'{e} de Gen\`{e}ve, 1211 Gen\`{e}ve 4, Switzerland}
\affiliation[2]{International School of Advanced Studies (SISSA), via Bonomea 265, 34136 Trieste, Italy}
\affiliation[3]{Bogolyubov Institute for Theoretical Physics, Metrologichna 14-b, 03143 Kyiv, Ukraine}
\affiliation[4]{Theoretical Physics Department, CERN, 1211 Geneva 23, Switzerland}
\abstract{
We consider the Topological String/Spectral theory duality  on  toric Calabi-Yau threefolds obtained from the resolution of the cone over the $Y^{N,0}$ singularity.  Assuming Kyiv formula, we demonstrate this duality in a special regime thanks to an underlying connection between spectral determinants of quantum mirror curves and the non-autonomous (q)-Toda system. 
We further exploit this link to connect small and large time expansions in Toda equations. In particular we provide an explicit expression for their  tau functions at large time in terms of a strong coupling version of irregular $W_N$ conformal blocks at $c=N-1$. These are related to a special class of multi-cut matrix models which describe  the strong coupling regime of four dimensional, $\mathcal{N}=2$ $SU(N)$  super Yang-Mills.
}

\maketitle

\section{Introduction and Summary}

String theory was originally developed using perturbative approaches and a full understanding of its non-perturbative effects remains as a challenge.
In the last decades, a lot of progress has been made in this direction with the help of dualities, in particular the AdS/CFT correspondence \cite{Maldacena:1997re}.  However, although this duality is very powerful and has numerous applications to various fields, many aspects of it remain conjectural. For instance, testing  AdS/CFT beyond the planar limit has been proven to be quite challenging.

One way to make further progress toward the understanding of non-perturbative effects is to study them in simpler string theory models, such as topological string theory. In this paper, we  focus on the topological string/spectral theory duality (TS/ST correspondence) which states that  topological string theory on toric  Calabi-Yau (CY) manifolds has a dual description in terms of one dimensional quantum mechanical operators \cite{Grassi:2014zfa,Codesido:2015dia,mz2}. As a result, the enumerative geometry of these CY manifolds emerge from the spectral properties of the corresponding operators. 
One interesting aspect of the TS/ST correspondence is the possibility of formulating it in terms of precise  statements. 
One example which plays an important role in this paper is the following conjectural identity 
\be
 \label{5dintro}{\sum_{ {\bs w} \in Q_{N-1}} \exp \left( \mathsf{J}_{N}(\bs{t}(\hbar)+2 \pi \ri  {\bs w}, {t_N}, \hbar) \right)=  \det \left(1+\sum_{i=1}^{N-1} \kappa_i {\rm A}_i^{\rm 5D}\right) ~. }
\ee
Let us briefly introduce \eqref{5dintro}, while leaving the technical details to \autoref{sec:TSSTcor}.
In this paper, we focus on a particular family of  toric CY manifolds. Such family is obtained from the resolution of the cone over the $Y^{N,0}$ singularity and it is denoted by $X_{N-1}$, where $N-1$ is the genus of its mirror curve. 

We study  topological string theory on $X_{N-1}$, with $N\geq 2$.   
The grand canonical potential $\mathsf{J}_{N}$ appearing on the l.h.s~of \eqref{5dintro}  is fully determined by the (refined) topological string partition function on  $X_{N-1}$, see \eqref{J5d} . It contains two main building blocks. The first one is the standard topological string partition function $F_{\rm GV}$, also known as the  Gopakumar-Vafa  function. The second one is  the five dimensional Nekrasov-Shatashvili  function $F_{\rm NS}$. The particular combination of $F_{\rm GV}$ and $F_{\rm NS}$ is chosen to make $\mathsf{J}_{N}$ a well-defined function \cite{hmo2, hmmo}, see  \eqref{J5d} and below.
From the point of view of topological string theory, $F_{\rm GV}$ is the perturbative partition function of the model, while $F_{\rm NS}$ encodes the non-perturbative effects.

The r.h.s of \eqref{5dintro} comes from the spectral theory side of the correspondence. Specifically, ${\rm A}_i^{\rm 5D}$, $i=1,\cdots,N-1,$ are traceclass non-commuting operators on the real line which are obtained via the quantisation of the mirror curve to the $X_{N-1}$ geometry, see \eqref{Ajd}. Since the  mirror curve has genus $N-1$, there are $N-1$ ways of quantization, leading to these $N-1$ non-commuting  operators, each characterized by a discrete spectrum  \cite{Codesido:2015dia}.  We also emphasize that the operators ${\rm A}_i^{\rm 5D}$ depend on the quantization parameter $\hbar$. The latter is related to the  string coupling constant $g_s$  via $\hbar\sim {1\over g_s}$. Hence the TS/ST correspondence is an example of a strong/weak coupling duality.

Equation \eqref{5dintro} has important applications in both topological string and spectral theory. For instance, the interplay between the two sides of the correspondence has provided new insights into the Nakajima-Yoshioka blowup equations \cite{Grassi:2016nnt}, enabling the extension of these equations to a larger class of CY geometries, see e.g.~\cite{Gu:2017ccq,Gu:2018gmy}.
 When $\hbar=2\pi$, interesting simplifications occur, as first noted in \cite{Codesido:2014oua}, and some proofs are presented in   \cite{Doran:2021hcy}.  Other interesting predictions of the TS/ST correspondence concerning the behavior of periods at the conifold point are  demonstrated in  \cite{Angel:2017aa,Doran:2021hcy}.
Some applications to the Hofstadter's butterfly were presented in \cite{Duan:2018dvj,Hatsuda:2016mdw}. 
 Several interesting connections between the TS/ST correspondence and the resurgence program have also been established, see \cite{Couso-Santamaria:2016vwq,Codesido:2017jwp,Gu:2021ize,Gu:2022fss,Gu:2022sqc,Grassi:2022zuk,Alim:2022oll,Alim:2021mhp, Rella:2022bwn}. 
There is also a vast number of applications in spectral theory and relativistic quantum integrable systems, see e.g.~\cite{ Grassi:2014zfa,Wang:2015wdy,Gu:2015pda,Marino:2017gyg, Hatsuda:2015qzx,Franco:2015rnr,Sciarappa:2017hds,Hatsuda:2018lnv}. We refer to \cite{Marino:2015nla} for a review and a more exhaustive list of references.

Despite of  numerous tests, a rigorous mathematical proof of  \eqref{5dintro}  is still missing. Making progress toward proving \eqref{5dintro} is one of the motivations behind the present work.
 The main tool that we will use is the relation between the TS/ST correspondence and ($q$)-isomonodromic deformation equations. This relation was first observed in \cite{Bonelli:2016idi}, and further developed in \cite{Bonelli:2017gdk,Bonelli:2022dse,Nosaka:2020tyv}.  The central player in 
this connection is the so-called Kyiv formula. 
Originally developed in the context of Painlev\'e equations \cite{Gamayun:2012ma,Gamayun:2013auu},   Kyiv formula gives generic solutions to  Painlev\'e  equations in a very explicit form  by using the self-dual Nekrasov functions (or $c=1$ Liouville conformal blocks) as building blocks. In \cite{Gavrylenko:2018ckn,Gavrylenko:2018fsm,Bonelli:2019yjd,Bershtein:2018srt,Bonelli:2022iob,Bonelli:2021rrg},  Kyiv formula was extended to higher-rank isomonodromic equations, such as non-autonomous Toda, see \eqref{eq:taunek}. Further generalizations were obtained in \cite{Bershtein:2016aef,Jimbo:2017ael,Matsuhira:2018qtx} for discrete multiplicative $q$-Painlev\'e equations and in \cite{Bershtein:2018srt} for (non-autonomous) $q$-Toda, see \eqref{tau1804}. Kyiv formula and some of its generalisations were proven in \cite{Bershtein:2014yia,ilte,Gavrylenko:2016zlf, Jeong:2020uxz, DelMonte:2020wty,Bershtein:2018zcz, Shchechkin:2020ryb,Nekrasov:2020qcq}  using various methods.

A strategy to approach \eqref{5dintro} is to show that both sides of the  \eqref{5dintro} satisfy the same $q$-isomonodromic  equation. To begin with, in \autoref{nnteq}, we show that the l.h.s~of \eqref{5dintro} solves the  $q$-Toda system provided we choose some specific initial conditions. More precisely such system is characterised by $N-1$ tau functions. The l.h.s of \eqref{5dintro} is one of such tau functions, the others can be obtained by performing a suitable translation of the argument, see \eqref{5ds}.  When $N=2$, the $q$-Toda system  reduces to  $q$-Painlev\'e $\rm III_3$\footnote{This equation correspond to  $A_7^{(1)'}$ surface type and $A_1^{(1)}$ symmetry type in Sakai's classification \cite {sakai}.} and we recover the result of \cite{Bonelli:2017gdk}. The next step to prove the TS/ST correspondence is to demonstrate  that the r.h.s~of \eqref{5dintro} also satisfies the $q$-Toda equations. However, we will not do so in this work. We will only demostrate \eqref{5dintro} in a specific limit:
the so called "dual" four dimensional limit \cite{Bonelli:2016idi,Bonelli:2017ptp}. From the point of view of the quantum mechanical operators, in this limit, we are sending $\hbar\to \infty$, hence probing an highly quantum regime. It is important to stress that, at the level of the quantum mirror curves, this limit is different from the "standard" four dimensional limit where one takes $\hbar\to 0$ \cite{kkv,Nekrasov:1996cz,Klemm:1996bj}, see \autoref{sec:dual4d}. 
In the language of isomonodromic deformations, the "dual" four dimensional limit corresponds to  the continuous limit. This is when the discrete $q$-Painlev\'e equations reduce to  standard differential Painlev\'e equations. Analogously,  $q$-Toda equations become non-autonomous Toda equations in this limit. Thus, to prove \eqref{5dintro} in this regime, our task is to demonstrate that the resulting r.h.s is the same solution to the non-autonomous Toda equation as the l.h.s.
When $N=2$, it was proven in \cite{Bonelli:2017gdk}. 
In this paper, we will extend these results to generic $N$.
As a summary, we find that in this four dimensional limit, \eqref{5dintro}  becomes 

\be \label{4dintro}\sum_{\boldsymbol{w}\in Q_{N-1}}\frac{T^{\frac12(\boldsymbol{\sigma}+\boldsymbol{w})^2}Z_{\rm inst}^{\rm 4d}(\boldsymbol{\sigma}+\boldsymbol{w},T)}{\prod\limits_{\boldsymbol{\alpha}\in \Delta}G\left(1+(\boldsymbol{\alpha},\boldsymbol{\sigma}+\boldsymbol{w})\right)}
=  \frac{T^{\frac{N^2-1}{24N}}}{N^{1/ 12}\E^{(N^2-1)\zeta'(-1)}\E^{N^2 T^{1\over N}}}\det \left(1+\sum_{k=1}^{N-1} x_k A_k\right),\ee
see equation  \eqref{eq:taunek} and \eqref{mresult}. We leave the definitions and technical details to \autoref{sec:dual4d}. For now, let us briefly explain how to show that the two sides of \eqref{4dintro} are the same solution to the non-autonomous Toda system. 

The l.h.s~of \eqref{4dintro} is the Zak transform of the four dimensional Nekrasov function, see \eqref{eq:taunek}. By using  Kyiv formula  we know that the l.h.s of \eqref{4dintro} solves  the non-autonomous Toda equations with specific initial conditions.
 Note, to fully specify the Toda system, $2(N-1)$ initial conditions are required. In \eqref{4dintro},  the time variable is denoted by $T$ and $\bs \sigma$ parametrizes $N-1$ initial conditions, while the remaining $N-1$  initial conditions are fixed to a specific value, as indicated in \eqref{eta0}. 
 
Now we move to the r.h.s~of \eqref{4dintro}. There is still a generalized Fredholm determinant, but this involves "simpler" operators $A_i$'s compared to $A_i^{\rm 5D}$'s on the r.h.s~of \eqref{5dintro}, see \eqref{Aop}. 
 In addition, the $x_k$ parameters  are related to the initial condition $\bs \sigma$ on the l.h.s via \eqref{mmap}
\be \label{mmapintrp}x_k=\sum_{1\leq i_1<i_2<\cdots<i_k\leq N}\prod_{m=1}^k \E^{2\pi \ri \sigma_{i_m}},\quad k=1,\cdots, { N-1}~.\ee
Equation \eqref{mmapintrp}  is closely related to the so-called mirror map which relates the K\"ahler moduli of $X_{N-1}$ to the complex moduli of its mirror partner, see \autoref{sec:dual4d}. In the context of isomonodromic equations, \eqref{mmapintrp} is the map relating monodromy data in the large  time expansion to monodromy data in the  small time expansion.
In our proof, we take advantage of an existing results of determinant solutions for the non-autonomous Toda system \cite{Tracy:1998aa,Widom:1997aa}. 
More precisely, we can establish an equivalence between the determinant in \eqref{4dintro} and the one  in \cite{Tracy:1998aa,Widom:1997aa}\footnote{
We want to emphasize  that the r.h.s of \eqref{5dintro} provides new solutions to the   $q$-Toda systems \eqref{qtoda} which  are $q$-deformations of the ones constructed by Tracy and Widom  \cite{Tracy:1998aa,Widom:1997aa}. A rigorous proof of this would be equivalent to a rigorous proof of \eqref{5dintro}, which is still lacking at the moment, despite several numerical and analytical tests having been performed, see eg \cite{Codesido:2015dia,Codesido:2016ixn,Doran:2021hcy,Bonelli:2017ptp,Marino:2017gyg,Hatsuda:2015qzx,Grassi:2016nnt,Bonelli:2017gdk,Grassi:2018bci}. }. Thus, we can prove that the r.h.s of \eqref{4dintro} is also a solution to the same non-autonomous Toda system as the l.h.s. Furthermore, we can  compare the initial conditions on the two sides of \eqref{4dintro}
and show that they are identical, see \autoref{sec:proof}. This completes our proof of \eqref{4dintro}.

Another interesting use of \eqref{4dintro} is to study the large time  expansion of the non-autonomous Toda system (i.e.~large $T$). This is the opposite regime   to  the one where Kyiv formula \eqref{eq:taunek}  is defined (i.e.~small $T$). More precisely, the spectral determinant  on the r.h.s of \eqref{4dintro} is exact in $T$, see \eqref{Aop} and \eqref{fdef};  therefore it provides a resummation of the Kyiv formula.
 Even though such resummation is obtained for the initial conditions characterized by \eqref{eta0}; by re-expanding it at large $T$, and using an appropriate analytic continuation, we manage to obtain an explicit expression for the complete $2(N-1)$  family of solutions at generic boundary conditions 
\begin{equation}\small{\ba
\label{anaconint}
\tau_j^{\infty}({\bs x}, {\bs \nu }, r)&= r^{(N^2-1)/12}\re^{\frac{r^2}{16}}\\
&\sum_{\mathbf{M}\in {\IZ}^{N-1}}(\hat{\zeta}^j\boldsymbol{x})^{\mathbf{M}+\boldsymbol{\nu}}\re^{\ri r (\mathbf{M}+\boldsymbol{\nu},\sin \frac{\pi \mathbf{k}}{N})}r^{-\frac12 (\mathbf{M}+\boldsymbol{\nu})^2}\re^{{\ri \pi \over 4}(\mathbf{M}+\boldsymbol{\nu})^2} C(\mathbf{M}+\boldsymbol{\nu})\sum_{\ell=0}^{\infty} \frac{D_\ell^{(N)}(\mathbf{M}+\boldsymbol{\nu})}{(-\ri r)^\ell}\ea}~.
\end{equation}
In the above equation $r=4N  T^{1\over 2N} \ri$ is the time variable and ${\bs x}, {\bs \nu }$ parametrise the boundary conditions. The coefficient $C(\bs{\nu})$ is a product of Barnes functions while $D_\ell^{(N)}(\bs{\nu})$ are polynomials of degree at most $3\ell$  in the $\nu_i$'s,  see discussion around \eqref{myC}, \eqref{eds}  and \eqref{anacon} for more details. In particular, the coefficients $D_\ell^{(N)}$ appearing in such expansion can be interpreted as a strong coupling version of irregular $W_N$ conformal blocks at $c=N-1$, parallel to what was suggested in \cite{Bonelli:2016qwg,Gavrylenko:2020gjb} for $N=2$ (see discussion around \eqref{stN}). The $D_\ell^{(N)}$ coefficients  are  related to the operators \eqref{ajkernel} appearing in the dual four dimensional limit. More precisely the fermionic traces of these operators can be rewritten in the form of $N-1$ multi-cut matrix model, see \eqref{mm4d}. The expansion of these matrix models around their Gaussian points directly give the  strong coupling blocks $D_\ell^{(N)}$, see \autoref{frommm}. In the context of  matrix models, we interpret  the sum in \eqref{anaconint} as a sum over filling fractions, parallel to \cite{Bonnet:2000dz,Eynard:2008yb,Eynard:2008he}. Likewise, in the framework of resurgence, the sum in \eqref{anaconint} can be interpreted in terms of trans-series, see \cite{Grassi:2018spf,Dunne:2019aqp} for related discussion in the case $N=2$.

This paper is structured as follows. In \autoref{smallt} we recall   known facts about the non-autonomous Toda system and its $q$-deformation, including  Kyiv formula for the their tau functions.
In  \autoref{sec:TSSTcor}, we review some aspects of the TS/ST correspondence by focusing  on the examples of the $\rm X_{N-1}$ geometries. 
In \autoref{nnteq}, we relate this duality to $q$-isomonodromic deformation equations. Specifically, we show that the l.h.s~of \eqref{5dintro} solves the $q$-Toda system with a particular choice of initial conditions. In  \autoref{sec:dual4d}, we exploit this connection to demonstrate the TS/ST in a specific regime. In \autoref{stexp}, we use the results of \autoref{sec:dual4d} to bridge small and large time expansion in the non-autonomous Toda system  and to provide an explicit expression for the large time expansion. In \autoref{sec:conc}, we conclude by listing some open problems.  
We have several appendices that serve different purposes.  Appendices \ref{app:groupth} -- \ref{appen:nek} contain definitions used in the main text, while appendices \ref{app:WL} -- \ref{appbilin} contain  some technical details on the  computations presented in the main text.

\section{Non-autonomous Toda equations}\label{smallt}
In this section, we summarize the non-autonomous (or radial) Toda equations and their $q$-deformed generalizations. We also review  Kyiv formula connecting the solutions to the non-autonomous Toda equations with the self-dual $SU(N)$ Nekrasov functions and fix notation.

\subsection{Various forms of Toda equations}

The radial, or non-autonomous Toda system,  is given by the set of coupled ordinary differential equations
\begin{equation}
\label{eq:Toda1}
\partial_{\mathfrak{r}}^2q_j+\frac1{\mathfrak{r}}\partial_{\mathfrak{r}}q_j=-\frac1{4}\re^{q_{j+1}-q_j}+\frac1{4}\re^{q_j-q_{j-1}}, \quad j=0,\ldots, N-1~,
\end{equation}
for some $N\in\mathbb{N}$, with the constraint $q_j=q_{N+j}$.
This equation is a consequence of a little bit more fundamental relation for the tau functions
\begin{equation}
\label{eq:Todatau1}
\partial_{\log\mathfrak{r}}^2\log\tau_j=-\frac{\mathfrak{r}^2}{4}\frac{\tau_{j+1}\tau_{j-1}}{\tau_j^2},
\end{equation}
with the replacements
\begin{equation}
\label{eq:5}
q_j=\log \left( \frac{\tau_j}{\tau_{j-1}} \right)
\end{equation}
where $\tau_j=\tau_{N+j}$.
For \(N=2\), \eqref{eq:Toda1} takes the form
\begin{equation}
\label{eq:1}
\left(\partial_{\mathfrak{r}}^2+\frac1{\mathfrak{r}}\partial_{\mathfrak{r}}\right)(q_1-q_2)=\sinh(q_1-q_2),
\end{equation}
 which is also known as the radial form of the Painlev\'e ${\rm III_3}$ equation. 
 Note that \eqref{eq:1} can also be viewed as the equation of motion of a particle in a \(\cosh\) potential with time-dependent viscous friction.
 So for \(\mathfrak{r}\in\mathbb{R}\), a generic real solution will go to infinity in the limit \(\mathfrak{r}\to\infty\). It's also possible to get a special solution going to zero in the same limit by adjusting the initial velocity.
Such one-parameter family of solutions corresponds to the Fredholm determinant studied in \cite{wu1,zamo,Bonelli:2016idi}.
This is still true for \(N>2\): there is a \((N-1)\)-parameter family of solutions which goes to \(0\) in the limit \(\mathfrak{r}\to\infty\) and those solutions correspond to more general determinants that we will introduce in \autoref{sec:dual4d}.

We can also study real solutions of this equation for purely imaginary \(\mathfrak{r}\), which is conveniently realized by introducing the new variable
\begin{equation}
\label{eq:2}
r=\ri \mathfrak{r}.
\end{equation}
In this case \(N=2\) equation will have the form
\begin{equation}
\label{eq:3}
\left(\partial_r^2+\frac1{r}\partial_r\right)(q_1-q_2)=-\sinh(q_1-q_2),
\end{equation}
which implies that its \(r\to\infty\) real asymptotics will be oscillating. The equation for the tau functions \eqref{eq:Todatau1} will have the obvious form
\begin{equation}
\label{eq:Todatau2}
\partial_{\log r}^2\log\tau_j=\frac{r^2}{4}\frac{\tau_{j+1}\tau_{j-1}}{\tau_j^2},
\end{equation}
and in the same way,
\begin{equation}
\label{eq:Toda2}
\partial_{r}^2q_j+\frac1{r}\partial_rq_j=\frac1{4}\re^{q_{j+1}-q_j}-\frac1{4}\re^{q_j-q_{j-1}}, \quad j=0,\ldots, N-1.
\end{equation}
The variables \(r\) and \(\mathfrak{r}\) are more suitable when studying  the limit \(r\to\infty\), however, it turns out that in the region \(r\to 0\), another variable is useful:
\begin{equation}
\label{eq:6}
T=\left( \frac{\mathfrak{r}}{4N} \right)^{2N}.
\end{equation}
Equation \eqref{eq:Todatau1} can be rewritten as 
\begin{equation}
\label{eq:Todatau3}
\partial_{\log T}^2\log\tau_j=-T^{1/N}\frac{\tau_{j+1}\tau_{j-1}}{\tau_j^2}.
\end{equation}
One can think naively that the ``true'' variable is \(T^{1/N}\), since the transformation \(T^{1/N}\to e^{2\pi i/N}T^{1/N}\) maps \eqref{eq:Todatau3} to seemingly different equation, but this transformation, accompanied by some change of tau variables, is actually a symmetry.
The easiest way to see this is to introduce the new tau variables
\begin{equation}
\label{eq:4}
\tau_j=T^{\frac{j(N-j)}{2N}} \tilde{\tau}_j.
\end{equation}
In these variables equation \eqref{eq:Todatau3} becomes
\begin{equation}
\label{eq:7}
\partial_{\log T}^2\log\tilde{\tau}_j=-T^{\delta_{j,0}}\frac{\tilde{\tau}_{j+1}\tilde{\tau}_{j-1}}{\tilde{\tau}_j^2},
\end{equation}
which is obviously single-valued in \(T\).

\subsection{$q$-deformed  Toda equations}
The non-autonomous Toda equations \eqref{eq:Toda1}, \eqref{eq:Todatau1} can be deformed into the so-called non-autonomous $q$-Toda system \cite[eq.~(3.7)]{Bershtein:2018srt} \cite[eq.~(62)]{Takasaki:2009ik}
\be\label{qtoda}
 {\mathcal T}_j( q z){\mathcal T}_j( q^{-1} z)=  {\mathcal T}_j( z)^2 - z^{1/N} {\mathcal T}_{j+1}( z){\mathcal T}_{j-1}(z)~,\quad j=0, \cdots N-1,\ee
with $ {\mathcal T}_j= {\mathcal T}_{N+j}$.  Sometimes we refer to the variable $z$ as the "time".
When $N=2$ this equation is the tau form of  $q$-Painlev\'e ${\rm III_3}$ corresponding to $A_7^{(1)'}$ surface (and $A_1^{(1)}$ symmetry)  in Sakai classification \cite {sakai}. We refer to \cite{Bershtein:2016aef,Bershtein:2018srt} for more details and references on the subject.
If we scale $z^{1/N}=R^{2} T^{1/N}$ and  $q=\re^{R}$ in \eqref{qtoda} and take the scaling limit $R\to 0$, we recover the equations for the non deformed Toda tau functions \eqref{eq:Todatau3}.

In the rest of the manuscript we will often omit "non-autonomous" and refer to \eqref{qtoda} simply as $q$-Toda system or $q$-Toda equations.

\subsection{The Kyiv formula}\label{sec:Kyiv}
Since the pioneering works \cite{Gamayun:2013auu,Gamayun:2012ma}, it has been known that the small \(T\) expansion for a generic  solution to
\eqref{eq:Todatau3} is given by \cite{Gavrylenko:2015wla,Gavrylenko:2016moe,Bershtein:2018srt,Bonelli:2022iob}
\begin{equation}\label{eq:taunek}
\tau_j(\boldsymbol{\eta},\boldsymbol{\sigma},T)=\sum_{\boldsymbol{w}\in \boldsymbol{\omega}_j+Q_{N-1}}\frac{\re^{2\pi \ri(\boldsymbol{\eta},\boldsymbol{w})}T^{\frac12(\boldsymbol{\sigma}+\boldsymbol{w})^2}}{\prod\limits_{\boldsymbol{\alpha}\in \Delta}G\left(1+(\boldsymbol{\alpha},\boldsymbol{\sigma}+\boldsymbol{w})\right)}Z_{\rm inst}^{\rm 4d}(\boldsymbol{\sigma}+\boldsymbol{w},T),\quad j=0,\cdots,N-1,
\end{equation}
where $G(\cdots)$ in the denominator stands for the Barnes G function and $Z_{\rm inst}^{\rm 4d}({\bs \sigma},T)$ is the self-dual Nekrasov function defined in \eqref{Zinst} (or the irregular $c=N-1$ $W_N$ conformal block). We use  
$\Delta$ for the root system of $SU(N)$, see \eqref{eq:23}, and $\bs \omega_j$ for the fundamental weights of $SU(N)$, see \eqref{eq:omegas}. The root lattice $Q_{N-1}$ is defined in \eqref{eq:22}. The sum over ${\bs \omega}$ in \eqref{eq:taunek} means a sum over all  vectors of the form \be {\bs \omega}={\bs \omega}_j+\sum_{i=1}^{N-1}n_i {\bs \alpha}_i, \quad n_i\in \IZ,\ee
where ${\bs \alpha}_i $ are the simple roots, see \eqref{simpler}. Hence in practice, the sum in \eqref{eq:taunek} is a sum over the $N-1$ integers $n_i$.
 The $2(N-1)$ variables $(\bs \sigma, \bs \eta)$ parametrise the space of initial conditions for \eqref{eq:Todatau3}. More precisely, they can be written in components as
\be \ba  {\bs \sigma}=\sum_{i=1}^N \sigma_i {\bs e_i} , \quad \sum_{i=1}^N \sigma_i=0\qquad \text{and} \qquad
  {\bs \eta}=\sum_{i=1}^N \eta_i {\bs e_i}, \quad \sum_{i=1}^N \eta_i=0 ,
\ea\ee 
where ${\bs e_i}$ are the weights of the fundamental representation, see \eqref{eidef}.  We also note that $Z_{\rm inst}^{\rm 4d}$ in the summand of \eqref{eq:taunek} has poles when $\sigma_i-\sigma_j\in \IZ$. However, when $\bs \eta=\bs 0$, the sum over $\bs \omega$ makes these poles removable\footnote{
This fact is an avatar of a more general phenomenon, namely that the isomonodromic equations have logarithmic solutions.
For example, for \(N=2\), one can consider the limit \(\sigma_1\to 0\), while keeping \(\eta_1/\sigma_1=\gamma\) finite.
The corresponding tau function is a well-defined series in \(t\) and \(\gamma\log t\).
Setting \(\eta_1=0\) from the beginning and then taking the limit \(\sigma_1\to 0\) corresponds to setting \(\gamma=0\) in the logarithmic tau function, and gives a well-defined object.
}.

The formula \eqref{eq:taunek}, as well as the analogous results in \cite{Bonelli:2019boe,Bonelli:2019yjd} for the case of the torus, is usually referred to as the Kyiv formula. 
When $N=2$ this result was proven in \cite{Bershtein:2014yia,ilte,Gavrylenko:2016zlf, Jeong:2020uxz, DelMonte:2020wty}  using various methods.
Several of these methods admit generalisations to  $N>2$ as well, see for instance \cite{Gavrylenko:2018ckn,Gavrylenko:2018fsm}.

A remarkable aspect of the Kyiv formula is that it can also be generalised to the framework of $q$-deformed isomonodromic equations which was first pointed out in  \cite{Bershtein:2016aef}, see also \cite{Jimbo:2017ael,Bershtein:2018zcz,Matsuhira:2018qtx,Bonelli:2020dcp,
Bonelli:2017gdk,Bershtein:2017swf,Semenyakin:2022cri,DelMonte:2022kxh, Mironov:2022xew,Mironov:2019pij,Moriyama:2021mux,Awata:2022idl} for other subsequent works. The generalization to the  $q$-Toda equations was studied in \cite{Bershtein:2018srt} where it was found that the 
generic solutions to such equations can be constructed using the five dimensional $SU(N)$ Nekrasov functions as building blocks. More precisely, the generic solution to \eqref{qtoda}  reads \cite[eq.~(3.4)-(3.6)]{Bershtein:2018srt} \be\label{tau1804} {\mathcal T}_j(\bs{s}, \widetilde{\bs{t}}, z, q)={F(\bs{\widetilde{t}},q,z)}\sum_{\boldsymbol{w}\in \boldsymbol{\omega}_j+Q_{N-1}} {\re}^{(\bs s,\bs w)} {Z}(\bs{\widetilde{t}}+\boldsymbol{w} \log q , q, z)
\ee 
where 
\be \ba  {\bs {\widetilde t}}=\sum_{i=1}^N {\widetilde{\sigma}}_i^{\rm 5d} {\bs e_i} , \quad \sum_{i=1}^N {\widetilde{\sigma}}_i^{\rm 5d} =0\qquad \text{and} \qquad
  {\bs s}=\sum_{i=1}^N{ \eta_i {\bs e_i}}, \quad \sum_{i=1}^N \eta_i=0 ,
\ea\ee 
parametrise the space of initial conditions. Moreover, 
\be\label{z1804}\ba  \\
 {Z}(\bs{\widetilde t}, q, z)=&{\re^{{\log z\over 2 (\log q)^2}\sum_{i=1}^N({\widetilde \sigma}_i^{\rm 5d})^2}\over \prod\limits_{\substack{1\leq i<j\leq N}} (\re^{{\widetilde \sigma}_i^{\rm 5d}-{\widetilde \sigma}_j^{\rm 5d}}q,q,q)_{\infty}(\re^{{\widetilde \sigma}_j^{\rm 5d}-{\widetilde \sigma}_i^{\rm 5d}}q,q,q)_{\infty}}  Z^{\rm 5d}_{\rm inst} (z^{-1},{\bs{\widetilde t}},\ri\epsilon_1,-\ri\epsilon_1)
 \ea \ee
where $q=\re^{\ri \epsilon_1}$ and $Z^{\rm 5d}_{\rm inst}$ is defined in \eqref{Zinst5d}. The function 
$ F({\bs {\widetilde  t}},q,z)$ is a $q$-periodic function in $z$ in the sense that
$ F({\bs{ \widetilde t}},q,qz)F({\bs{ \widetilde t}},q,q^{-1}z)=F({\bs {\widetilde t}},q,z)^2$. Formula \eqref{tau1804} for $N=2$ was proven in \cite{Matsuhira:2018qtx,Bershtein:2018zcz,Shchechkin:2020ryb}.

\subsection{Symmetries of the Kyiv formula and $tt^*$ equations}\label{ttstar}

We can easily check using the definition \eqref{Zinst} that there are the following equalities
\begin{equation}
\label{eq:14}
Z_{\mathrm{inst}}^{\mathrm{4d}}(\boldsymbol{\sigma},T)=Z_{\mathrm{inst}}^{\mathrm{4d}}(-\boldsymbol{\sigma},T), \qquad Z_{\mathrm{inst}}^{\mathrm{4d}}(\boldsymbol{\sigma},T)=Z_{\mathrm{inst}}^{\mathrm{4d}}\left(s(\boldsymbol{\sigma}),T\right),
\end{equation}
where \(s\) is an arbitrary element of the Weyl group (permutation group \(S_N\)).

Other useful identities are, see also \autoref{app:groupth}
\begin{equation}
\label{eq:15}
s \left( \boldsymbol{\omega}_j+Q_{N-1} \right)= \boldsymbol{\omega}_j+Q_{N-1},\qquad
-\left( \boldsymbol{\omega}_j+Q_{N-1} \right)= \boldsymbol{\omega}_{N-j}+Q_{N-1}.
\end{equation}
Using these identities we can check the following relations \eqref{eq:taunek}:
\begin{align}
\label{eq:17}
\tau_j(s(\boldsymbol{\eta}),s(\boldsymbol{\sigma}),T)&=\tau_j(\boldsymbol{\eta},\boldsymbol{\sigma},T),\\
\tau_j(-\boldsymbol{\eta},-\boldsymbol{\sigma},T)&=\tau_{N-j}(\boldsymbol{\eta},\boldsymbol{\sigma},T),\\
\tau_j(\boldsymbol{\eta}+\boldsymbol{\alpha}_i,\boldsymbol{\sigma},T)&=\tau_j(\boldsymbol{\eta},\boldsymbol{\sigma},T),\\
\tau_j(\boldsymbol{\eta},\boldsymbol{\sigma}+\boldsymbol{\alpha}_i,T)&=\re^{-2\pi \ri(\boldsymbol{\eta},\boldsymbol{\alpha}_i)}\tau_j(\boldsymbol{\eta},\boldsymbol{\sigma},T),\\
\tau_j(\boldsymbol{\eta},\boldsymbol{\sigma}+\boldsymbol{\omega}_k,T)&=\re^{-2\pi \ri(\boldsymbol{\eta},\boldsymbol{\omega}_k)}\tau_{j+k}(\boldsymbol{\eta},\boldsymbol{\sigma},T).
\end{align}
In the context of thermodynamic Bethe Ansatz \cite{cfiv}, people  usually consider system \eqref{eq:Toda1} with the extra condition  
\begin{equation}
\label{eq:reflection}
q_j+q_{N-1-j}=0 \quad \Longrightarrow \quad \frac{\tau_j}{\tau_{N-2-j}}=\frac{\tau_{N-1-j}}{\tau_{j-1}}
\end{equation}
and call the resulting system the radial reduction of the $tt^*$ equations \cite{its2,its1}\footnote{
In the case $N=3$ it reduces to Painlev\'e $\rm III_2$.}. We refer to \cite{Bonelli:2022iob} for a recent discussion on the relation between  $tt^*$ equations and Kyiv formula.
To fulfill the extra condition \eqref{eq:reflection}, let us consider the relation
\begin{equation}
\label{eq:83}
\frac{\tau_j(\boldsymbol{\eta},\boldsymbol{\sigma})}{\tau_j(-\boldsymbol{\eta},-\boldsymbol{\sigma}-\boldsymbol{\omega}_2-\boldsymbol{v})}=
\re^{2\pi\ri (\boldsymbol{\eta},\boldsymbol{\omega}_2+\boldsymbol{v})}\frac{\tau_j(\boldsymbol{\eta},\boldsymbol{\sigma})}{\tau_{N-j-2}(\boldsymbol{\eta},\boldsymbol{\sigma})},
\end{equation}
where \(\boldsymbol{v}\in Q_{N-1}\).
We see that \eqref{eq:reflection} can be fulfilled by imposing extra conditions
\begin{equation}
\label{eq:reflection1}
(\boldsymbol{\eta},\boldsymbol{\omega}_2+\boldsymbol{v})=0,\quad s(\boldsymbol{\eta}) + \boldsymbol{\eta}=0, \quad s(\boldsymbol{\sigma}) + \boldsymbol{\sigma} = \boldsymbol{\omega}_2 + \boldsymbol{v},
\end{equation}
where \(s\in S_N\). To find the manifold of solutions of \eqref{eq:reflection1} of maximal possible dimension, we should take \(s\) to be a product of \([N/2]\) transpositions.
 We can choose for simplicity \(s=(1,N)(2,N-2)\ldots\), so \(s^2=1\), and \(s(\boldsymbol{\omega}_2+\boldsymbol{v})=\boldsymbol{\omega}_2+\boldsymbol{v}\).
Equation \eqref{eq:reflection1} can then be solved by
\begin{align}
\label{eq:16}
\boldsymbol{\eta}&=(\eta_1,\eta_2,\ldots ,-\eta_2,-\eta_1),\\
\boldsymbol{\sigma}&=
\left( \frac12-\frac1{N}+\tilde{\sigma}_1,-\frac1{N}+\tilde{\sigma}_2,\ldots , -\frac1{N}-\tilde{\sigma}_2,\frac12-\frac1{N}-\tilde{\sigma}_1 \right).
\end{align}
It is also interesting to see the relation between different symmetric functions of \(\re^{2\pi\ri \sigma_i}\), which will be identified with other variables \(x_k\) \eqref{mmap}. Consider their generating function
\begin{equation}
\label{eq:18}
f(z,\boldsymbol{\sigma})=\sum_{k=0}^Nz^kx_k(\boldsymbol{\sigma})=\prod_{i=1}^N(1+z \re^{2\pi \ri \sigma_k})=z^Nf(z^{-1},-\boldsymbol{\sigma}).
\end{equation}
We know that \(\re^{-2\pi \ri\sigma_k}=\re^{4\pi \ri/N}\re^{2\pi \ri\sigma_{N-k}}\), and therefore \(f(z,\boldsymbol{\sigma})=z^Nf(\re^{4\pi\ri/N}z^{-1},\boldsymbol{\sigma})\),
so
\begin{equation}
\label{eq:19}
x_{N-k}=\re^{4\pi\ri k/N}x_k.
\end{equation}
\section{The TS/ST duality and the $q$-Toda system}
\label{sec:TSST}

In the first part of this section, we  review the TS/ST correspondence in the example of the so-called $X_{\rm N-1}$ CY geometries. These toric Calabi-Yau manifolds are used to geometrically engineer the 4d $\mathcal{N}=2$ $SU(N)$ supersymmetric gauge theories or the 5d $\mathcal{N}=1$ $SU(N)$ supersymmetric gauge theories \cite{kkv,Nekrasov:1996cz,Klemm:1996bj}.
In the second part of this section  we establish a connection between the TS/ST duality on such geometries and the $q$-Toda system \eqref{qtoda}.

\subsection{The TS/ST correspondence}\label{sec:TSSTcor}
 The $X_{\rm N-1}$ CY manifold is the resolution of the cone over the Sasaki--Einstein manifold $Y^{N,0}$.  When $N=2$ , the $X_{1}$ is the well known local $\IF_0$ (or local $\mathbb{P}^1\times\mathbb{P}^1$). Focusing on the symplectic structure, the $X_{\rm N-1}$ geometry has a K\"{a}hler moduli space parametrized by $N$ K\"{a}hler moduli denoted by $t_i$, $i=1,\dots, N$.

By mirror symmetry, we know that the K\"{a}hler moduli space of a CY manifold is identified with the complex moduli space of its mirror CY manifold. Thus the mirror curve to $X_{\rm N-1}$ (see e.g.~\cite{Brini:2009nbd} and reference there)
\be\label{5dmirror} \E^{p}+\E^{-p+(-N+2)x}+\sum_{i=1}^{N-1}\kappa_{N-i}\E^{(i-N+1)x}+\xi\E^{(-N+1)x}+\E^{x}=0,\quad x,p\in \mathbb{C},\ee
is parametrized by the complex moduli $\kappa_i$ with $i=1,\cdots, N-1$. Note in convention, we also refer to $\xi$ as a complex modulus. In this paper, we also introduce the following parameters
\be \label{Hidef}H_{i}=\kappa_{i}\xi^{-{i\over N}}, \qquad \kappa_i=\re^{\mu_i}, \quad i=1,\dots, N-1.\ee
This replacement of complex moduli is convenient for describing the mirror map which relates the complex moduli of a CY manifold to the K\"{a}hler moduli of its mirror CY manifold. More precisely the mirror map takes the following form
 \be\label{5dqmcl} t_i=\sum_{j=1}^{N-1} C_{ij}\log\left(H_j\right)+\mathcal{O}(H_j^{-1}), \quad i=1,\dots, N-1,\ee
 where $C_{ij}$ is the $(N-1)\times (N-1)$ Cartan matrix of $SU(N)$, see \eqref{eq:21}. 
  We also define
 \be \label{TN}t_N=  \log\left( \xi\right). \ee
In the context of the TS/ST correspondence we  promote the mirror map \eqref{5dqmcl} to a quantum mirror map, $t_i (\hbar),\quad i=1,\cdots,N-1$ \cite{acdkv}\footnote{The quantum mirror map differs from the classical mirror map in \eqref{5dqmcl} by terms of the form $\mathcal{O}(H_j^{-1})$. In particular the leading piece $\sum\limits_{j=1}^{N-1} C_{ij}\log\left(H_j\right)$ is the same for both the classical and quantum mirror map.}, where $\hbar$ is the quantization parameter.
Examples of quantum mirror map for $N=2$ and $N=3$ can be found in \eqref{mirrorx2} and \eqref{mirrorsu31} respectively.  
From the point of view of the underlying $SU(N)$ supersymmetric gauge theory that is geometrically engineered by $X_{N-1}$, the quantum mirror map is expressed using  the  NS limit of the Wilson loops in the purely $k$th-antisymmetric representations of $SU(N)$ \cite{Losev:2003py,Bullimore:2014awa,Flume:2004rp,Fucito:2015ofa,Sciarappa:2017hds}, see \cite[pg.~14]{Grassi:2018bci} for details and more references. We summarise some of these results in  \autoref{app:WL}.
It is convenient to define
\be \label{tdef}{\bs t}(\hbar)= \sum_{j=1}^{N-1} t_{j}(\hbar) {\bs \omega}_j\ee
where again ${\bs \omega}_j, ~ j=1,\cdots,N-1$ are the fundamental weights of the $SU(N)$ weight lattice. 

The topological string/spectral theory duality for the $X_{N-1}$ geometry gives the following identity \cite{Codesido:2015dia, Codesido:2016ixn, Bonelli:2017ptp}  
\be
 \label{5d}{\sum_{ {\bs w} \in Q_{N-1}} \exp \left( \mathsf{J}_{N}(\bs{t}(\hbar)+2 \pi \ri  {\bs w}, {t_N}, \hbar) \right)=  \det \left(1+\sum_{i=1}^{N-1} \kappa_i A_i^{\rm 5D}\right)~.}
\ee
In this subsection, our goal is to explain this formula and the quantities appearing in it. We will discuss about proofs and the connection to  $q$-Toda equations in \autoref{nnteq}.  

Appearing on the r.h.s~of \eqref{5d},
the operators $A_j^{\rm 5D}$ are traceclass quantum mechanical operators depending on $\xi$ and $\hbar$. They are defined via the quantization of the mirror curve \eqref{5dmirror}. 
More precisely,
 \be \label{Ajd} {\rm A}_j^{\rm 5D}=\rho_{1, N-2,\xi}{\rm Q}_j,\ee
where
\be {\rm Q}_j=\re^{ -(j-1) \hat x},\ee
and
\be  \label{rad}\ba&  \rho_{1, N-2,\xi}=\left(\re^{\hat p}+\re^{-\hat p +(-N+2)\hat x}+\xi \re^{(-N+1)\hat x} +\re^{\hat  x}\right)^{-1},  \quad [\hat x, \hat p]= \ri \hbar.\ea\ee 
In the definition of ${\rm A}_j^{\rm 5D}$, $\hat x$ and $\hat p$ are the standard position and momentum operators in quantum mechanics satisfying the canonical commutation relation.  We refer to \cite[Sec.~3]{Bonelli:2017ptp} for a more detailed  discussion of the relation between these operators and the quantization of the mirror curve \eqref{5dmirror}.
As demonstrated in \cite[Sec.~3]{Bonelli:2017ptp}, 
the kernel of $A_i^{\rm 5D}$  can be expressed explicitly in terms of the Faddeev's quantum dilogarithm $\fad$:  
\be \label{op5d}A_j^{\rm 5D}(p,p') = \re^{-\ri  \pi b^2( j-1)^2/N^2}\re^{-4\pi ( j-1)b p'/N } \rho_{1, N-2,\xi} (p, p'+\ri {b (j-1)\over N}) , \ee
where $b^2= {N\hbar\over 2\pi }$ and $\rho_{1, N-2,\xi} (p,p') $ is the kernel of \eqref{rad} in the momentum representation. This  reads \cite{Codesido:2016ixn}
\be \label{op5d2}\rho_{1, N-2,\xi} (p,p')=
 {{\overline { f_{\rm 5d}(p)}}  f_{\rm 5d}(p')  \over 2 b \cosh\left( \pi {p-p' \over b}+\frac{\ri \pi  (N-2)}{2 N} \right)} \ee
where \be\label{op5d3}
f_{5d}(x)={\fad(x-{1\over 2 \pi b}\log \xi + \frac{\ri b}{2 N}) \over \fad(x-\frac{\ri b (N-1)}{2 N})} \re^{ \frac{\pi  b (N-1)}{N} x } \re^{ -{1\over 2 N }\log \xi}.\ee
The conventions for the Faddeev's quantum dilogarithm $\fad$ are the same as in \cite{Codesido:2016ixn}.
The grand potential $ \mathsf{J}_{N}$ on the l.h.s~of \eqref{5d} is fully determined using the (refined) topological string partition functions on $X_{\rm N-1}$. It takes the following form \cite[Sec.~5.1]{Grassi:2018bci}
\be\label{J5d}
\ba\mathsf{J}_N(\bs{t}(\hbar),{t_N},\hbar) &=A_N({ t_N}, \hbar)+ F_{\rm p}\left( \frac{2\pi}{\hbar}t_N, \frac{2\pi}{\hbar}\bs{t}(\hbar), \frac{4\pi^2}{\hbar} \right)+ \sum_{i=1}^{N}\frac{t_i(\hbar)}{2\pi}\frac{\partial}{\partial t_i} F_{\rm NS}({t}_N,\bs{t}(\hbar),\hbar) \\
	&+ \frac{\hbar^2}{2\pi}\frac{\partial}{\partial \hbar}
	\left( \frac{F_{\rm NS}({t}_N,\bs{t}(\hbar),\hbar)}{\hbar} \right)+    F_{\rm GV}\left( \frac{2\pi}{\hbar}{t}_N + \pi \ri N,  \frac{2\pi}{\hbar}{\bs t}(\hbar), \frac{4\pi^2}{\hbar} \right). 
\ea
\ee

Specifically, 
\begin{itemize}
\item The parameters $\kappa_i$, $i=1, \cdots N-1$ on the l.h.s~of \eqref{5d} are interpreted  as the complex moduli of the mirror curve \eqref{5dmirror}. They are  related to  $\bs{t}(\hbar)$  on the r.h.s~of \eqref{5d} via the quantum mirror map, as discussed above. Likewise $\xi$ and $t_N$ are related as in \eqref{TN}.  
The shift 
\be {\bs t}(\hbar)+2 \pi \ri {\bs w}, \quad {\bs w}=\sum_{i=1}^{N-1} n_k {\bs \alpha}_k~,\quad n_k \in \IZ\ee
can also be expressed, in a component-wise manner, as
\be\label{tshifts} t_i(\hbar)+ 2\pi \ri \sum_{j=1}^{N-1} C_{ij}n_j~, \quad i=1,\dots ,N-1 .\ee
Via the quantum mirror map, this shift of K\"{a}hler moduli is translated to a shift of complex moduli
\be \mu_i\to \mu_i+2\pi\ri n_i, \quad i=1,\dots ,N-1 \ee
where $\mu_i$'s are defined by
\be\kappa_i=\E^{\mu_i},\quad i=1,\cdots,N-1.\ee

\item $F_{\rm p} (t_N, \bs{t}, g_s)$ is the perturbative part of the free energy; it is a polynomial of degree 3 in $t_i$'s and it is given by  \cite[eq.~(5.6)]{Grassi:2018bci}: \be
\label{fpert}
F_{\rm p} ({t}_N, \bs{t}, g_s)= {1\over 6g_s^2} \sum_{\balpha \in \Delta_+} (\bs{t}, \balpha)^3 + {t_N \over 2 N g_s^2 } \sum_{\balpha \in \Delta_+} (\bs{t}, \balpha)^2 + 
{1\over 6}\left( 1-{4 \pi^2 \over g_s^2} \right) (\bs{t} , {\bs \rho} ),
\ee
where $\balpha \in \Delta_+$ denotes the positive roots of $SU(N)$ and 
\be {\bs \rho}= {1\over 2} \sum_{\balpha \in \Delta_+} \balpha,\ee
see also \autoref{app:groupth}. 
\item The conventional topological string free energy on  $X_{\rm N-1}$  is denoted by $F_{\rm GV} (t_N, \bs{t}, g_s)$. 
For the purpose of our work, it is useful to factor out the $t_N$ independent part $\mathcal{F}_{\rm GV}(\bs{t}, g_s)$ from $F_{\rm GV} (t_N, \bs{t}, g_s)$. So we define $\mathcal{F}_{\rm GV}(\bs{t}, g_s)$ by
 \cite[eq.~(5.7) and eq.~(5.8)]{Grassi:2018bci}
\be \label{splitgv5d}F_{\rm GV}( t_N, \bs{t}, g_s)= \mathcal{F}_{\rm GV}(\bs{t}, g_s)+\mathcal{O}\left(\re^{-t_N} \right),\ee
so
\be\ba
\label{Ffgv}
 \mathcal{F}^{\rm GV} (\bs{t},g_s)=&-2 \sum_{\balpha \in \Delta_+}\sum_{v \ge 1} {1\over v} {1\over 4 \sin^2\left( {g_s v \over 2} \right)} \re^{- v (\balpha, \bs{t})}\\
&= -2  \sum_{\balpha \in \Delta_+} \log ~(\re^{\ri g_s} \re^{-(\balpha, \bs{t})},\re^{\ri g_s},\re^{\ri g_s})_{\infty} .\ea\ee
The remaining $\mathcal{O}\left(\re^{-t_N} \right)$ part of \eqref{splitgv5d} is simply the instanton part of the self-dual Nekrasov free energy 
\be  \label{gv5d}F_{\rm GV}( t_N, \bs{t}, g_s)- \mathcal{F}_{\rm GV}(\bs{t}, g_s)=\log{Z_{\text{inst}}^{\rm 5d}(\re^{t_N}, \bs{t},\ri g_s,-\ri g_s)},  \ee
where $Z_{\text{inst}}^{\rm 5d}$ is defined in \eqref{Zinst5d}.

\item We denote by $F_{\rm NS}(t_N, \bs{t},\hbar)$  the  Nekrasov-Shatashvili partition function  for the five dimensional, $\mathcal{N}=1$ $SU(N)$  SYM theory ($\epsilon_2=0, \epsilon_1=\hbar$) \cite{ns,Nekrasov:2002qd,ikv}. Parallel to \eqref{splitgv5d}, we decompose it to the $t_N$ dependent and $t_N$ independent parts as well  \cite[eq.~(5.7) and eq.~(5.8)]{Grassi:2018bci}
\be\label{5dnssplit} F_{\rm NS} ( {t}_N, \bs{t}, \hbar)= \mathcal{F}_{\rm NS} (\bs{t}, \hbar)  +\mathcal{O}\left(\re^{-t_N} \right) ,\ee
where the $t_N$ independent part $\mathcal{F}_{\rm NS} (\bs{t}, \hbar)$ is
\be
\label{Ffns}
\mathcal{F}_{\rm NS}(\bs{t}, \hbar)=-\sum_{\balpha \in \Delta_+} \sum_{w\ge 1} {1\over w^2} \cot\left( {\hbar w \over 2}\right) \re^{-w (\balpha, \bs{t})} .
\ee
And the $\mathcal{O}\left(\re^{-t_N} \right)$ part is given by 
\be  \label{nsfull}F_{\rm NS}( {t}_N, \bs{t}, \hbar)- \mathcal{F}_{\rm NS}(\bs{t}, \hbar)=\ri \lim_{\epsilon_2\to 0}\epsilon_2 \log{Z_{\text{inst}}^{\rm 5d}(\re^{t_N}, \bs{t},\ri\hbar,\epsilon_2)}  \ee
where $Z_{\text{inst}}^{\rm 5d}$ is the the instanton part of the NS free energy defined in \eqref{Zinst5d}.
\item  The shift by $\ri \pi N$ in the last term of \eqref{J5d} is due to the pole cancellation mechanism (HMO cancellation mechanism \cite{hmo2, hmmo}). It guarantees  that $\mathsf{J}_N(\bs{t}(\hbar),{t_N},\hbar)$ is well defined at any real value of $\hbar$, even though $ F_{\rm NS}$ and $ F_{\rm GV}$ individually are not.  
\item\label{item:an} As the last point, $A_N({ t_N}, \hbar)$ is an overall normalisation constant which includes  the so-called constant map contribution. A closed form expression for $A_N({ t_N}, \hbar)$ is known for $N=2$ \cite{yhum,kmz}. We will derive the $A_N({ t_N}, \hbar)$ for generic $N$  at the end of \autoref{nnteq}.
\end{itemize}

 When the underlying  mirror curve is of genus one, the vanishing of the Fredholm determinant determines the spectrum of a corresponding two-particle relativistic quantum integrable system, see \cite{Grassi:2014zfa}. 
Analogously, the mirror curve of genus $N-1$ has a natural connection with  $N$-particle relativistic quantum integrable system. In this case,
the vanishing of the generalised Fredholm determinant is obviously not enough to determine the full spectrum of the integrable system, since the latter requires $N-1$ quantization conditions. Nevertheless, it was pointed out in \cite{Hatsuda:2015qzx, Franco:2015rnr,Sun:2016obh} that these $N-1$ quantization conditions can still be obtained from the vanishing of a single generalised Fredholm determinant, provided that we rotate the moduli $\kappa_i$  appropriately for each of them. This is done by introducing the so-called ${\bs r}$-fields which add a phase to the complex moduli $\kappa_i$, see \cite[eq.~(1.4)]{Sun:2016obh}.  It was shown in \cite{Grassi:2016nnt} that these shifts are intrinsically related to Nakajima-Yoshioka blowup equations.    

More precisely for the $X_{N-1}$ geometry we have 
\be
 \label{5dj}\sum_{ {\bs w} \in Q_{N-1}} \exp \left( \mathsf{J}_{N}(\bs{t}+2 \pi \ri  {\bs w}+ \pi \ri \bs{r}^{(j,d)}, {t_N+{\ri \pi r_N^{(j,d)}}}, \hbar) \right)  =\det \left(1+\sum_{\ell=1}^{N-1}~ {\kappa}_\ell^{(j,d)} A_\ell^{\rm 5D}\right),
\ee
where  \cite{Grassi:2016nnt}
\be  \bs{r}^{(j,d)}=\sum_{i=1}^{N-1} r_i^{(j,d)} {\bs \omega_i},\ee
and
\be\label{rfield}
r_i^{(j,d)} = \begin{cases}
0 \ , & i\leq N-2 \\
2j \ , & i=N-1 \\
N  - 2d \ , & i=N\ ,\end{cases}
\ee
for $0< j, d< N$. 
By using
$ (\bs C^{-1})_{\ell,N-1}=\ell/N $, this means that the effect of such $\bs r$ fields is to rotates $\kappa_\ell$ by 
\be \label{kappash} \kappa_{\ell}\to  \kappa_\ell^{(j,d)}=\re^{2 \ri \pi j  \ell/N } \re^{- \ri \pi (N-2d) \ell/N }\kappa_{\ell}. \ee
 It was shown in \cite{Grassi:2016nnt}  that there are $(N-1)^2$ inequivalent choices for the $\bs r$ fields. As we will discuss later, some of these choices correspond to the shifts in the $q$-Toda tau functions, $\mathcal{T}_j$'s ,as compared to $\mathcal{T}_0$.
For this purpose, it is convenient to rewrite \eqref{5d} as
\be
 \label{5ds}\boxed{\sum_{ {\bs \omega} \in {\bs \omega}_j+ Q_{N-1}} \exp \left( \mathsf{J}_{N}(\bs{t}(\hbar)+2 \pi \ri  {\bs w}, {t_N}, \hbar) \right)=  \det \left(1+\sum_{\ell=1}^{N-1} \re^{-2\ri \pi j \ell/N} \kappa_\ell A_\ell^{\rm 5D}\right) ~,\\}
\ee
where we used  \eqref{Hidef} and \eqref{5dqmcl} to show that
\be \label{shiwi}\bs{t}(\hbar)\to  \bs{t}(\hbar) + 2 \pi \ri {\bs \omega}_j\ee 
 is equivalent to
\be \kappa_{\ell}\to \re^{-2\ri \pi j \ell/N}\kappa_{\ell} .\ee
Note that this shift is of the same type as  \eqref{kappash}\footnote{This corresponds to a specific choice of $(N-1)$ $\bs r$-fields among the $(N-1)^2$ possibilities. }.

\subsection{The relation to $q$-Toda equations}\label{nnteq}

When $N=2$, it was found  in \cite{Bonelli:2017gdk} that \eqref{5d} solves the $q$-Painlev\'e ${\rm III_3}$ equation in the tau form if we choose the initial condition in a specific way.
Remarkably, this implies that the Fredholm determinant of the quantum mirror curve to local $\IF_0$ is a solution  to $q$-Painlev\'e ${\rm III_3}$.
In \cite{Bonelli:2017gdk}, it was also discussed that this Fredholm determinant can be seen as a $q$-deformation of the Fredholm determinant solution to Painlev\'e ${\rm III_3}$ obtained in  \cite{wu1} within the framework of the 2d Ising model.

 The result of \cite{Bonelli:2017gdk} was generalised to other $q$-Painlev\'e equations and other CY geometries in \cite{Bonelli:2022dse}. 
 We will now see how to generalise these results to  $q$-Toda equations, which correspond to $q$-isomonodromic problems of higher rank \footnote{The statement that $q$-Toda equations correspond to $q$-isomonodromic problems of higher rank was discussed in some talks and still unpublished notes for a long time, but the idea is very simple: we can deautonomize \cite{Bershtein:2018srt} the Lax pairs from \cite{Fock:2014ifa}. We hope it will be published in the near future.}. In particular we show that the l.h.s~of \eqref{5ds} solves \eqref{qtoda}. 
 In the view of a mathematical proof of the TS/ST correspondence, it would be important to demonstrate  that spectral determinant appearing on the r.h.s~of \eqref{5ds} solves \eqref{qtoda} as well. 

To show that the l.h.s~of \eqref{5ds} solves \eqref{qtoda} it is useful to first introduce a few definitions.
We denote the elliptic Gamma function by
 \be\label{elliptg} \Gamma(X,t,q)= {  (X^{-1} t q, t, q)_{\infty} \over  (X,t , q)_{\infty}},
\ee 
where \be   (X , t, q)_{\infty}=\prod_{i,j=0}^{\infty}\left(1-Xq^i t^j\right).\ee
One can easily show that \eqref{elliptg} satisfies \be  \label{gammd} {\Gamma(Xq,q,q) \Gamma(Xq^{-1},q,q)\over \Gamma(X,q,q)^2}=- q X^{-1}.\ee
Using the special functions reviewed above, we further make 2 definitions which will be used to rewrite the l.h.s~of \eqref{5ds}. We define 
\be \ba \Delta_1(x, \hbar)=& {\re^{-{1\over 12 \pi \hbar} (x)^3}\re^{-{\ri\over 8 \pi }(x)^2}\over  \Gamma(\re^{-{2 \pi \over \hbar}x}  q,q,q)}~,\ea \ee
and 
\be\ba  \log \Delta_2( \bs{t}(\hbar),{t}_N,\hbar)=& \frac{\hbar^2}{2\pi}\frac{\partial}{\partial \hbar}\left( \frac{F_{\rm NS}({t}_N, \bs{t}(\hbar),\hbar)}{\hbar} \right) +\sum_{i=1}^{N}\frac{t_i(\hbar)}{2\pi}\frac{\partial}{\partial t_i} F_{\rm NS}({t}_N, \bs{t}(\hbar),\hbar)
\\
&+{1\over 6}\left( 1-{\hbar^2 \over 4 \pi^2} \right){2\pi \over \hbar} (\bs{t} (\hbar), {\bs \rho} ).\ea\ee
The functions $\Delta_1\left(\left(\bs{t}(\hbar) , {\bs \alpha}\right), \hbar\right)$ and $\Delta_2\left( \bs{t}(\hbar),{t}_N,\hbar\right)$ have some nice properties which enable us to rewrite the l.h.s~of \eqref{5ds} in a from which is suitable to compare with \eqref{tau1804}, \eqref{z1804}. By using \eqref{gammd}  we get
\be \label{recurdelta}{\Delta_1(x-2\pi \ri, \hbar) \Delta_1(x+2\pi \ri, \hbar)\over \Delta_1(x, \hbar)^2 }=1~.\ee
Applying \eqref{recurdelta} recursively, we further obtain
\be\label{deltashift} \ba  \prod_{{\bs \alpha}\in \Delta_+}  {\Delta_1\left(\left(\bs{t}(\hbar) +2 \pi \ri {\bs w},{\bs \alpha}\right) , \hbar\right)\over \Delta_1\left(\left(\bs{t}(\hbar) , {\bs \alpha}\right), \hbar\right)}=
&\prod_{{\bs \alpha}\in \Delta_+}\left({\Delta_1\left(\left({\bs{t}(\hbar) }, {\bs \alpha}\right), \hbar\right)\over \Delta_1\left(\left(\bs{t}(\hbar) , {\bs \alpha}\right)-2 \pi \ri, \hbar\right)}\right)^{( \bs{w},{\bs \alpha} )}, 
\ea\ee 
where \be \bs{w}=\sum_{k=1}^{N-1} n_k {\bs \alpha}_k, \quad n_k\in \IZ .\ee 
It is also straight forward to verify that
\be\label{delta2s} \Delta_2\left(\bs{t}(\hbar), {t_N}-2\pi \ri, \hbar\right)  \Delta_2\left(\bs{t}(\hbar), {t_N}+2\pi \ri, \hbar\right) = \Delta_2\left(\bs{t}(\hbar), {\xi}, \hbar\right) ^2. \ee

We already summarize in \autoref{sec:Kyiv} that \eqref{tau1804} solves the  $q$-Toda system \eqref{qtoda}. Using $\Delta_1((\bs{t}(\hbar) , {\bs \alpha}), \hbar)$ and $\Delta_2( \bs{t}(\hbar),{t}_N,\hbar)$ that we have introduced and their properties \eqref{delta2s} and \eqref{deltashift}, we can recast the l.h.s~of \eqref{5ds} into the ${\mathcal T}_j(\bs{s}_0, \widetilde{\bs{t}}, z, q)$ function defined in \eqref{tau1804}. We find
\be\label{detjt}\boxed{  \sum_{\bs \omega\in{\bs \omega}_j +Q_{N-1} }\exp \left( \mathsf{J}_{N}(\bs{t}(\hbar)+2 \pi \ri  {\bs \omega},t_N, \hbar) \right) = \re^{A_N(t_N, \hbar)}  {\mathcal T}_j(\bs{s}_0, \widetilde{\bs{t}}, z, q) } \ee
where \(A_N(h_N,\hbar)\) is given by \eqref{ANg},
\be\label{dic}
 \ba 
q&=\re^{\ri 4\pi^2\over \hbar},\\
z &=\re^{-\ri \pi N }\re^{-{2\pi \over \hbar} t_N},\\
\widetilde{\bs{t}}&={2\pi \over \hbar}\bs{t}(\hbar) ,\ea\ee
 \be \label{s0d}{\bs s}_0={\left( 1-{\hbar^2 \over 4 \pi^2} \right){2\pi^2 \ri \over 3 \hbar}{\bs \rho} }+\left( \sum_{k=1}^{N-1}{  \ri \frac{ \partial}{\partial { t_k}} F_{\rm NS}({t}_N, \bs{t}(\hbar),\hbar){\bs \alpha}_k}\right) +\sum_{{\bs \alpha}\in \Delta_+}{\bs \alpha}\log\left( {\Delta_1(({\bs{t}(\hbar)}, {\bs \alpha}), \hbar)\over \Delta_1(({\bs{t}(\hbar)}, {\bs \alpha})-2 \pi \ri, \hbar)}\right)~ \ee
and
\be \label{fdefs}F(\widetilde{\bs{t}}, q,z )= \Delta_2(\bs{t}(\hbar), {\xi}, \hbar) \left(\prod_{ {\bs \alpha}\in \Delta_+}\Delta_1((\bs{t} (\hbar), {\bs \alpha}), \hbar) \right).\ee
So $\sum\limits_{\bs \omega\in{\bs \omega}_j +Q_{N-1} }\exp \left( \mathsf{J}_{N}(\bs{t}(\hbar)+2 \pi \ri  {\bs \omega},t_N, \hbar) \right) $ appearing on the l.h.s~of \eqref{detjt} is a solution to the  $q$-Toda equation in the tau form, provided that we fix the initial conditions ${\bs s}$ in  \eqref{tau1804} to be \eqref{s0d} and the overall periodic function $F(\widetilde{\bs{t}}, q,z )$ in  \eqref{tau1804} to be \eqref{fdefs}. 

Furthermore, if we assume the TS/ST correspondence \eqref{5ds},   it follows from  \eqref{detjt}   that\be\label{dettoda}  \re^{-A_N(t_N, \hbar)}  \det \left(1+\sum_{\ell=1}^{N-1} \re^{-2 \ri \pi j \ell/N}\kappa_{\ell}A_{\ell}^{\rm 5D}\right) \ee
 also solves the  $q$-Toda equations, where the initial conditions are parametrised by $\kappa_i$, the parameter $t_N$ is the time and $q=\re^{4\pi^2\ri\over \hbar}$. Recall that $A_{\ell}^{\rm 5D}$ depends both on $t_N$ and $q$, see \eqref{op5d}-\eqref{op5d3}.

In particular, we identify $\re^{A_N(t_N, \hbar)}$ with the so-called $q$-"algebraic"\footnote{$A_N(t_N, \hbar)$ itself is not algebraic. Nevertheless it can be viewed as  the $q$-deformation of algebraic solution to non-autonomous Toda, see \autoref{4danl}.} solution of the  $q$-Toda equations \eqref{qtoda}. As mentioned in \autoref{sec:TSSTcor}, for generic $N$, the TS/ST correspondence does not give a closed form expression for this quantity. 
However, from our analysis above, we know that $\re^{A_N(t_N, \hbar)}$ corresponds to  a $q$-"algebraic" solution of the $q$-Toda equations (up to possible $q$-periodic functions). So we can make an educated guess for it.
More precisely, we propose that 
\be\label{ANg} \re^{A_N(t_N, \hbar)}={ \re^{{N\over 2}A_c\left(\frac{\hbar }{\pi }\right)-\frac{1}{2} A_c\left(\frac{N \hbar }{\pi }\right)+{ \left(N^2-1\right)}\frac{\pi }{12 N \hbar }t_N}  \over Z_{\rm coni}\left(\frac{2\pi}{ N\hbar}t_N,  \hbar N\right) Z_{\rm coni}^{\rm np}\left(\frac{2 \pi}{N \hbar }t_N,{4 \pi^2\over N \hbar}\right)}, \ee
where the function $A_c(\hbar)$ was introduced in  \cite{Hatsuda:2014vsa} and reads
\be 
A_{\rm c}(k)= \frac{2\zeta(3)}{\pi^2 k}\left(1-\frac{k^3}{16}\right)
+\frac{k^2}{\pi^2} \int_0^\infty \frac{x}{\re^{k x}-1}\log(1-\re^{-2x})\rd x.
\ee
Moreover,   \be\label{zcono} \ba  Z_{\rm coni}(t,\hbar)=&
 \left(-\re^{-t}\re^{\ri {4\pi^2\over \hbar}}, \re^{\ri {4\pi^2\over \hbar}},\re^{\ri {4\pi^2\over \hbar}} \right)_{\infty},\\
 Z_{\rm coni}^{\rm np}(t,{g_s})=& \exp\left[{ {1\over 2 \pi \ri}{\partial\over \partial g_s}\left( g_s \mathcal{F}_{\rm NS}^{\rm coni}({2\pi \over g_s}, {2 \pi t \over g_s}\right)}\right]
, \ea\ee
 where 
 \be  \mathcal{F}_{\rm NS}^{\rm coni}(g_s, t )={1\over 2 \ri}\sum_{\ell\geq 1}{1\over \ell^2\sin (\ell g_s)}\re^{-\ell t}.\ee
When $N=2$, \eqref{ANg} agrees with the known expression in \cite{yhum,kmz,Bonelli:2017gdk}. \footnote{The polynomial part in $\log \xi$ here is different from what we have in \cite{Bonelli:2017gdk} because we have a different definition of $F_p$ in \eqref{fpert}. In particular $t\Big|_{\rm there}= (t+\log \xi)\Big|_{\rm here}$}

To be more precise, our proposal \eqref{ANg} is made based on requiring the following good properties:
\begin{enumerate}
\item It satisfies 
\be\label{als} {\re^{-A_N(t_N +2 \pi \ri, \hbar)} \re^{-A_N(t_N -2 \pi \ri, \hbar)}\over \re^{-2A_N(t_N, \hbar)}} =1+\re^{-{2 \pi \over \hbar N}t_N}=1-\re^{-{2 \pi \over \hbar N}(t_N+\ri \hbar N/2)} = 1- z^{1/N}.\ee
Hence it is indeed a $q$-"algebraic" solution to the  $q$-Toda equations \eqref{qtoda}.
\item It is well defined at $q=1$ and the HMO cancellation mechanism \cite{hmo2} is satisfied. This is a necessary requirement to make contact with the spectral determinant and the TS/ST correspondence.
\item In the dual four dimensional limit,  \eqref{ANg} reproduces the correct results. This point will be made clear in \autoref{sec:dual4d}, where we discuss this limit from the expectation of the TS/ST correspondence, see also \autoref{app:oneloop}.
\item For $N=2$ it reproduces the known result.
\end{enumerate}
From the point of view of the TS/ST correspondence, one should test the proposal \eqref{ANg} more in details, e.g.~numerically.  We would like to stress that the arguments presented above also give a good guideline  to determine this type of normalisation constants for other toric CY manifolds. It would be interesting to explore this further. However, we will leave them to future work.

\section{The dual four dimensional limit}\label{sec:dual4d}

Ideally, we would like to give a rigorous  proof of \eqref{5ds} in full generality. 
As the first step toward a full proof,  we  demostrate  \eqref{5ds} in a special regime. This is the so called   {\it{dual}} four dimensional limit introduced in \cite{Bonelli:2016idi,Bonelli:2017ptp}. This means that we scale the parameters in the following way: 
 \be\label{dual4dcomplex}   \ba & \hbar={  1 \over { \epsilon}  \beta}, \quad \log \xi ={1 \over 2\pi \beta { \epsilon} } \left(a { \epsilon}  \beta -\log(\beta^{2N}T) \right),  \\
 &\log \kappa_j=-  { j \over 2 \beta { \epsilon}  \pi N} \log \left(\beta^{2N} T\right)+\log  \left(x_j\right) +{j a\over 2 \pi N}\ea\ee
 and take 
 \be \beta \to 0. \ee
The parameter $\epsilon$ can be reabsorbed into a redefinition of other parameters. Hence we can set without loss of generality $\epsilon={1\over 4\pi^2}$. Let us stress that, at the level of the mirror curve, this limit is {\it not} the standard four dimensional limit \cite{kkv, kkv,Nekrasov:1996cz,Klemm:1996bj}. In the standard four dimensional limit one take
\be \hbar= \beta , \quad \log \xi=a\beta -\log (\beta^{2N} T), \quad \log \kappa_j= -  { j \over N} \log \left(\beta^{2N} T\right)+\mathcal{O}(\beta^0) \ee
and send $\beta \to 0$. 
In such  limit, the quantum mirror curve becomes the Baxter equation for the $SU(N)$ quantum Toda system, see \cite{Grassi:2018bci} and reference there.
 Instead, in the dual four dimensional limit, the relevant operators that we obtain are the ones in equation \eqref{Aop}. 
 
 Let us take $N=2$ as an example.
In the standard four dimensional limit, we obtain the modified Mathieu operator 
 \be \hat p^2 +T^{1/2} \cosh \hat x .\ee
 In the dual four dimensional limit,  we obtain the following difference operator \cite{Bonelli:2016idi}
 \be \label{n24d}  \re^{-4 T^{1\over 4} \cosh (\hat x)}\frac{1}{2\cosh(\frac{\hat{p}}{2})}  \re^{-4 T^{1\over 4} \cosh (\hat x)}  .\ee

 \subsection{The result}
 In the $\beta\rightarrow 0$ limit, with the scaling specified by \eqref{dual4dcomplex}, the equation \eqref{5ds} becomes
 
 \be \label{mresult}\boxed{\sum_{\boldsymbol{w}\in \boldsymbol{\omega}_j+Q_{N-1}}\frac{T^{\frac12(\boldsymbol{\sigma}+\boldsymbol{w})^2}Z_{\rm inst}^{\rm 4d}(\boldsymbol{\sigma}+\boldsymbol{w},T)}{\prod\limits_{\boldsymbol{\alpha}\in \Delta}G\left(1+(\boldsymbol{\alpha},\boldsymbol{\sigma}+\boldsymbol{w})\right)}
=  \frac{T^{\frac{N^2-1}{24N}}}{N^{1/ 12}\E^{(N^2-1)\zeta'(-1)}\E^{N^2 T^{1\over N}}}\det \left(1+\sum_{k=1}^{N-1} {\rm e}^{-2\pi \ri k  j/N}x_k A_k\right)}\ee
where $ j=0,\cdots,N-1$ and
 $A_k$'s are the following  $N-1$ non-commuting traceclass operators  on the real line
   \be\label{Aop}\ba A_k= \E^{\frac{2k-N}{2 N}\hat{p}} f(\hat{x})\frac{1}{2\cosh(\frac{\hat{p}}{2})}f(\hat{x}), \quad k=1,\cdots,N-1~,
\ea\ee
where  $[\hat x, \hat p]= 2 \pi \ri$ and
\be \label{fdef}f (x) = \exp \left(-2 N T^{1\over 2N} \cosh (x)\right) . \\
\ee
 The kernel of $A_k$ is 
 \be\label{ajkernel} A_k(x,y)=\frac{f\left(x+\frac{\ri\pi(2k-N)}{N}\right)f\left(y\right)}{4\pi\cosh\left(\frac{x-y}{2}+\frac{\ri\pi(2k-N)}{2N}\right)}.\ee
Component-wise, the parameters ${\sigma}_i$'s and  $x_j$'s appearing on the two sides of \eqref{mresult} are related via 
\be \label{mmap}x_k=\sum_{1\leq i_1<i_2<\cdots<i_k\leq N}\prod_{m=1}^k \E^{2\pi \ri \sigma_{i_m}},\quad k=1,\cdots, { N-1}~,\ee
where we also recall that \be\label{recsum} \sum_{i=1}^N \sigma_i=0~.\ee
Equation \eqref{mmap} can be understood as the  map that links monodromy data of the short time expansions ($T$ is small - $\sigma_i$) to  monodromy data of the large time expansions ($T$ is large - $\rm x_i$). Likewise,  \eqref{mresult} presents a solution to the Toda connection problem in a case where half of the monodromy parameters are fixed to the specific value   
\be\label{eta0} \bs{\eta}=\bs 0.\ee
In \autoref{sec:TS/ST}, we show explicitly the derivation of \eqref{mresult}
from \eqref{5ds}. In \autoref{sec:proof}, assuming Kyiv formula \eqref{eq:taunek}, we provide a proof for \eqref{mresult}.

\subsection{Derivation from TS/ST}
\label{sec:TS/ST}
In this section, we show explicitly that  \eqref{5ds} reduces to \eqref{mresult} after implementing the limit $\beta\rightarrow 0$, where the 5d quantities in \eqref{5ds} are scaled as described in \eqref{dual4dcomplex}.
We first restrict ourselves to the $j=0$ case of \eqref{mresult}. 
The generalization to \eqref{mresult} is then straight forward and we will discuss it at the end of this section. 

Let us start with the r.h.s of \eqref{5ds} with $j=0$. In \cite{Bonelli:2017ptp}, the dual 4d limit was implemented at the level of the spectral determinant and it was found that 
\be\label{limitdet} \det \left(1+\sum_{i=1}^{N-1} \kappa_i A_i^{\rm 5D}\right)~\xrightarrow[~\eqref{dual4dcomplex}~]{\beta\rightarrow 0}~ \det \left(1+\sum_{i=1}^{N-1} x_i A_i\right) \ee
where $A_i$ are as in \eqref{Aop}.
The motivation of \cite{Bonelli:2017ptp} for using the spectral determinants was that these determinants can be expressed using matrix models which capture the strong coupling regime of the four dimensional $\mathcal{N}=2$ $SU(N)$ gauge theories near the magnetic point. 
This is also an important reason why we are interested in \eqref{mresult}. This point will be made clear in \autoref{stexp}, where we use the corresponding matrix models to provide the large time solutions to the non-autonomous Toda equations.
In \cite{Bonelli:2017ptp}  the connection between the determinant on the r.h.s of \eqref{limitdet} and Toda equations was observed for the so-called "one-period phase"; that is when $x_i=0$ for $i=1,\cdots, N-2$ and $x_{N-1}\neq 0$. In \autoref{sec:proof} we will extend this connection to generic values of the parameters $x_i$.

%
Let us now implement the limit on the l.h.s~of \eqref{5ds} with $j=0$. Our approach to take the dual 4d limit is mostly parallel to what was done for $N=2$ in \cite{Bonelli:2016idi}, and thus we are rather brief, focusing mainly on showing the key steps. More details can be found in \autoref{app:oneloop}. We first note that, by
using the explicit expression of the quantum mirror map coming from 5d Wilson loops (see  \autoref{app:WL}),  in the dual 4d limit $\beta\rightarrow 0$ with the parametrizaiton \eqref{dual4dcomplex},
the quantum K\"{a}hler parameters $t_i(\hbar)$'s become constant. We denote their limits in terms of $\sigma_i$'s, which are defined in the following way:
\be\label{ti4} t_i(\hbar)~\xrightarrow[~\eqref{dual4dcomplex}~]{\beta\rightarrow 0}~   {2 \pi \ri} (\sigma_i-\sigma_{i+1}), \quad {i=1,\dots N-1}. \ee
Furthermore, using \autoref{app:WL}, it is easy to see that in this limit, the quantum mirror map simply reduces to the following relation
\be\label{4dmm} x_k=\sum_{1\leq i_1<i_2<\cdots<i_k\leq N}\prod_{m=1}^k \E^{2\pi\ri\sigma_{i_m}},\quad k=1,\cdots,  N-1.\ee
Equation \eqref{4dmm} is due to the fact that, in this limit, the instantons corrections in the Wilson loops \eqref{WKinst} vanish, because 
\be \label{tn4}
 t_N\approx-{2 \pi \over \beta} \log (\beta^{2N} T)\to \infty.\\
\ee
It follows from \eqref{ti4} that, after taking the limit $\beta\rightarrow 0$ with the parametrization \eqref{dual4dcomplex}, the shifts by integers on the l.h.s~of \eqref{5d} are mapped to shifts of $\sigma_i$'s 
\be   \label{shiftalpha}\sigma_i-\sigma_{i+1}~\to~\sigma_i-\sigma_{i+1}+\sum_{j=1}^{N-1}C_{i j}n_j.\ee

Moreover, it can be readily verified that
the specified scaling in \eqref{ti4} and \eqref{tn4}  implies that the only part of $F_{\rm NS}$  that survives in this regime is the $t_N$ independent part, namely $ \mathcal{F}_{\rm NS} (\bs{t}, \hbar)
$ in \eqref{5dnssplit}. In brief, as part of the grand potential \eqref{J5d}, this piece  combined with  the $t_N$ independent part coming from \eqref{splitgv5d}, namely $ \mathcal{F}_{\rm GV}(\frac{2\pi}{\hbar}\bs{t}, \frac{4\pi^2}{\hbar})
$,  
is responsible for the two Barnes functions appearing in the summand of the tau function \eqref{eq:taunek}. More details can be found in  \autoref{app:oneloop}.

Besides, the $t_N$ dependent part of  $F_{\rm GV}\left( \frac{2\pi}{\hbar}{t}_N+\ri \pi N, \bs{t}(\hbar) ,  \frac{4\pi^2}{\hbar} \right)$ in this limit  gives the instanton part of the four dimensional Nekrasov function in \eqref{eq:taunek}. Indeed, even if ${t}_N$, $\bs{t}(\hbar)$ and $\hbar$ do not scale as in the usual geometric limit, the rescaled variables \be \frac{2\pi}{\hbar}{t}_N +\ri \pi N,\quad \frac{2\pi}{\hbar} \bs{t}(\hbar), \quad \frac{4\pi^2}{\hbar}  \ee
have the typical scaling behaviour that we encounter in the geometric engineering construction \cite{kkv}.

The polynomial part $F_p$ and  $A_N({ t_N}, \hbar)$ give rise to the overall factor
\be N^{1/ 12}\E^{(N^2-1)\zeta'(-1)}\E^{N^2 T^{1\over N}}T^{-\frac{N^2-1}{24 N}} \ee
as well as the perturbative part $ T^{{1\over 2}(\bs \sigma+\bm n)^2}$  in the summand of \eqref{eq:taunek}. The details are shown in \autoref{app:oneloop}.

In summary, we get \eqref{limitdet} for the r.h.s of \eqref{5d} and
\be \ba \sum_{ {\bs w} \in Q_{N-1}} \exp \left( \mathsf{J}_{N}(\bs{t}(\hbar)+2 \pi \ri  {\bs w}, {t_N}, \hbar) \right) ~\xrightarrow[~\eqref{dual4dcomplex}~]{\beta\rightarrow 0}&~ N^{1/ 12}\E^{(N^2-1)\zeta'(-1)}\E^{N^2 T^{1/N}}T^{-\frac{N^2-1}{12}}\\
\times&\sum_{\boldsymbol{w}\in Q_{N-1}}\frac{T^{\frac12(\boldsymbol{\sigma}+\boldsymbol{w})^2}Z_{\rm inst}^{\rm 4d}(\boldsymbol{\sigma}+\boldsymbol{w},T)}{\prod\limits_{\boldsymbol{\alpha}\in \Delta}G\left(1+(\boldsymbol{\alpha},\boldsymbol{\sigma}+\boldsymbol{w})\right)} \ea \ee 
for the l.h.s.
Hence the dual 4d limit of \eqref{5d} can be rewritten as

 \be \label{mresult0}{\sum_{\boldsymbol{w}\in Q_{N-1}}\frac{T^{\frac12(\boldsymbol{\sigma}+\boldsymbol{w})^2}Z_{\rm inst}^{\rm 4d}(\boldsymbol{\sigma}+\boldsymbol{w},T)}{\prod\limits_{\boldsymbol{\alpha}\in \Delta}G\left(1+(\boldsymbol{\alpha},\boldsymbol{\sigma}+\boldsymbol{w})\right)}
=  \frac{T^{\frac{N^2-1}{24N}}}{N^{1/ 12}\E^{(N^2-1)\zeta'(-1)}\E^{N^2 T^{1\over N}}}\det \left(1+\sum_{k=1}^{N-1} x_k A_k\right)}~.\ee
In particular, the maps between monodromies   \eqref{mmap} relating parameters on the two sides of \eqref{mresult0} are nothing but the dual 4d limit of the quantum mirror maps ${t}_i(\hbar)$'s.

Having \eqref{mresult0} in mind, we now show how to get the main result \eqref{mresult}. Using \eqref{eq:taunek},
we have $\tau_j(\bs{0}, \bs{\sigma} , T)=\tau_0(\bs{0}, \bs{\sigma}+\bs \omega_j, T)$.
According to the  maps \eqref{mmap}, it is straight forward to show that shifting $\bs \sigma\rightarrow \bs{\sigma}+\bs \omega_j$
is equivalent to shifting $x_k\rightarrow x_k{\rm e}^{-2\pi \ri k  j/N},\ \forall k$. Hence, by applying the shift of $\bs \sigma$ and the equivalent shifts of $x_k$'s on the l.h.s and r.h.s, \eqref{mresult0} trivially implies \eqref{mresult}.

\subsection{A proof}\label{sec:proof}

In this section, we prove \eqref{mresult} for $N\geq 3$\footnote{For $N=2$ this was proven in \cite{Bonelli:2016idi}.}. More specifically, we show  that both sides of \eqref{mresult} satisfy the non-autonomous Toda equation \eqref{eq:Todatau1} with the same initial conditions. The uniqueness  of the solution  then implies the equality between the two sides.
We also note that \eqref{mresult} can also be tested numerically with high precision, see discussion in \autoref{sec:num}.

In order to show that the r.h.s~of \eqref{mresult} is a solution to the non-autonomous Toda equation \eqref{eq:Todatau1}, we use the result of \cite{Tracy:1998aa,Widom:1997aa}, where it was already proven that \eqref{eq:Toda1} is solved by\footnote{\label{convpfdet} We note that   $q_j$ in our convention corresponds to $q_{N-j+1}$ in the convention of \cite{Tracy:1998aa}.}
\be\label{qjproven} q_j= \log \det \left(1-\lambda \sum_{k=1}^{N-1} c_k \re^{2\pi \ri j k/N}G_k\right)-\log\det \left(1-\lambda \sum_{k=1}^{N-1} c_k \re^{2\pi \ri (j-1) k/N}G_k\right)\ee
where $G_k$ is an integral operator on $\IR_+$ with kernel
\be G_k(u,v)={\re^{-N T^{1\over 2N} \left[(1-\re^{2 \pi \ri k/N})u+(1-\re^{-2 \pi \ri k/N})u^{-1}\right]}\over v- u \re^{2 \pi \ri k/N}},\ee
and $\mathfrak{r}$ in \eqref{eq:Toda1} is related to $T$ by \eqref{eq:6}. Hence we want to demonstrate  that
\be\label{step1}\det \left(1+\sum_{k=1}^{N-1} x_k A_k\right)=\det \left(1-\lambda \sum_{k=1}^{N-1} c_k G_k\right),\ee
where  \be  \label{step10}\lambda c_k= -{ x_k \over 2 \pi \ri}\re^{\ri \pi k/N}.\ee
To show \eqref{step1} we first convert $A_k(x,y)$ defined in \eqref{ajkernel} into the following kernel on $\IR_+$
\be \small 2\pi \ri  \re^{-\pi \ri k/N}  A_k(x,y)\rd y= {\rd v}\sqrt{u\over v}~{\re^{N T^{1\over 2N} \left[\re^{2\ri \pi k/N}u +\re^{-2\ri \pi k/N}u^{-1}\right]}\re^{-N T^{1\over 2N}\left[v +v^{-1}\right]} \over v-u \re^{2 \pi \ri k/N}},\ee
where we set  $u=\re^{x}$, $v=\re^{y}$.
Recall that, for any traceclass set of one dimensional quantum mechanical operators ${\rm O}_k$'s,  we have
\be\label{detzn}\det \left(1+\sum_{k=1}^{N-1} x_k  {\rm O}_k\right)=\sum_{M_1, \dots, M_{N-1} \geq 0} Z({\bs M}) x_1^{M_1}\cdots  x_{N-1}^{M_{N-1}} \ee
where ${\bs M}=\left(M_1, \cdots, M_{N-1}\right)$ and $Z(\boldsymbol{M})$'s are the fermionic spectral traces given by
\be Z(\boldsymbol{M}) ={1\over M_1! \cdots M_{N}!} \int {\rm det}_{m,n} \left(R(u_m, u_n)\right)\rd^N u\ee
where \be R(u_m, u_n)={\rm O}_k(u_m, u_n),\quad \text{if}\quad \sum_{s=0}^{k-1}M_s\leq m\leq \sum_{s=1}^{k}M_s .\ee
This can be written explicitly as 
\be \label{gmat} \ba Z({\bs M})=&{1\over M_1! \cdots M_{N-1}! } \sum_{\sigma \in S_{M}}(-1)^\sigma\int  \rd ^M x \left(\prod_{i=1}^{M_1} {\rm O_1}(x_{\sigma(i)},x_i) \right)\\
&\left(\prod_{i=1+M_1}^{M_1+M_2} {\rm O_2}(x_{\sigma(i)},x_i)\right) \cdots
  \left( \prod_{i=1+\cdots +M_{N-2}}^{M_1+\cdots +M_{N-1}}{\rm O_{N-1}}(x_{\sigma(i)},x_i)\right)\\
\ea\ee
We refer to \cite[Sec.~2.3]{Codesido:2015dia} for more details. 
Hence, using the expressions \eqref{detzn} and \eqref{gmat}, we immediately have
\be\det \left(1+\sum_{k=1}^{N-1} x_k  A_k\right)=\det \left(1+\sum_{k=1}^{N-1} {x_k \re^{\pi \ri k/N}\over 2 \pi \ri}  G_k\right).\ee
which is precisely what is stated in \eqref{step1} and \eqref{step10}.

Moreover the solution to \eqref{eq:Toda1} is uniquely fixed by the initial conditions.
 The initial conditions for \eqref{qjproven} have already been computed explicitly in \cite{Tracy:1998aa}. In particular, it was shown that
\be \label{small}\det \left(1-\lambda \sum_{k=1}^{N-1} c_k G_k\right)\approx b (N T^{1\over 2N})^a\left( 1+\mathcal{O}(T^{1\over 2N})\right).\ee
where  
\be\label{abdef} \ba
a=&{1\over N}\sum_{k=1}^N a_k^2-{(N+1)(2N+1)\over 6}~,\\
b=&{\prod_{|j|<N} G\left({j\over N}+1\right)^{N-|j|}\over\prod_{0\leq \ell, k\leq N-1} G\left({a_\ell-a_k\over N}+1\right)}~.
 \ea\ee
The $a_k$'s in the first line of \eqref{abdef}
are defined as the solutions to
\be \label{soly0}\sin (\pi a_k) +\lambda \pi \sum_{j=1}^{N-1}   c_j \re^{{2 \pi \ri j\over N }(a_k-1)} \re^{- \pi \ri a_k}=0,\ee
where $a_k$ depends analytically on $\lambda$ and $a_k\Big|_{\lambda=0}=k,$ for $ k=1,\cdots,N$. 
Plug in $\lambda c_j$ as in \eqref{step10}, we have
\be \label{soly}\sin (\pi a_k) - {1\over 2 \ri }\sum_{j=1}^{N-1}\re^{\ri \pi j/N}   x_j \re^{{2 \pi \ri j\over N }(a_k-1)} \re^{- \pi \ri a_k}=0.\ee
The solutions to \eqref{soly} are precisely given by \be \label{asigma}a_k= -N \sigma_{N-k+1} +{N+1\over 2}\ee
where $ \sigma_k$ satisfy \eqref{mmap} and \eqref{recsum}. In other words, equation \eqref{soly} and \eqref{mmap} are equivalent. 
By using \eqref{asigma} and the multiplicative formula of the Barnes function, we write \eqref{abdef} as
\be \ba a=&N{\bs \sigma}^2+{1\over 12}(1-N^2)~,\\
b=&{N^{1/ 12} \E^{(N^2-1)\zeta'(-1)} }\prod_{1\leq \ell,k\leq N} {1\over G(1+{ \sigma_\ell-\sigma_k}) }~.
\ea\ee
Hence on the spectral theory side, thanks to the work of \cite{Tracy:1998aa,Widom:1997aa}, we can prove that 
\be \tau_j^{\rm spectral}=   N^{-1/ 12}\E^{-(N^2-1)\zeta'(-1)}\E^{-N^2 T^{1\over N}}T^{\frac{N^2-1}{24N}}\det \left(1+\sum_{k=1}^{N-1} {\rm e}^{-2\pi \ri k  j/N}x_k A_k\right) \ee
satisfy non-autonomous Toda equations\footnote{To be more precise, as said in \autoref{convpfdet}, our $\tau_j^{\rm spectral}$ should be $\tau_{N-j}$ in \cite{Tracy:1998aa}.} with initial conditions uniquely specified  by \be \label{spectic}\tau_j^{\rm spectral}\approx T^{\frac12(\boldsymbol{\sigma}+\boldsymbol{w}_j)^2}{1\over \prod_{\boldsymbol{\alpha}\in \Delta}G\left(1+(\boldsymbol{\alpha},\boldsymbol{\sigma}+\boldsymbol{w}_j)\right)}\ee
where $\bs \sigma$ and $x_k$'s are related as in \eqref{mmap}.

We now move to the l.h.s~of \eqref{mresult}. By Kyiv formula  \eqref{eq:taunek}, we know that the l.h.s~of \eqref{mresult} satisfies non-autonomous Toda equations. The fact that the initial conditions are the same as \eqref{spectic} , follows immediately from \eqref{eq:taunek} by setting $\bs{ \eta=0}$, see also \autoref{app:hamil}.  The uniqueness  of the solution then implies the equality of the two sides of \eqref{mresult}.

\section{The large time expansion }
\label{stexp}

The solution given in \eqref{eq:taunek} is an explicit and generic solution for the non-autonomous Toda system \eqref{eq:Todatau1} at small time, namely it is a convergent expansion at small $T$\footnote{
The proof of this fact for \(N=2\) is written in \cite[Proposition 1]{Its:2014lga}.
The proof for the arbitrary \(N\) can be obtained as a trivial generalization.}.
A natural question to ask is whether we can find an analogous solution at large time, i.e.~around $T= \infty$. 
To address this question, we  generalise the ideas of \cite{Gavrylenko:2020gjb} from $N=2$ to generic $N$. Our approach consists of two steps.
First, we use our result \eqref{mresult} to obtain the large time expansion of the tau function at the specific initial conditions compatible with \eqref{eta0}.
 Then we perform analytic continuation to generic values of   initial conditions.

\subsection{From matrix models}\label{frommm}

Several examples indicate that the generic large time expansion of Painlev\'e equations \cite{Bonelli:2016qwg,Its:2016jkt,Lisovyy:2018mnj,Nagoya:2015cja}   can be obtained by expanding the corresponding matrix model around its Gaussian point, without taking double scaling limits or adding external sources \cite{Grassi:2018spf, Dunne:2019aqp,Gavrylenko:2020gjb}.
Here, we will extend this procedure to non-autonomous Toda systems. The relevant matrix models are derived from the fermionic spectral traces $Z({\bs M})$ in \eqref{gmat} for the  operators \eqref{ajkernel}. Thanks to  Cauchy identity, we can express  \eqref{gmat} as a matrix model with $N-1$ cuts.  Specifically  we have  \cite{Bonelli:2017ptp}, 
\be\ba
\label{mm4d}Z({\bs M})=&\frac{1}{M_1!\cdots M_{N-1}!}\int\limits_{\mathbb{R}^M}\frac{d^M x}{(2\pi)^M}\prod_{j=1}^{N-1}\prod_{r_{j-1}\leq i_j\leq r_j}\E^{-4N T^{1\over 2N}\sin(\frac{\pi j}{N})\cosh(x_{i_j})}\\
&\times \frac{\prod\limits_{1\leq i<j\leq M}2\sinh\left(\frac{x_i-x_j}{2}+\frac{1}{2}(d_i-d_j)\right)2\sinh\left(\frac{x_i-x_j}{2}+\frac{1}{2}(f_i-f_j)\right)}{\prod\limits_{i,j=1}^M2\cosh\left(\frac{x_i-x_j}{2}+\frac{1}{2}(d_i-f_j)\right)} ,
\ea\ee
where \be r_0=1,\quad r_j=\sum_{i=1}^{j}M_i \\
\quad j=1,2,\cdots .\ee 
We also define
\be\ba d_j&=-\frac{(N-1-k)\ri\pi}{N},  \\
f_j&=-\frac{(N-2)\ri\pi}{N}-d_j,\ea\ee
where \be r_{k-1}\leq j\leq r_{k} .\ee
These matrix models are also interesting from a gauge-theoretic perspective, as they provide a way to move beyond the large-radius frame and explore the monopole point of Seiberg-Witten theory, which is entirely inaccessible from the Nekrasov function; see \cite{Bonelli:2016idi,Bonelli:2017ptp}.

We have
\be\label{ressumingZnek}\ba  Z({\bs M})=&{1\over (2\pi \ri)^{N-1}}\frac{N^{1/ 12}\E^{(N^2-1)\zeta'(-1)}\E^{N^2 T^{1\over N}}}{T^{\frac{N^2-1}{24N}}}\\
&\ \times\oint_{\gamma} \rd x_1\cdots\oint_{\gamma} \rd x_{N-1} { \prod_{i=1}^{N-1}\left(x_i({\bs \sigma})\right)^{{-}1-M_i} } {T^{\frac12(\boldsymbol{\sigma})^2} Z_{\rm inst}^{\rm 4d}(\boldsymbol{\sigma},T)
 \over \prod_{\boldsymbol{\alpha}\in \Delta}
G\left(1+(\boldsymbol{\alpha},\boldsymbol{\sigma})\right)},
\ea\ee
where $Z({\bs M})$ is defined in \eqref{mm4d} and $\gamma$ denotes a sufficiently small loop around 0.

The matrix model \eqref{mm4d} admits a natural expansion at large $T$, which is the expansion around the Gaussian point. More precisely, we have
\be\ba\label{expofmmZ} &Z_N({\bs M})
\sim\E^{-4N T^{\frac{1}{2N}}({\bf M},\sin(\frac{\pi {\bf k} }{N}))}\left(4N T^{\frac{1}{2N}}\right)^{-\frac{1}{2}{\bs M}^2}C(\mathbf{M}) ~\mathcal{E}^{\infty}(\mathbf{M}) ,
\ea\ee
where $\sim$  means that the r.h.s is an asymptotic expansion of the l.h.s\footnote{The $D_\ell^{(N)}\left({\bs M}\right)$ coefficients in \eqref{eds} are expected to grow factorially at large $\ell$.}.  In \eqref{expofmmZ} we use the notation  
\be\label{vecsinM}({\bf M},\sin(\frac{\pi {\bf k} }{N}))=\sum_{k=1}^{N-1} M_k  \sin(\frac{\pi k }{N})\ee
 and \be
\label{myC}
C(\mathbf{M})=\prod_{k=1}^{N-1}(2\pi)^{-\frac{M_k}{2}}G(M_k+1)\left(\sin \frac{\pi k}{N} \right)^{-\frac{3}{2}M_k^2}2^{-M_k^2}\prod_{1\leq j<k\leq N-1}\left(\frac{\sin\frac{(j-k)\pi}{2N}}{\sin\frac{(j+k)\pi}{2N}}\right)^{2M_jM_k} ~.
\ee
Moreover
\be\label{eds}\mathcal{E}^\infty(\mathbf{M}) =1+\sum_{\ell\geq 1}\left(\frac{1}{T^{\frac{1}{2N}}}\right)^\ell\left(\frac{1}{4N}\right)^\ell D_\ell^{(N)}\left({\bs M}\right).\ee
The $D_\ell^{(N)}\left({\bs M}\right)$ coefficients in \eqref{eds} can be systematically generated by expanding the matrix model \eqref{mm4d} around its Guassian point\footnote{A useful information on $D^{(N)}_\ell\left({\bs M}\right)$ is that it is an order at most  $3\ell$ polynomial in $M_i$'s. 
This follows from the fact that the matrix model admit a genus expansion of the form
\be \log Z_N^{\text{4d}}(M_1,\cdots,M_{N-1})=\sum_{g\geq 0}g_s^{2g-2}F_g^D(T_1,\cdots,T_{N-1}),\ee
where $g_s^{-1}=T^{\frac{1}{2N}}{4N}\sin({\pi \over N})$, $T_i=M_i g_s$ and $F_g^D(T_1,\cdots,T_{N-1})$ is a polynomial in $T_i$'s. 
%
}. 
For example,
\be\ba\label{Einfty1} D_1^{(N)}({\bs M})=&\sum_{l=1}^{N-1}\frac{\left(1-3\text{csc}^2\left(\frac{l \pi}{N}\right)\right)}{12 \sin\left(\frac{\pi l}{N}\right)}M_l(M_l^2-1)-\sum_{l=1}^{N-1}\frac{1}{24 \sin\left(\frac{\pi l}{N}\right)}M_l(1+2M_l^2)\\
&+\left(\sum\limits_{1\leq l<l'\leq N-1}\frac{\sin \left(\frac{\pi  l}{N}\right) \sin \left(\frac{\pi 
   l'}{N}\right)}{\left(\cos \left(\frac{\pi  l}{N}\right)-\cos \left(\frac{\pi 
   l'}{N}\right)\right)^2}\left({M_{l'}}\frac{M_{l}^2}{\sin\left(\frac{\pi l}{N}\right)}+{M_{l}}\frac{M_{l'}^2}{\sin\left(\frac{\pi l'}{N}\right)}\right)\right)\\
\ea\ee
See  \autoref{app:dnres}  for some explicit expressions of $D_\ell^{(N)}$'s at higher order $N$'s.

Combining \eqref{mresult}, \eqref{detzn}, \eqref{expofmmZ}, we conclude that the tau function has the following expansion around infinity
\be\label{taustrong} \ba \tau_0(\bs{0}, \bs{\sigma}, T)
&\sim  N^{-1/ 12}\E^{-(N^2-1)\zeta'(-1)}\E^{-N^2 T^{1\over N}}T^{\frac{N^2-1}{24N}}\\
&\times \sum_{\bs M\geq {\bs 0}} \boldsymbol{x}^{\mathbf{M}}\E^{-4N T^{\frac{1}{2N}}({\bf M},\sin(\frac{\pi {\bf k} }{N}))}\left(4N T^{\frac{1}{2N}}\right)^{-\frac{1}{2}{\bs M}^2}C(\mathbf{M}) ~\mathcal{E}^{\infty}(\mathbf{M})\ea\ee
where we use the notation $ \boldsymbol{x}^{\mathbf{M}}= \prod_{i=1}^{N-1}x_i^{M_i}$ and $\bs \sigma$ and $\bs x$ are related by \eqref{mmap}. As before $\sim$ in \eqref{taustrong} means that the r.h.s is an asymptotic expansion of the l.h.s . In particular, when expanding $\tau_j$ at small $T$ we have convergent expansion, while at large $T$ we have asymptotic behaviour. 

\subsection{Solving equations around infinity}
\label{proofsol}
In this section, we derive \eqref{taustrong} starting from the bilinear equations \eqref{eq:Todatau1}.

Equations \eqref{eq:Todatau1} have a trivial solution, in the sense that all $\tau_j$'s collapse to one simple function:
 \be\label{trivialtau}\tau_j=\mathfrak{r}^{(N^2-1)/12}\re^{-\frac{\mathfrak{r}^2}{16}},\quad \forall j.\ee
It's natural to factor out this trivial solution from \(\tau_j\) and define \(\Xi_j\) as \footnote{The $\mathfrak{r}$ independent coefficient is added for convenience, since we want to make connection to our solution \eqref{taustrong}.}

\be \tau_j= N^{-1/ 12}\E^{-(N^2-1)\zeta'(-1)} \left({\mathfrak{r}\over { 4N}}\right)^{(N^2-1)/12}\re^{-\frac{\mathfrak{r}^2}{16}}\Xi_j.\ee 
In this subsection, instead of the original non-autonomous Toda equations \eqref{eq:Todatau1}, we study the equivalent equations for $\Xi_j $:
\begin{equation}
\label{eq:XiEquation}
(\partial_{\log \mathfrak{r}})^2\log\Xi_j=\frac{\mathfrak{r}^2}{4}-\frac{\mathfrak{r}^2}{4}\frac{\Xi_{j+1}\Xi_{j-1}}{\Xi_j^2}.
\end{equation}
For the purpose of finding a solution, it is easier for us to use the bilinear form of \eqref{eq:XiEquation},
\begin{equation}
\label{eq:33}
\Xi_j\partial_{\mathfrak{r}}^2\Xi_j-(\partial_{\mathfrak{r}}\Xi_j)^2+\Xi_j \frac1{\mathfrak{r}}\partial_{\mathfrak{r}}\Xi_j={ 1\over 4}\Xi_j^2-{ 1\over 4}\Xi_{j+1}\Xi_{j-1} .
\end{equation}
To check \eqref{taustrong}, we use the following ansatz \begin{equation}
\label{eq:XiExpansion}
\Xi_j(\boldsymbol{\varkappa},\mathfrak{r})=\sum_{\mathbf{M}\geq 0}\Xi_j(\mathbf{M},\mathfrak{r})\boldsymbol{\varkappa}^{\mathbf{M}},\qquad \quad\Xi_j({\bs 0},\mathfrak{r})=1,
\end{equation} 
where we turn on  $(N-1)$ initial conditions, \(\boldsymbol{\varkappa}\), by perturbing the trivial solution \eqref{trivialtau}.
As usual, we start by solving for the first nontrivial term in this expansion. Note that because our choice of initial conditions $\bs\varkappa$ is generic, it is more convenient if we group $\Xi_j(\mathbf{M},\mathfrak{r})\boldsymbol{\varkappa}^{\mathbf{M}},\ \forall M_1+\cdots+M_{N-1}=1$ all together, and solve them as a whole first. Following this idea, it is natural to use
\begin{equation}
\label{eq:34}
\Xi_j(\boldsymbol{\varkappa},\mathfrak{r})=\sum_{{M}\geq 0} \Xi_j^{( M)}(\boldsymbol{\varkappa}),
\end{equation}
where  \(\Xi_j^{(0)}=1\) and $\Xi_j^{( M)}(\boldsymbol{\varkappa})$ is defined by
\be \label{mya}\Xi_j^{(M)}(\boldsymbol{\varkappa})=\sum_{M_1+\dots +M_{N-1}=M}\boldsymbol{\varkappa}^{\mathbf{M}}\Xi_j(\mathbf{M},\mathfrak{r}) .\ee
Plug in \eqref{eq:34}, the leading order term for $\boldsymbol{\varkappa}^{\mathbf{M}}$, i.e. the first nontrivial term with minimal $M$, in \eqref{eq:33} is
\begin{equation}
\label{eq:35}
\partial_{\mathfrak{r}}^2\Xi^{(1)}_j+\frac1{\mathfrak{r}}\partial_{\mathfrak{r}}\Xi^{(1)}_j=\frac14\left(2\Xi^{(1)}_j-\Xi^{(1)}_{j+1}-\Xi^{(1)}_{j-1}\right).
\end{equation}
Equation \eqref{eq:35} can be solved by diagonalizing the r.h.s. To do this, we introduce the following transformation

\begin{equation}
\label{eq:BesselEq}
\Xi_j^{(1)}(\boldsymbol{\varkappa},\mathfrak{r})=\sum_{k\geq0}^{N-1}\re^{\frac{-2\pi \ri kj}{N}}\tilde{\Xi}_k^{(1)}(\boldsymbol{\varkappa},\mathfrak{r}),\quad \tilde{\Xi}_0^{(1)}(\boldsymbol{\varkappa},\mathfrak{r})\equiv 0.
\end{equation}
Using \eqref{eq:BesselEq}, \eqref{eq:35} is converted into
\begin{equation}
\label{eq:37}
\left(\partial_{\mathfrak{r}}^2+\frac1{\mathfrak{r}}\partial_{\mathfrak{r}}-\sin^2\frac{\pi k}{N}\right)\tilde{\Xi}_k^{(1)}=0.
\end{equation}
It's straightforward to solve these second order ODEs. And we get 
\begin{equation}\begin{gathered}
\label{eq:Bessels}
\Xi_j^{(1)}(\boldsymbol{\varkappa},\mathfrak{r})=\sum_{k=1}^{N-1}\re^{-\frac{2\pi \ri k j}{N}}\varkappa_k K_0\left(\mathfrak{r}\sin\frac{\pi k}N\right)=\\
=
\sqrt{\frac{\pi}{2}}\sum_{k=1}^{N-1}\re^{-\frac{2\pi \ri k j}{N}}\varkappa_k \re^{-\mathfrak{r}\sin\frac{\pi k}N}\frac1{\sqrt{\mathfrak{r}\sin\frac{\pi k}N}}\left(1-\frac1{8\mathfrak{r}\sin\frac{\pi k}N}+\ldots \right) .
\end{gathered}
\end{equation}
Notice that for \(k=0\), we can naively get \(\tilde{\Xi}_0^{(1)}=c_0+c_1\log \mathfrak{r}\). But such a solution is a trivial zero mode corresponding to the simultaneous rescaling of the tau functions:
\begin{equation}
\label{eq:20}
\Xi_j\mapsto e^{\epsilon c_0 + \epsilon c_1 \log \mathfrak{r}}\Xi_j,
\end{equation}
which is a symmetry of \eqref{eq:XiEquation}. This means that we can fix this symmetry and without loss of generality, put \(\tilde{\Xi}_0^{(1)}=0\).

Another remark is that in principle, \eqref{eq:BesselEq} has a second solution \(I_0\). But we exclude it from consideration, since it grows exponentially when \(\mathfrak{r}\to\infty\). To be more precise, if \(I_0\) is included, neither can we treat \eqref{eq:XiExpansion} as a series in \(\boldsymbol{\varkappa}\), nor can we view \eqref{eq:XiExpansion} as an asymptotic series at large \(\mathfrak{r}\).

By investigating the dependence on $j$, we notice that $\Xi_{j}^{(1)}\left(\boldsymbol{\varkappa},\mathfrak{r}\right)$'s have the following symmetry:
\begin{equation}
\label{eq:40}
\Xi_{j+1}^{(1)}\left(\boldsymbol{\varkappa},\mathfrak{r}\right)=\Xi_j^{(1)}\left(\hat{\zeta}\boldsymbol{\varkappa},\mathfrak{r}\right),
\end{equation}
where we are using
\be\label{zetaaction} \hat{\zeta}^j\boldsymbol{\varkappa}= \left(\re^{-\frac{2\pi \ri j}{N}}{\varkappa}_1,\re^{-\frac{2\pi \ri j \cdot 2}{N}}{\varkappa}_2,\ldots , \re^{-\frac{2\pi \ri j \cdot(N-1)}{N}}{\varkappa}_{N-1}\right).\ee
These relations among $\Xi_j^{(1)},\quad j=0,\cdots,N-1,$ imply that all the $\Xi_j^{(1)}$ can be expressed in terms of just one function, which we choose to be \(\Xi^{(1)}(\boldsymbol{\varkappa},\mathfrak{r})\equiv\Xi^{(1)}_0(\boldsymbol{\varkappa},\mathfrak{r})\). The functions $\Xi_j^{(k)}$ satisfy the  recurrence relation obtained from plugging in \eqref{eq:34} into \eqref{eq:33}, thus they are expressed in terms of $\Xi_j^{(1)}$. As a result, they are related by the shift \eqref{zetaaction} as well. This enables us to express the whole series $\Xi_j,\forall j$ in terms of just $\Xi_0$ as
\begin{equation}
\label{eq:41}
\Xi_j \left(\boldsymbol{\varkappa},\mathfrak{r} \right)=\Xi\left( \hat{\zeta}^j\boldsymbol{\varkappa},\mathfrak{r} \right).
\end{equation}
This matches with our expectation as stated near the end of \autoref{sec:TS/ST}. Moreover, we check that \eqref{eq:33} itself is satisfied when \eqref{eq:41} holds.

Relations \eqref{eq:41} simplify the coupled bilinear equations \eqref{eq:33} to a single bilinear equation
\begin{equation}
\label{eq:43}
\Xi(\boldsymbol{\varkappa},\mathfrak{r})\partial_{\mathfrak{r}}^2\Xi(\boldsymbol{\varkappa},\mathfrak{r})-\left(\partial_{\mathfrak{r}}\Xi(\boldsymbol{\varkappa},\mathfrak{r})\right)^2+\Xi(\boldsymbol{\varkappa},\mathfrak{r}) \frac1{\mathfrak{r}}\partial_{\mathfrak{r}}\Xi(\boldsymbol{\varkappa},\mathfrak{r})-\frac{1}{4}\Xi(\boldsymbol{\varkappa},\mathfrak{r})^2+\frac{1}{4}\Xi(\hat{\zeta}\boldsymbol{\varkappa},\mathfrak{r})\Xi(\hat{\zeta}^{-1}\boldsymbol{\varkappa},\mathfrak{r})=0.
\end{equation}
Keeping \eqref{eq:41}  in mind, from now on, we will only work with \eqref{eq:43}. To check the explicit expression \eqref{taustrong}, which is obtained from the matrix model, we use the ansatz
\begin{equation}
\label{eq:44}
\Xi(\boldsymbol{\varkappa},\mathfrak{r})=\sum_{\mathbf{M}} \boldsymbol{\varkappa}^{\mathbf{M}}\re^{-\mathfrak{r} ({\bf M},\sin(\frac{\pi {\bf k} }{N}))}\mathfrak{r}^{-Q(\mathbf{M})}B(\mathbf{M},\mathfrak{r}).
\end{equation}
We want to show that $Q(\mathbf{M})$ and $B(\mathbf{M},\mathfrak{r})$ are in the same form of the ones in \eqref{taustrong}.

We now use \(\mathbf{M}\in \frac12\mathbb{Z}^{N-1}\) and extract the coefficient of  \(\boldsymbol{\varkappa}^{2\mathbf{M}}\) in \eqref{eq:43}:
\begin{equation}
\begin{gathered}
\label{eq:45}
\sum_{\mathbf{M}'+\mathbf{M}''=2\mathbf{M}}
B(\mathbf{M}',\mathfrak{r})B(\mathbf{M}'',\mathfrak{r})
\mathfrak{r}^{-Q(\mathbf{M}')-Q(\mathbf{M}'')}
\left(\left(\sum\limits_{k=1}^{N-1} M''_k\sin \frac{\pi k}{N}+\frac{Q(\mathbf{M}'')}{\mathfrak{r}}\right)^2-\right.\\\left.-\left(\sum\limits_{k=1}^{N-1} M'_k\sin \frac{\pi k}{N}+\frac{Q(\mathbf{M}')}{\mathfrak{r}}\right)\left(\sum\limits_{k=1}^{N-1} M''_k\sin \frac{\pi k}{N}+\frac{Q(\mathbf{M}')}{\mathfrak{r}}\right)+
\right.\\\left.
+\frac{1}{4}\re^{\frac{2\pi \ri}{N}\sum\limits_{k=1}^{N-1}k(M_k'-M_k'')}-\frac{1}{4}+\frac{Q(\mathbf{M}'')}{\mathfrak{r}^2}-\frac{Q(\mathbf{M}'')+\mathfrak{r}\sum\limits_{k=1}^{N-1} M_k''\sin \frac{\pi k}{N}}{\mathfrak{r}^2}
\right)+
\\
+
\sum_{\mathbf{M}'+\mathbf{M}''=2\mathbf{M}}B(\mathbf{M}',\mathfrak{r})\partial_{\mathfrak{r}}B(\mathbf{M}'',\mathfrak{r})\mathfrak{r}^{-Q(\mathbf{M}')-Q(\mathbf{M}'')}\left(-2\frac{Q(\mathbf{M}'')+\mathfrak{r}\sum\limits_{k=1}^{N-1} M_k''\sin \frac{\pi k}{N}}{\mathfrak{r}}+\right.\\\left.+\frac1{\mathfrak{r}}+2\frac{Q(\mathbf{M}')+\mathfrak{r}\sum\limits_{k=1}^{N-1} M_k'\sin \frac{\pi k}{N}}{\mathfrak{r}}\right)
+\\
+\sum_{\mathbf{M}'+\mathbf{M}''=2\mathbf{M}}\left(
B(\mathbf{M}',\mathfrak{r})\partial_{\mathfrak{r}}^2B(\mathbf{M}'',\mathfrak{r})-\partial_{\mathfrak{r}}B(\mathbf{M}',\mathfrak{r})\partial_{\mathfrak{r}}B(\mathbf{M}'',\mathfrak{r})
\right)\mathfrak{r}^{-Q(\mathbf{M}')-Q(\mathbf{M}'')}=0.
\end{gathered}
\end{equation}
We can rewrite this equation in a symmetric way:
\begin{equation}
\begin{gathered}
\label{eq:45}
\sum_{\mathbf{M}'+\mathbf{M}''=2\mathbf{M}}
B(\mathbf{M}',\mathfrak{r})B(\mathbf{M}'',\mathfrak{r})
\mathfrak{r}^{-Q(\mathbf{M}')-Q(\mathbf{M}'')}
\left(\left(\sum\limits_{k=1}^{N-1} (M''_k-M_k')\sin \frac{\pi k}{N}+\frac{Q(\mathbf{M}'')-Q(\mathbf{M}')}{\mathfrak{r}}\right)^2-\right.\\\left.-
\sin^2\left(\sum\limits_{k=1}^{N-1} \frac{\pi k}{N}(M_k''-M_k')\right)-\frac{2\sum\limits_{k=1}^{N-1} M_k\sin \frac{\pi k}{N}}{\mathfrak{r}}
\right)+
\\
+2
\sum_{\mathbf{M}'+\mathbf{M}''=2\mathbf{M}}\left(B(\mathbf{M}',\mathfrak{r})\partial_{\mathfrak{r}}B(\mathbf{M}'',\mathfrak{r})-\partial_{\mathfrak{r}}B(\mathbf{M}',\mathfrak{r})B(\mathbf{M}'',\mathfrak{r})\right)\mathfrak{r}^{-Q(\mathbf{M}')-Q(\mathbf{M}'')}\times
\\\times
\frac{Q(\mathbf{M}')-Q(\mathbf{M}'')+\mathfrak{r}\sum\limits_{k=1}^{N-1} (M_k'-M_k'')\sin \frac{\pi k}{N}}{\mathfrak{r}}+
\\+\sum_{\mathbf{M}'+\mathbf{M}''=2\mathbf{M}}\left(B(\mathbf{M}',\mathfrak{r})(\partial_{\mathfrak{r}}^2+\frac1{\mathfrak{r}}\partial_{\mathfrak{r}})B(\mathbf{M}'',\mathfrak{r})+(\partial_{\mathfrak{r}}^2+\frac1{\mathfrak{r}}\partial_{\mathfrak{r}})B(\mathbf{M}',\mathfrak{r})B(\mathbf{M}'',\mathfrak{r})\right)\mathfrak{r}^{-Q(\mathbf{M}')-Q(\mathbf{M}'')}
-\\-
2\sum_{\mathbf{M}'+\mathbf{M}''=2\mathbf{M}}\left( \partial_{\mathfrak{r}}B(\mathbf{M}',\mathfrak{r})\partial_{\mathfrak{r}}B(\mathbf{M}'',\mathfrak{r}) \right) \mathfrak{r}^{-Q(\mathbf{M}')-Q(\mathbf{M}'')}=0
\end{gathered}
\end{equation}
It is more convenient to work with a single summation over integers which reformulates the double summations in integers $\mathbf{M}'$ and $\mathbf{M}''$ with the constraint $\mathbf{M}'+\mathbf{M}''=2\mathbf{M}$. So we introduce the following parameterization
\begin{equation}
\label{eq:46}
\mathbf{M}'=\mathbf{M}+\frac12\boldsymbol{\varepsilon}+\boldsymbol{\Delta},\quad \mathbf{M}''=\mathbf{M}-\frac12\boldsymbol{\varepsilon}-\boldsymbol{\Delta},
\end{equation}
where ${\bs\Delta}\in\mathbb{Z}^{N-1}$ and \(\boldsymbol{\varepsilon}\in \mathbb{Z}^{N-1}/2\mathbb{Z}^{N-1}\) is the remainder of the division of \(2\mathbf{M}\) by 2. The inclusion of  \(\boldsymbol{\varepsilon}\) is simply to make ${\bs\Delta}$ an integer vector. Note that \eqref{eq:45} is the coefficient of a fixed $\mathbf{M}$ that we choose, and $\boldsymbol{\varepsilon}$ is completely determined by $\mathbf{M}$. So the resulting single summation is only over ${\bs\Delta}$.

For simplicity, we use the explicit knowledge of \eqref{taustrong} and assume
\begin{equation}
\label{eq:47}
Q(\mathbf{M})=\frac12 \mathbf{M}^2=\frac12(\mathbf{M},\mathbf{M})=\frac12\sum_{k=1}^{N-1}M_k^2.
\end{equation}
Plugging in \eqref{eq:47}, the combinations of $Q(\mathbf{M})$ in \eqref{eq:45} become
\begin{equation}
\label{eq:49}
Q(\mathbf{M}')+Q(\mathbf{M}'')= \mathbf{M}^2+\frac1{4}\boldsymbol{\varepsilon}^2 +(\boldsymbol{\Delta},\boldsymbol{\Delta}+\boldsymbol{\varepsilon}),
\quad Q(\mathbf{M}')-Q(\mathbf{M}'')=(\boldsymbol{\varepsilon}+2\boldsymbol{\Delta},\mathbf{M}).
\end{equation}
Parallel to \eqref{vecsinM}, it is also useful to introduce the following vectors:
\begin{equation}
\label{eq:50}
\frac{\pi\mathbf{k}}{N}=\left(\frac{\pi}{N},\frac{2\pi}N,\ldots,\frac{\pi(N-1)}N\right),\quad \sin\frac{\pi\mathbf{k}}{N}=\left(\sin\frac{\pi}{N},\sin\frac{2\pi}N,\ldots,\sin\frac{\pi(N-1)}N\right).
\end{equation}
After expressing \eqref{eq:45} with the substitution \eqref{eq:49} and the notations \eqref{eq:50} and \eqref{vecsinM}, we further rearrange the terms by their $\mathfrak{r}$ dependence:
\begin{equation}\begin{gathered}
\label{eq:bilinearInfinity1}
\sum_{\boldsymbol{\Delta}}\mathfrak{r}^{-(\boldsymbol{\Delta},\boldsymbol{\Delta}+\boldsymbol{\varepsilon})}
B(\mathbf{M}',\mathfrak{r})B(\mathbf{M}'',\mathfrak{r})
\left(\left(\boldsymbol{\varepsilon}+2\boldsymbol{\Delta},\sin \frac{\pi \mathbf{k}}{N} \right)^2-\sin^2 \left(\boldsymbol{\varepsilon}+2\boldsymbol{\Delta},\frac{\pi \mathbf{k}}{N}\right)\right)+
\\
+2\sum_{\boldsymbol{\Delta}}\mathfrak{r}^{-1-(\boldsymbol{\Delta},\boldsymbol{\Delta}+\boldsymbol{\varepsilon})}
B(\mathbf{M}',\mathfrak{r})B(\mathbf{M}'',\mathfrak{r})
 \left(\left(\boldsymbol{\varepsilon}+2\boldsymbol{\Delta},\mathbf{M}\right)\left(\boldsymbol{\varepsilon}+2\boldsymbol{\Delta},\sin \frac{\pi \mathbf{k}}{N} \right)-\left(\sin\frac{\pi \mathbf{k}}{N},\mathbf{M} \right) \right)+
\\
+2\sum_{\boldsymbol{\Delta}}\mathfrak{r}^{-(\boldsymbol{\Delta},\boldsymbol{\Delta}+\boldsymbol{\varepsilon})}
\left(B(\mathbf{M}',\mathfrak{r})\partial_{\mathfrak{r}}B(\mathbf{M}'',\mathfrak{r})-\partial_{\mathfrak{r}}B(\mathbf{M}',\mathfrak{r})B(\mathbf{M}'',\mathfrak{r})\right)
\left(\boldsymbol{\varepsilon}+2\boldsymbol{\Delta},\sin\frac{\pi \mathbf{k}}{N}\right)+
\\
+
\sum_{\boldsymbol{\Delta}}\mathfrak{r}^{-2-(\boldsymbol{\Delta},\boldsymbol{\Delta}+\boldsymbol{\varepsilon})}
B(\mathbf{M}',\mathfrak{r})B(\mathbf{M}'',\mathfrak{r})
\left(\boldsymbol{\varepsilon}+2\boldsymbol{\Delta},\mathbf{M}\right)^2
+
\\
+
2\sum_{\boldsymbol{\Delta}}\mathfrak{r}^{-1-(\boldsymbol{\Delta},\boldsymbol{\Delta}+\boldsymbol{\varepsilon})}
\left(B(\mathbf{M}',\mathfrak{r})\partial_{\mathfrak{r}}B(\mathbf{M}'',\mathfrak{r})-\partial_{\mathfrak{r}}B(\mathbf{M}',\mathfrak{r})B(\mathbf{M}'',\mathfrak{r})\right)
\left(\boldsymbol{\varepsilon}+2\boldsymbol{\Delta},\mathbf{M}\right)
+
\\
+
\sum_{\boldsymbol{\Delta}}\mathfrak{r}^{-(\boldsymbol{\Delta},\boldsymbol{\Delta}+\boldsymbol{\varepsilon})}
\left(B(\mathbf{M}',\mathfrak{r})(\partial_{\mathfrak{r}}^2+\frac1{\mathfrak{r}}\partial_{\mathfrak{r}})B(\mathbf{M}'',\mathfrak{r})+(\partial_{\mathfrak{r}}^2+\frac1{\mathfrak{r}}\partial_{\mathfrak{r}})B(\mathbf{M}',\mathfrak{r})B(\mathbf{M}'',\mathfrak{r})\right)
-\\-
2\sum_{\boldsymbol{\Delta}}\mathfrak{r}^{-(\boldsymbol{\Delta},\boldsymbol{\Delta}+\boldsymbol{\varepsilon})}\left(\partial_{\mathfrak{r}}B(\mathbf{M}',\mathfrak{r})\partial_{\mathfrak{r}}B(\mathbf{M}'',\mathfrak{r})\right)=0
\end{gathered}
\end{equation}
Inspired by \eqref{taustrong}, we use the following ansatz for $B(\mathbf{M},\mathfrak{r})$,
\begin{equation}
\label{eq:ansatz1}
B(\mathbf{M},\mathfrak{r})=C(\mathbf{M})D(\mathbf{M},\mathfrak{r})=C(\mathbf{M})\sum_{k=0}^{\infty}\mathfrak{r}^{-k}D_k(\mathbf{M}),\quad D_0(\mathbf{M})=1.
\end{equation}
The resulting equations for $C(\mathbf{M})$ and $D_k (\mathbf{M})$ will exactly reproduce the result in \eqref{taustrong}. We leave the detailed computations to \autoref{appbilin}.
 
\subsection{Continuation at generic initial conditions}
Following what was done in  \cite[Sec.~5]{Gavrylenko:2020gjb} for the case $N=2$, we now analytically continue \eqref{taustrong} to generic values of the initial conditions.

As we showed explicitly in \autoref{proofsol} above, the r.h.s~of \eqref{taustrong}
indeed solves the bilinear relation \eqref{eq:Todatau1} with $\mathfrak{r}=4N T^{\frac{1}{2N}}$.
In particular, \(D_\ell(\mathbf{M})\)'s satisfy the bilinear relations \eqref{eq:bilinearInfinity1} with the ansatz \eqref{eq:ansatz1} implemented.
In \autoref{proofsol}, we solved \(D_\ell(\mathbf{M})\)'s for arbitrary positive integer \(\mathbf{M}\), and it is obvious that the \(D_\ell(\mathbf{M})\)'s we obtained are still valid when \(\mathbf{M}\)'s have non-integer components, since \(D_\ell(\mathbf{M})\) are polynomials.
In this case, the Barnes functions \(G(M_k+1)\) in \eqref{myC} will be replaced by \(G(M_k+\nu_k+1)\) and will no longer vanish for negative \(M_{k}\).
So in order to avoid exponentially growing behavior, we need to consider the solution for purely imaginary \(\mathfrak{r}\)'s, \(\mathfrak{r}=-ir\):

\begin{equation}
\label{eq:31}
(\partial_{\log r})^2\log\tau_j=\frac{r^2}{4}\frac{\tau_{j+1}\tau_{j-1}}{\tau_j^2} .
\end{equation}
This leads to the following Ansatz \footnote{We use the upper script   $\tau_j^{\infty}$ because we are not careful about the normalisation factor. In particular if we want to identify  $\tau_j^{\infty}$  with $\tau_j$ for generic initial conditions, the overall normalisation  should be chosen more carefully.}
\begin{equation}\boxed{\ba
\label{anacon}
\tau_j^{\infty}({\bs x}, {\bs \nu }, r)&= r^{(N^2-1)/12}\re^{\frac{r^2}{16}}\\
&\sum_{\mathbf{M}\in {\IZ}^{N-1}}(\hat{\zeta}^j\boldsymbol{x})^{\mathbf{M}+\boldsymbol{\nu}}\re^{\ri r (\mathbf{M}+\boldsymbol{\nu},\sin \frac{\pi \mathbf{k}}{N})}r^{-\frac12 (\mathbf{M}+\boldsymbol{\nu})^2}\re^{{\ri \pi \over 4}(\mathbf{M}+\boldsymbol{\nu})^2} C(\mathbf{M}+\boldsymbol{\nu})\sum_{\ell=0}^{\infty} \frac{D_\ell^{(N)}(\mathbf{M}+\boldsymbol{\nu})}{(-\ri r)^\ell}\ea}
\end{equation}
where $r$ and $T$ are related as in \eqref{eq:2} and \eqref{eq:6}. We also use \eqref{zetaaction}. 
 The variables $\bs x, \bs\nu$  in \eqref{anacon} are the monodromies  of the linear system at infinity and they parametrize the generic choice of the boundary conditions.
  The map between 
 $ {\bs x}, {\bs \nu }$ and ${\bs \eta}, \bs \sigma$ reduces to \eqref{mmap} if $\eta=0$ and $\nu=0$.
 It is an interesting question to generalise the  map \eqref{mmap} for generic values of initial conditions. This may be achieved geometrically by phrasing the problem in terms of Fock-Goncharov and Fenchel-Nielsen coordinates, parallel to \cite{Coman:2020qgf}. We hope to report on this in the future.
 
 From the point of view of the supersymmetric gauge theory, the summand in \eqref{anacon}, namely
 \be \label{stN} r^{(N^2-1)/12}\re^{\frac{r^2}{16}}\\
r^{-\frac12 (\boldsymbol{\nu})^2}\re^{{\ri \pi \over 4}(\boldsymbol{\nu})^2} C(\boldsymbol{\nu})\sum_{\ell=0}^{\infty} \frac{D_\ell^{(N)}(\boldsymbol{\nu})}{(-\ri r)^\ell}
\ee
  is interpreted as a strong coupling analogue of the self-dual $SU(N)$ Nekrasov function.
Physical interpretation of its two arguments in the Seiberg-Witten limit \(\epsilon\to 0\)  is
  \be \nu \sim {a_D\over \epsilon}, \quad r\sim T^{1\over 2N} \sim {\Lambda\over \epsilon},\ee
where \(a_D\) is a dual Seiberg-Witten period, \(\Lambda\) is the coupling constant, and \(\epsilon\) is the self-dual \(\Omega\)-background parameter.
Expansion \eqref{stN} is exact in the dual period $\nu$ and \(\epsilon\), but it is given as an asymptotic series in $r^{-1}$ ($\sim$ inverse of the gauge coupling).  Hence, keeping  AGT in mind \cite{Alday:2009aq,Wyllard:2009hg},  we interpret $D_\ell^{(N)}(\boldsymbol{\nu})$  as a strong coupling version of irregular $W_N$ conformal blocks at $c=N-1$.

\section{Outlook}\label{sec:conc}

We conclude by listing some open problems and future directions closely related to the content of this paper.
\begin{itemize}
\item The TS/ST correspondence is one of the string dualities which can be formulated by precise  statements. One example of such statement is \eqref{5ds}, which relate enumerative geometry of toric CY manifolds to the spectral properties of a class of quantum mechanical operators. In this paper, we managed to demonstrate \eqref{mresult}, which is the dual 4d limit of \eqref{5ds}. The next challenge would be to provide a proof for \eqref{5ds}. For this purpose, a key step is to demonstrate that the determinant appearing on the r.h.s~of  \eqref{5ds} solves the $q$-Toda system \eqref{qtoda}.  
We would like to explore more in this direction in our future work.

\item On the other hand, \eqref{5ds} also has implications on the physics side. Indeed \eqref{5ds} 
makes a precise prediction for non-perturbative effects in topological string theory. It would   be important to provide a physical realisation of such effects. 

\item Within the formulation of the TS/ST correspondence, a concrete open question is to systematically determine the overall normalisation of the grand potential. For the $X_{N-1}$ geometries this  factor is denoted by $A_N({ t_N}, \hbar)$ in \eqref{J5d}. We propose that such factor corresponds to a certain $q$- "algebraic" solution for the the underlying $q$-deformed isomonodromic system, suitably completed by non-perturbative effects, see end of \autoref{nnteq}. These non-perturbative effects can be determined by examining  the structure of the Borel plane of the selected  $q$- "algebraic" solution, similar to what was done for the resolved conifold partition function in \cite{Alim:2021mhp,Grassi:2022zuk}. It would be interesting to test this proposal in more details  for other geometries as well.

\item Additionally, it would be interesting to establish a direct connection between the tau functions arising in the context of the TS/ST correspondence, i.e.~\eqref{5ds} and the  tau functions defined via \cite{Bridgeland:2016nqw,Bridgeland:2023eka}. When $N=2$ the $q$-"algebraic" solution $A_{N}(t_N,q)$ is the same as the tau function of \cite{tb}, obtained in the context of the resolved conifold.  This is however a very special case in which the only interesting free parameter, $\kappa$, is fixed at a specific value, namely $\kappa=0$. More in general, the relation still needs to be understood.

\item Another intriguing direction to explore  involves developing a deeper physical understanding of the operators appearing in the dual four dimensional limit, namely \eqref{n24d} and \eqref{Aop}. 
These operators correspond to difference equations which are solved by a certain type of surface defects in the self-dual phase of the four dimensional $\mathcal{N}=2$ $SU(N)$ SYM \cite{matijn-alba}. However, at present it is  unclear how to construct such operators directly from the physics of four dimensional gauge theory. 

\item Equations \eqref{mresult} and \eqref{mmap} provide an explicit solutions  for the connection problem of non-autonomous Toda equations \eqref{eq:Todatau3}  
when half of the initial conditions are fixed, that is when \eqref{eta0} is set\footnote{Relation \eqref{mmap} connects parameters of the two tau functions, whereas \eqref{mresult} defines the connection constant.}. 
Extending this to generic initial conditions would be of significant interest. For instance we expect a generalisation of
 the  map \eqref{mmap} to have a geometrical  interpretation via  Fock-Goncharov and Fenchel-Nielsen coordinates, parallel to \cite{Coman:2020qgf}. Relatedly, we note that the connection problem for non-autonomous  Toda with the extra requirement \eqref{eq:reflection}, was  studied very recently in \cite{Guest:2023spy}.
We would like to explore more in this direction in our future work.

\item
 Focusing on the relation between non-autonomous Toda equations and 4d $\mathcal{N}=2$ $SU(N)$  SYM, a more fundamental question one can ask is why this relation exists. One approach is to study the surface defects coupled to the bulk 4d theories \cite{Cecotti:1991me, Bonelli:2022iob}. The non-autonomous Toda equations are the $tt^*$ equations for the surface defects viewed as 2d theories. On one hand, the 2d $tt^*$ equations have been generalized to higher dimensions \cite{Cecotti:2013mba,Gaiotto:2010okc}. On the other hand, in \autoref{nnteq}, we show that solutions of q-deformed Toda equations are closely related to 5d $\mathcal{N}=1$ $SU(N)$  SYM theories which are geometrically engineered by $X_{N-1}$ toric CY manifolds. Above all, we expect the q-deformed Toda equations to be obtained from the $tt^*$ equations of the 3d defects coupled to the bulk 5d SYMs.
 We would like to explore more in this direction in our future work.
 
 \item From the point of view of isomonodromic deformations, a challenging question is to find a combinatorial expression, similar to the one for the coefficients of the instanton expansion in Nekrasov functions, for the coefficients $D^{(N)}_\ell$ characterising the large time expansion of  Toda tau functions, see \eqref{anacon}. It would be interesting to delve deeper  into the connection with the matrix models \eqref{mm4d} to see if we can use this representation to get such combinatorial expressions.

\item  The non-autonomous Toda system  with the additional constraints \eqref{eq:reflection}
also appear  in the context of the TBA equations \cite{cfiv}. If $N=3$, this constraints reduces the Toda system to Painlev\'e $\rm III_2$.  It would be interesting to explore this aspect in more details especially from the point of view of the operator theory.

\item Finally, it would be interesting to use the spectral determinant representation of the $q$-Toda tau functions, i.e.~the determinant appearing on the r.h.s~of \eqref{5ds}, to study the large time expansion in $q$-Toda equations, parallel to what we did in \eqref{anacon} for the non-autonomous Toda system.

\end{itemize}

\section*{
Acknowledgements} We would like to thank Mikhail Bershtein, Andrea Brini, Giulio Bonelli, Fabrizio Del Monte, Saebyeok Jeong, Marcos Mari\~no, Andrew Neitzke and Alessandro Tanzini for discussions. We are particularly thankful to Mikhail Bershtein and  Marcos Mari\~no  for a careful reading of the manuscript. 
We would like to thank the referee for their valuable feedback, which helped us improve this manuscript. The work of AG, PG and QH is partially supported by the Swiss National Science Foundation Grant No.~185723,  the NCCR SwissMAP and  the SNSF measures in support of Ukrainian scientists. PG would also like to thank the Defense Forces of Ukraine for protecting his relatives and friends.

\appendix

\section{Root system and conventions}\label{app:groupth}

Throughout the paper, our notation for the root vectors is \(\mathfrak{gl}(N)\)-like. That is to say, they are represented by the vectors in \(\mathbb{C}^N\) with the extra restriction that the sum of components is zero.

Let  $\left\{\hat {\bs e}_i\right\}_{i=1}^N$ be the standard Euclidean orthonormal basis, i.e. $(\hat {\bs e}_i, \hat {\bs e}_j)=\delta_{ij}$. The weights of the fundamental representation in our notation are
\be\label{eidef} {\bs e}_i= \hat {\bs e}_i-{1\over N}\mathbf{e}, \quad  i=1, \cdots N \ee
where 
  \(\mathbf{e}=\sum_{i=1}^N \hat {\bs e}_i\).  Thus,
\be( {\bs e}_i,  {\bs e}_j)=\delta_{ij}-{1\over N}.\ee 
 The scalar product is given by the usual Euclidean formula
\begin{equation}
\label{eq:13}
(\boldsymbol{u},\boldsymbol{v})=\sum_{i=1}^Nu_iv_i.
\end{equation}
 The set of positive \(\Delta_+\) and  negative  \(\Delta_-\) roots are defined by
\be\label{eq:23}\ba  \Delta_+=&\left\{ \boldsymbol{\alpha}_{ij}= {\bs e}_i- {\bs e}_j~ | ~1\leq i<j\leq N\right\},\\
 \Delta_-=&\left\{ \boldsymbol{\alpha}_{ij}= {\bs e}_i- {\bs e}_j~ | ~1\leq j<i\leq N\right\}. \ea\ee
The set of all roots is then $\Delta= \Delta_+ \cup  \Delta_- $.
The simple positive roots are 
\be  \label{simpler}\boldsymbol{\alpha}_i= {\bs e}_i- {\bs e}_{i+1}, \quad i=1, \cdots,N-1.\ee
The scalar products \(C_{ij}=(\boldsymbol{\alpha}_i,\boldsymbol{\alpha}_j)\) of the simple roots are thus given by the following matrix (which in this case coincides with the Cartan matrix):
\begin{equation}
\label{eq:21}
C=
\begin{pmatrix}
2 & -1 & 0 &\ldots &0\\
-1 &2  & -1& \ldots& 0\\
0 &-1& \ddots & \ddots&\vdots\\
\vdots&&\ddots &\ddots&-1\\
0&\cdots&\cdots&-1&2
\end{pmatrix}
\end{equation}
The lattice generated by the fundamental roots is called \emph{root lattice} and is denoted by \(Q_{N-1}\):
\begin{equation}
\label{eq:22}
Q_{N-1}=\mathbb{Z}\langle \boldsymbol{\alpha}_1,\ldots, \boldsymbol{\alpha}_{N-1} \rangle = \mathbb{Z}^N_0\subset \mathbb{Z}^N,
\end{equation}
where \(\mathbb{Z}^N_k\) with the subscript $k$ means the subset of all vectors such that the sum of components equals \(k\).
The fundamental weights $\left\{\boldsymbol{\omega}_i\right\}_{i=1}^{N-1}$ form the dual basis for the fundamental roots:
\begin{equation}
\label{eq:24}
(\boldsymbol{\omega}_i,\boldsymbol{\alpha}_j)=\delta_{ij}.
\end{equation}
This allows one to write down the explicit formula for them as
\be \label{eq:omegas}{\bs {\omega}_i}=\sum_{j=1}^i {\bs e}_j, \quad i=1,\dots N-1 . \ee
We also have
\be {\bs e}_1={\bs {\omega}_1}, \quad  {\bs e}_i={\bs {\omega}_i}-{\bs {\omega}_{i-1}},\quad 2\leq i\leq N-1, ~\quad  {\bs e}_N=-{\bs {\omega}_{N-1}}~.\ee
It is also natural to define \(\boldsymbol{\omega}_0=\bm{0}\), where $\bm{0}\cdot \bm{v}=\bm{v}\cdot \bm{0}=0,\  \forall \bm{v}\in \mathbb{C}^N$.
These vectors form the weight lattice \(P_{N-1}\) 
\begin{equation}
\label{eq:28}
P_{N-1}=\mathbb{Z}\langle \boldsymbol{\omega}_1,\ldots , \boldsymbol{\omega}_{N-1}\rangle \in \frac1{N}\mathbb{Z}^N~.
\end{equation}
We can also check that
\begin{equation}
\label{eq:26}
P_{N-1}/Q_{N-1}=\{\boldsymbol{\omega}_1,\ldots ,\boldsymbol{\omega}_{N-1}\}=\mathbb{Z}/N\mathbb{Z},
\end{equation}
where the generator of \(\mathbb{Z}/N\mathbb{Z}\) is \(\boldsymbol{\omega}_1\), since \((\boldsymbol{\omega}_k-k\boldsymbol{\omega}_1)\in Q_{N-1}\).

We see from the \eqref{eq:omegas} that interpretation of \(P_N\) is the following.
We can define orthogonal projection \(\Pi: \mathbb{C}^N\to \mathbb{C}^N_0\):
\begin{equation}
\label{eq:39}
\Pi \cdot\mathbf{v}=\mathbf{v}-\frac1{N}(\mathbf{e},\mathbf{v}),
\end{equation}
then
\begin{equation}
\label{eq:67}
\Pi\cdot\mathbb{Z}^N=P_{N-1},
\end{equation}
or in particular,
\begin{equation}
\label{eq:69}
\Pi\cdot \mathbb{Z}^N_{k+lN}=\boldsymbol{\omega}_k+Q_{N-1},\quad k=0,\ldots , N-1, \quad l\in \mathbb{Z}.
\end{equation}
It is also useful to introduce the Weyl vector 
\begin{equation}
\label{eq:73}
\boldsymbol{\rho}=\sum_{k=1}^{N-1}\boldsymbol{\omega}_k={1\over2}\sum_{\bs \alpha \in \Delta_+}{\bs \alpha}=\sum_{i=1}^N(N-i){\bs e}_i=(\frac{N-1}{2},\frac{N-3}{2},\ldots ,\frac{1-N}{2}).
\end{equation}
It's characteristic property following from the definition is that
\begin{equation}
\label{eq:76}
(\boldsymbol{\rho},\boldsymbol{\alpha}_i)=1.
\end{equation}
There is also Weyl group generated initially by reflections with respect to planes orthogonal to \(\boldsymbol{\alpha}_i\), but it actually coincides with the permutation group \(S_N\), which just permutes vectors' components.

There is an obvious, but sometimes useful observation that
\begin{equation}
\label{eq:78}
\forall s\in S_N: s\cdot \left( \boldsymbol{\omega}_k+Q_{N-1} \right)=\boldsymbol{\omega}_k+Q_{N-1},
\end{equation}
which implies in particular that
\begin{equation}
\label{eq:79}
s\cdot\boldsymbol{\omega}_k-\boldsymbol{\omega}_k\in Q_{N-1}, \qquad s\cdot\boldsymbol{\rho}-\boldsymbol{\rho}\in Q_{N-1}.
\end{equation}

Integer lattice \(\mathbb{Z}^N\) can be described as a subset of \(\mathbb{C}^N\) such that the component-wise exponential of a vector equals to \((1,\ldots ,1)\):
\begin{equation}
\label{eq:80}
\mathbb{Z}^N=\{\mathbf{v}\in \mathbb{C}^N|\re^{2\pi \ri \mathbf{v}}=(1,\ldots ,1)\}.
\end{equation}
It is clear that shifted root latices have the same description. Namely,
\begin{equation}
\label{eq:81}
\boldsymbol{\omega}_k+Q_{N-1}=\{\mathbf{v}\in \mathbb{C}^N|\re^{2\pi \ri \mathbf{v}}=(\omega^{-k},\ldots ,\omega^{-k})\},
\end{equation}
where
\begin{equation}
\label{eq:82}
\omega=\re^{2\pi \ri/N}.
\end{equation}

\section{Hamiltonian and tau functions}\label{app:hamil}

The system \eqref{eq:Toda2} is Hamiltonian with a Hamiltonian
\begin{equation}
H=\frac{T(p)}{r}+r U(q)=\sum_j \frac{p_j^2}{2r}+\frac{r}{4}\sum_j\re^{q_{j+1}-q_j}.
\end{equation}
The Hamiltonian equations of motion are
\begin{equation}
r\partial_rq_j=p_j, \qquad \partial_r r\partial_r q_j=-r \frac{\partial U(q)}{\partial q_j}.
\end{equation}
We would like to understand the relation between the tau functions and the Hamiltonian. In order to do this let us first notice the relation
\begin{equation}
\partial_rr\partial_r\log\prod_j\tau_j=rU(q).
\end{equation}
The second identity is
\begin{equation}
\partial_rH=-\frac{T(p)}{r^2}+U(r).
\end{equation}
Using it we get
\begin{equation}
\partial_rr\partial_r\log\prod_j\tau_j=\frac12(r\partial_rH+H)=\frac12\partial_r(rH).
\end{equation}
This leads us to the statement
\begin{equation}
\partial_r\log\prod_j\tau_j=\frac{c}{r}+\frac12 H.
\end{equation}
To find constant \(c\) we use that solution to the system \eqref{eq:Todatau3} is given by \eqref{eq:taunek}.
If we choose
 \be\bs  \sigma=-\epsilon\bs \rho+\epsilon^2\bs \delta ,\ee where \(\epsilon\) is some small positive number, \(\bs \rho\) is in \eqref{eq:73} and   \(\bs \delta\) is an arbitrary  vector,  then  the leading term in \eqref{eq:taunek} is the one with \(\bs w=\bs \omega_j\).
In this case
\begin{equation}
\tau_j\sim \frac{\re^{2 \pi \ri  (\bs \eta,\bs \omega_j)}T^{\frac12(\bs \sigma+\bs \omega_j)^2}}{\prod_{\bs \alpha\in \Delta}G\left(1+(\bs \alpha,\bs \sigma+\bs \omega_j)\right)}\sim r^{N(\bs \sigma+\bs \omega_j)^2}.
\end{equation}
Let us estimate contribution to the potential term:
\begin{equation}
r\frac{\tau_{j+1}\tau_{j-1}}{\tau_j^2}\sim r^{1+N \left(\bs \omega_{j-1}^2-2\bs\omega_j^2+\bs\omega_{j+1}^2+2(\sigma,\bs\omega_{j+1}-2\bs\omega_j+\bs\omega_{j-1}) \right)}=r^{-1-2N(\bs\sigma,\bs\alpha_j)}.
\end{equation}
Since \((\bs \alpha,\bs\sigma)<0\), this term can be neglected as compared to kinetic one.

Asymptotics of solution \(q_j\) is
\begin{multline}
q_j=\log\frac{\tau_j}{\tau_{j-1}}\sim N \log r \left( 2(\bs\sigma,\bs\omega_j-\bs\omega_{j-1})+\bs\omega_j^2-\bs\omega_{j-1}^2 \right)=\\
=\log r \left(2N(\bs\sigma,\bs e_j)+1-2j+N\right)=2N\log r\left(\bs\sigma+N^{-1}\bs\rho,\bs e_j\right).
\end{multline}
Therefore asymptotics of the Hamiltonian is 
\begin{equation}
H\sim 2N^2 \frac{(\bs\sigma+N^{-1}\bs\rho)^2}{r}.
\end{equation}
Now compute the asymptotics of \(\prod_j\tau_j\):
\begin{multline}
\partial_r\log\prod_j\tau_j=\frac{N}{r}\left(N\bs\sigma^2+2\left(\bs\sigma,\sum_j\bs\omega_j\right)+\sum_j\bs\omega_j^2\right)=\\
=\frac{N^2}{r}(\bs\sigma+N^{-1}\bs\rho)^2+\frac{N}{r}\sum_j\bs\omega_j^2-\frac{1}{r}\bs\rho^2=\frac{N^2}{r}(\bs\sigma+N^{-1}\bs\rho)^{-1}+\frac{N^3-N}{12r}
\end{multline}
Comparing the asymptotics we get finally
\begin{equation}
\label{eq:logDerivative1}
\partial_r\log\prod_j\tau_j=\frac{N^3-N}{12r}+\frac12 H=\frac{\bs\rho^2}{r}+\frac12 H.
\end{equation}

To get individual tau functions, and not their product, we rewrite
\begin{equation}
\tau_j=\tau_0\re^{\sum_{k=1}^jq_j}=\tau_0\re^{(\bs\omega_j,\bs q)},
\end{equation}
which gives
\begin{equation}
\prod_j\tau_j=\tau_0^N\re^{\sum_{k=1}^N(N-k)q_k}=\tau_0^N\re^{(\bs\rho,\bs q)}.
\end{equation}
So finally
\begin{equation}
\tau_j=\left(\prod_k\tau_k\right)^{1/N}\re^{(\bs\omega_j-\frac{1}{N}\bs\rho,\bs q)}.
\end{equation}
Its logarithmic derivative is given therefore by
\begin{equation}
\partial_r\log\tau_j=\frac{N^2-1}{12r}+\partial_r(\bs \omega_j-\frac1{N}\bs \rho, \bs q)+\frac1{2N}H.
\end{equation}

\section{The Nekrasov partition functions}
\label{appen:nek}
In this section, we review the definitions of Nekrasov functions for $SU(N)$ super Yang-Mills (SYM) in four and five dimension \cite{Losev:1997tp,Moore:1997dj,Lossev:1997bz,Nekrasov:2002qd}. In five dimension, such partition function agree, upon a suitable dictionary, with the refined topological string partition function of the $X_{N-1}$ geometry \cite{ikv,Taki:2007dh}.

In the definition of Nekrasov functions we use Young diagrams. Our conventions are as follows.
We denote a Young tableau by
\be Y=(y_1,y_2,\cdots),\ee
and a vector of  Young tableaux by
\be\label{vecY}{\bf{Y}}=(Y_1,\cdots,Y_N).\ee
We further define
\be  l(Y)=\sum_i y_i,\quad  l({\bf{Y}})=\sum_{i=1}^N l(Y_i).\ee
For a box $s=(i,j)$ (not necessarily in the partition $Y$),
\be h_Y(s)=y_i-j,\quad v_Y(s)=y_j^t-i,\ee
where $y_j^t$ is a component in the transposed Young tableau $Y^t=(y_1^t,y_2^t,\cdots)$.

\subsection{Nekrasov function in four dimension}\label{4ddef}

The Nekrasov function of four dimensional $\mathcal{N}=2$ $SU(N)$ super Yang-Mills theories in the  self-dual limit $\epsilon_1=-\epsilon_2=1$
is 
\be\label{znek4d} Z^{\text{4d}}_N({\bs \sigma},T)=Z_{\text{pert}}^{\rm 4d}({\bs \sigma},T)Z_{\text{inst}}^{\rm 4d}({\bs \sigma},T),\ee
where ${\bs \sigma}=\sum_{i=1}^N\sigma_i {\bs e}_i$ with $\sum_{i=1}^N\sigma_i=0$. Moreover,
\be Z_{\text{pert}}^{\rm 4d}({\bs \sigma},T)= {T^{\frac12(\boldsymbol{\sigma},\boldsymbol{\sigma})}
\over {\prod_{\boldsymbol{\alpha}\in \Delta}G\left(1+(\boldsymbol{\alpha},\boldsymbol{\sigma})\right)}}=\prod_{1\leq i,j\leq N} T^{{ (\sigma_i-\sigma_j)^2 \over 4N}}{1\over G(1+{ \sigma_i-\sigma_j}) }
\ee
and 
\be\label{Zinst} Z_{\text{inst}}^{\rm 4d}({\bs \sigma},T)=\sum_{\bf{Y}}T^{l(\bf{Y})}\mathcal{Z}_{\mathbf{Y}}({\bs \sigma},1, -1),\ee
with 
\be\ba \mathcal{Z}_{\mathbf{Y}}({\bs \sigma},\epsilon_1, \epsilon_2)=&\prod\limits_{I,J=1}^N\prod\limits_{s\in Y_I}\frac{1}{\sigma_I-\sigma_J-\epsilon_1v_{Y_J}(s)+\epsilon_2(h_{Y_I}(s)+1)}\\
&\times\prod\limits_{s\in Y_J}\frac{1}{\sigma_I-\sigma_J+\epsilon_1(v_{Y_I}(s)+1)-\epsilon_2h_{Y_J}(s)},
\ea\ee
The sum in \eqref{Zinst} is over a vector of Young tableaux, see \eqref{vecY}. 
For example, when $N=3$ we have
\be \label{exn3}Z_{\text{inst}}^{\rm 4d}({\bs \sigma},T)=1-2\frac{\sigma_1^2+\sigma_2^2+\sigma_3^2-\sigma_1\sigma_2-\sigma_2\sigma_3-\sigma_3\sigma_1}{(\sigma_1-\sigma_2)^2(\sigma_2-\sigma_3)^2(\sigma_3-\sigma_1)^2}T+\cdots.\ee
It was shown in \cite{Its:2014lga,Arnaudo:2022ivo} that \eqref{Zinst} is a convergent series in $T$ provided $\sigma_I -\sigma_J \notin \IZ $. 

\subsection{Nekrasov function in five dimension}
The instanton part of the Nekrasov function in five dimensional $\mathcal{N}=1$ $SU(N)$ super Yang-Mills  is defined as  
\be\label{Zinst5d} Z_{\text{inst}}^{\rm 5d}(\xi
, \bs{t},\epsilon_1,\epsilon_2 )=\sum_{\bf{Y}}\left(\E^{\frac{N}{2}(\epsilon_1+\epsilon_2)} \xi \right)^{-\ell(\bf{Y})}\mathcal{Z}_{\mathbf{Y}}^{\rm 5d}(\bs{t},\epsilon_1,\epsilon_2),\ee
where the sum is over a vector of Young tableaux, see \eqref{vecY}, and
\be\label{Z5dins} \mathcal{Z}_{\boldsymbol{Y}}^{\rm 5d}({\bs t},\epsilon_1,\epsilon_2)=\prod_{I,J=1}^N \prod_{s \in Y_I} {1\over 1- Q_{JI} \re^{\epsilon_1 v_{Y_J}(s)} \re^{-\epsilon_2 (h_{Y_I}(s)+1)}} \prod_{s\in Y_J}
{1\over 1- Q_{JI} \re^{-\epsilon_1 (v_{Y_I}(s)+1)} \re^{\epsilon_2 h_{Y_J}(s)}}.
\ee
The relation between $Q_{IJ}$ and ${\bs t}$ is 
\be Q_{IJ}={\rm e}^{{\sigma^{\rm 5d}_I}-{\sigma^{\rm 5d}_J}}\ee
where 
\be\label{tsigma} \ba  {\bs t}=\sum_{i=1}^N   {\sigma^{\rm 5d}_i}{\bs e_i} , \qquad \sum_{i=1}^N \sigma_i^{\rm 5d} =0,
\ea\ee 
or in components, 
\be t_i= \sigma_i^{\rm 5d}-\sigma_{i+1}^{\rm 5d}.\ee
We can also write ${\bs t}$ using the fundamental weight of the $SU(N)$. In this case we have
\be \label{tdefapp}{\bs t}= \sum_{j=1}^{N-1} t_{j}{\bs \omega}_j.\ee

Note that \eqref{Zinst5d} is not well defined when the $\epsilon_i$'s parameters are real since \eqref{Z5dins} has a dense set of poles in this slice.  This is a crucial difference between four dimension and five dimension. We refer to \cite{Marino:2015nla} for more details and references on this point. If the $\epsilon_i$'s parameters are in some suitable region of the complex plane (away from the real axis), then \eqref{Zinst5d} is convergent as a series in $\xi^{-1}$ provided that we also require $\log Q_{I J}\notin \E^{\IZ \epsilon_1+\IZ \epsilon_2}$, see   \cite{Bershtein:2016aef,Felder:2017rgg}.

\section{Wilson loops and quantum mirror maps}\label{app:WL}
\subsection{Definitions}
The quantum mirror maps for the $X_{N-1}$  geometry can also be  understood from a purely $SU(N)$ gauge theoretic point of view in terms of Wilson loops \cite{Losev:2003py,Bullimore:2014awa,Flume:2004rp,Fucito:2015ofa,Sciarappa:2017hds}. 
 We follow closely the presentation of \cite{Grassi:2018bci}. We use the same definitions for  Young diagrams and related quantities as the ones in \autoref{appen:nek}.

 We define
\be {\rm Ch}_{\boldsymbol{Y}}({\bs t}, \epsilon_1, \epsilon_2)=  \left(\sum_{I=1}^N \re^{\sigma^{\rm 5d}_I}\right)- (1-\re^{\epsilon_1}) (1-\re^{\epsilon_2})\left(\sum_{I=1}^N \re^{\sigma^{\rm 5d}_I}\sum_{(k,l)\in Y_I}\re^{ (k-1) \epsilon_1 + (l-1)\epsilon_2}\right)~,\ee
where $\bs t$ and $\sigma_I^{\rm 5d}$ are related as in \eqref{tsigma}. We also define
\be {\rm Ch}_{\boldsymbol{Y}}^j({\bs t}, \epsilon_1, \epsilon_2)={\rm Ch}_{\boldsymbol{Y}}({\bs t}, \epsilon_1, \epsilon_2)\Big|_{\sigma^{\rm 5d}_I  \rightarrow j \sigma^{\rm 5d}_I, \epsilon_i  \rightarrow j \epsilon_i}~,\ee
\be
{\rm Ch}_{\bs{k},\boldsymbol{Y}}({\bs t}, \epsilon_1, \epsilon_2)= \prod_{j\ge 1} \left( {\rm Ch}_{\boldsymbol{Y}}^j ({\bs t}, \epsilon_1, \epsilon_2) \right)^{k_j}, 
\ee
where  $\bs{k}=(k_1, k_2, \cdots)$,  $k_i\geq 0$. 
The five dimensional Wilson loop in the representation $\mathcal{R}$ of the gauge group $SU(N)$ is 
\be\label{WR}\ba  
W_{\mathcal{R}} (\xi, {\bs t}, \epsilon_1, \epsilon_2)=&{1\over Z_{\text{inst}}^{\rm 5d}(\xi, \bs{t},\epsilon_1,\epsilon_2 )} \\
&\times \sum_{\boldsymbol{Y}}  \left(\xi \E^{\frac{N}{2}(\epsilon_1+\epsilon_2)} \right)^{-\ell (\boldsymbol{Y})}  {\rm Ch}_{\mathcal{R},\boldsymbol{Y}}({\bs t}, \epsilon_1, \epsilon_2){\mathcal{Z}}_{\boldsymbol{Y}}^{\rm 5d}({\bs t}, \epsilon_1, \epsilon_2),\ea\ee
where  
\be
{\rm Ch}_{\mathcal{R},\boldsymbol{Y}}=\sum_{\bs{k}} {\chi_\mathcal{R}(\bs{k}) \over z_{\bs{k}}} {\rm Ch}_{\bs{k},\boldsymbol{Y}}, 
\ee 
\be z_{\bs k}= \prod_{j\ge 1} k_j! j^{k_j} \ee
and
$\chi_\mathcal{R}(\bs{k})$ is the character of $\bs{k}$ in the representation $\mathcal{R}$.
We will consider the purely $k{\rm th}$-antisymmetric representations which are represented by  the vertical Young tableaux with $k$ boxes. The corresponding Wilson loop is denoted by
\be W_{k} (\xi, {\bs t},\epsilon_1, \epsilon_2)~.\ee
At the leading order in $\xi^{-1}$, we have
\be \label{WKinst}W_{k} (\xi, {\bs t},\epsilon_1, \epsilon_2)=\sum\limits_{1\leq i_1< \cdots < i_k\leq N}\prod\limits_{m=1}^k\E^{\sigma^{\rm 5d}_{i_m}} +\mathcal{O}(\xi^{-1}),\ee
where we refer to $\mathcal{O}(\xi^{-1})$ as the instanton corrections.
In this paper, we are interested in the NS limit, i.e. $\epsilon_2\to 0$ limit, of the Wilson loops,  since in this limit,  $ W_{k}$'s are the inverse of the quantum mirror maps. More precisely we have 
\be H_{k}=W_k (\xi, {\bs t}(\hbar),\ri \hbar, 0), \ee
where ${\bs t}(\hbar)$ is the quantum mirror map and
 $H_{k}$ are the complex moduli defined in equation \eqref{Hidef}.
We give two explicit examples below for the  $X_{N-1}$ geometries where $N=2$ and ${N=3}$.
\subsection{Example: $N=2$}
The $X_1$ geometry is also  known as local $\IF_0$ (or local $\mathbb{P}_1\times \mathbb{P}_1$). It is related to  five dimensional, $\mathcal{N}=1$ $SU(2)$   super Yang-Mills theory.
The quantum mirror map $t_1(\hbar)$ for this geometry 
has been computed in  \cite{acdkv}. We have
 \be\label{mirrorx2}\ba -t_1(\hbar)=&-2 \log (H_1)+\frac{2 (\xi +1)}{H_1^2 \xi }+\frac{3 \xi ^2+4 \xi  \cos (\hbar )+8 \xi +3}{H_1^4 \xi ^2}\\
&+\frac{4 (\xi +1) \left(5 \xi ^2+18 \xi  \cos (\hbar )+3 \xi  \cos (2 \hbar )+19 \xi +5\right)}{3 H_1^6 \xi ^3}+\mathcal{O}\left(\left(\frac{1}{H_1}\right)^7\right). \ea\ee
where $H_1, \xi$ are the complex moduli of the curve \eqref{5dmirror}, and $H_1$ is defined in \eqref{Hidef}.

In this case, the only relevant Wilson loop is the one in the fundamental representation; by using the definition above,  we get
\be \label{su2w1}W_1(\xi, {\bs t},\ri \hbar,0)=Q_{21}^{1/2}+Q_{21}^{-1/2}-\xi^{-1} \re^{\ri \hbar}\left(\frac{\sqrt{Q_{21}} (Q_{21}+1)}{\left(-Q_{21}+\re^{\ri \hbar }\right) \left(-1+Q_{21} \re^{\ri \hbar }\right)}\right)+\mathcal{O}\left({1\over \xi^2}\right)\ee
where $Q_{21}=\re^{\sigma^{\rm 5d}_2-\sigma^{\rm 5d}_1}=\re^{-2\sigma^{\rm 5d}_1}$.
It is easy to test order by order in $\xi^{-1}$  that  we indeed have
\be H_1 =W_1(\xi, {\bs t}(\hbar),\ri \hbar,0).\ee
In this sense the Wilson loop is the inverse of the quantum mirror map.

\subsection{Example: $N=3$}

The $X_2$ geometry is related to the five dimensional, $\mathcal{N}=1$ $SU(3)$ super Yang-Mills theory. The quantum mirror maps $t_i(\hbar)$, $i=1,2,$ have been computed for instance in \cite{Hatsuda:2015qzx} and reads
 \be\small\label{mirrorsu31}\ba
t_1(\hbar) =&-\log \left(\frac{H_2}{H_1^2}\right)+\frac{126 H_1^5}{5 H_2^{10}}+\frac{35 H_1^4}{4 H_2^8}+\frac{10
   H_1^3}{3 H_2^6}+\frac{3 \left(H_2^3-10\right) H_1^2}{2
   H_2^7}+\frac{\left(H_2^3-4\right) H_1}{H_2^5}\\
   &-\frac{1}{H_2^3}-\frac{2
   H_2}{H_1^2}+\frac{2}{H_1^3}-\frac{3 H_2^2}{H_1^4}+\frac{8 H_2}{H_1^5}-\frac{20
   H_2^3}{3 H_1^6}+\frac{30 H_2^2}{H_1^7}-\frac{35 H_2^4}{2 H_1^8}-\frac{252 H_2^5}{5
   H_1^{10}}\\
   &+\frac{1}{\xi }\left(\frac{4 \cos (\hbar )}{H_1^3}-\frac{2 \cos (\hbar )}{H_2^3}+\frac{4 H_2 (5 \cos
   (\hbar )+\cos (3 \hbar ))}{H_1^5}-\frac{2 H_1 (5 \cos (\hbar )+\cos (3 \hbar
   ))}{H_2^5}\right)\\
   &+\mathcal{O}(\frac{1}{\xi^2})
   \\
t_2(\hbar)  =&-\log
    \left(\frac{H_1}{H_2^2}\right)-\frac{252 H_1^5}{5 H_2^{10}}-\frac{35 H_1^4}{2 H_2^8}-\frac{20
    H_1^3}{3 H_2^6}-\frac{3 H_1^2}{H_2^4}+\frac{30 H_1^2}{H_2^7}-\frac{2
    H_1}{H_2^2}+\frac{8
    H_1}{H_2^5}\\
    &+\frac{2}{H_2^3}-\frac{1}{H_1^3}+\frac{\left(H_1^3-4\right)
    H_2}{H_1^5}+\frac{10 H_2^3}{3 H_1^6}+\frac{3 \left(H_1^3-10\right) H_2^2}{2
    H_1^7}+\frac{35 H_2^4}{4 H_1^8}+\frac{126 H_2^5}{5 H_1^{10}}\\
    &+\xi^{-1}\left(\frac{4 \cos (\hbar )}{H_2^3}{-\frac{2 \cos (\hbar )}{H_1^3}-\frac{2 H_2 (5 \cos
    (\hbar )+\cos (3 \hbar ))}{H_1^5}+\frac{4 H_1 (5 \cos (\hbar )+\cos (3 \hbar
    ))}{H_2^5}}\right)\\
    &+\mathcal{O}(\frac{1}{\xi^2}) \\
    \\
\ea\ee
In this case the two relevant Wilson loops are

\be\small\ba\label{su3w1} 
&W_{1} (\xi, {\bs t},\ri\hbar,0)=\sum\limits_{1\leq i\leq 3}\E^{\sigma_{i}^{\rm 5d}} +\\
&\frac{\xi^{-1}\E^{\frac{3}{2}\ri\hbar} \left(\E^{2 \sigma_3^{\text{5d}}} \left(\E^{\sigma^{\text{5d}} _1+\sigma _2^{\text{5d}}} \left(\E^{3\ri\hbar}+1\right)-\E^{2 \sigma _1^{\text{5d}}}\E^{\ri\hbar} (\E^{\ri\hbar}+1)-\E^{2 \sigma _2^{\text{5d}}} \E^{\ri\hbar} (\E^{\ri\hbar}+1)\right)\right)}{\left(\E^{\sigma _1^{\text{5d}}} \E^{\ri\hbar}-\E^{\sigma _2^{\text{5d}}}\right) \left(\E^{\sigma _1^{\text{5d}}} \E^{\ri\hbar}-\E^{\sigma _3^{\text{5d}}}\right) \left(\E^{\sigma _1^{\text{5d}}}-\E^{\sigma _2^{\text{5d}}} \E^{\ri\hbar}\right)
   \left(\E^{\sigma _2^{\text{5d}}} \E^{\ri\hbar}-\E^{\sigma _3^{\text{5d}}}\right) \left(\E^{\sigma _1^{\text{5d}}}-\E^{\sigma _3^{\text{5d}}} \E^{\ri\hbar}\right) \left(\E^{\sigma _2^{\text{5d}}}-\E^{\sigma _3^{\text{5d}}} \E^{\ri\hbar}\right)}\\
&+\frac{ \xi^{-1}\E^{\frac{3}{2}\ri\hbar} \left(\left(\E^{\sigma _1^{\text{5d}}}+\E^{\sigma _2^{\text{5d}}}\right) \left(\E^{3\ri\hbar}+1\right)-\E^{2 \left(\sigma _1^{\text{5d}}+\sigma _2^{\text{5d}}\right)} \E^{\ri\hbar}
   (\E^{\ri\hbar}+1)\right)}{\left(\E^{\sigma _1^{\text{5d}}} \E^{\ri\hbar}-\E^{\sigma _2^{\text{5d}}}\right) \left(\E^{\sigma _1^{\text{5d}}} \E^{\ri\hbar}-\E^{\sigma _3^{\text{5d}}}\right) \left(\E^{\sigma _1^{\text{5d}}}-\E^{\sigma _2^{\text{5d}}} \E^{\ri\hbar}\right)
   \left(\E^{\sigma _2^{\text{5d}}} \E^{\ri\hbar}-\E^{\sigma _3^{\text{5d}}}\right) \left(\E^{\sigma _1^{\text{5d}}}-\E^{\sigma _3^{\text{5d}}} \E^{\ri\hbar}\right) \left(\E^{\sigma _2^{\text{5d}}}-\E^{\sigma _3^{\text{5d}}} \E^{\ri\hbar}\right)}\\ 
   &+\mathcal{O}\left(\frac{1}{\xi^{2}}\right)
\ea\ee

\be\small\ba\label{su3w2} \small &W_{2}(\xi, {\bs t},\ri\hbar, 0)=\E^{\sigma_1^{\rm 5d}+\sigma_2^{\rm 5d}}+\E^{\sigma_1^{\rm 5d}+\sigma_3^{\rm 5d}}+\E^{\sigma_2^{\rm 5d}+\sigma_3^{\rm 5d}}+\\
&\frac{\xi^{-1} \E^{\frac{3}{2}\ri\hbar} \left(\E^{\sigma _1^{\rm 5d}+\sigma _2^{\rm 5d}} \left(\E^{3\ri\hbar}+1\right)+\E^{\sigma _3^{\rm 5d}} \left(\E^{\sigma _1^{\rm 5d}}+\E^{\alpha
   _2}\right) \left(\E^{3\ri\hbar}+1\right)-(\E^{2 \sigma _1^{\rm 5d}}+\E^{2 \sigma _2^{\rm 5d}}+\E^{2 \sigma _3^{\rm 5d}} )\E^{\ri\hbar} (\E^{\ri\hbar}+1)\right)}{\left(\E^{\sigma _2^{\rm 5d}}-\E^{\sigma _1^{\rm 5d}+\ri\hbar} \right)
   \left(\E^{\sigma _3^{\rm 5d}}-\E^{\sigma _1^{\rm 5d}+\ri\hbar} \right) \left(\E^{\sigma _3^{\rm 5d}}-\E^{\sigma _2^{\rm 5d}+\ri\hbar}\right) \left(\E^{\sigma _2^{\rm 5d}+\ri\hbar} -\E^{\sigma _1^{\rm 5d}}\right) \left(\E^{\sigma _3^{\rm 5d}+\ri\hbar}
   -\E^{\sigma _1^{\rm 5d}}\right) \left(\E^{\sigma _3^{\rm 5d}+\ri\hbar} -\E^{\sigma _2^{\rm 5d}}\right)}\\
   &+\mathcal{O}\left(\frac{1}{\xi^2}\right)
\ea\ee
where $H_1,H_2, \xi$ are the complex moduli of the curve, where $H_1$ and $H_2$ are defined in \eqref{Hidef}. In this example,
\be \sigma^{\rm 5d}_1=\frac{2}{3} t_1(\hbar)+\frac{1}{3} t_2(\hbar), \ee
\be \sigma^{\rm 5d}_2=-\frac{1}{3} t_1(\hbar)+\frac{1}{3} t_2(\hbar), \ee
\be \sigma^{\rm 5d}_3=-\frac{1}{3} t_1(\hbar)-\frac{2}{3} t_2(\hbar). \ee
Plug in \eqref{mirrorsu31} for $\sigma_i$'s on the r.h.s~of  \eqref{su3w1} and  \eqref{su3w2}. We test order by order in $\xi^{-1}$ that 
\be H_2 =W_2(\xi, {\bs t}(\hbar),\ri\hbar,0),\ee
\be H_1 =W_1(\xi, {\bs t}(\hbar),\ri\hbar,0).\ee

\section{The dual 4d limit: some details}\label{app:oneloop}

In this appendix, we work out some details of the dual 4d limit (defined by the $\beta\rightarrow 0$ limit with the scaling \eqref{dual4dcomplex}) of the 5d topological string grand potential which appears on the l.h.s~of \eqref{5d}.

\subsection{The one-loop part and Barnes functions}\label{barnes}

In this part we show that the two $t_N$ independent quantities \eqref{Ffgv} and \eqref{Ffns} combine to give the Barnes functions appearing in  \eqref{eq:taunek}. At the level of the grand potential, the contribution of the two is
\be
\ba \mathfrak{F}^{\text{1loop}}( \bs{t}(\hbar),\hbar)&=\sum_{i=1}^{N-1}\frac{t_i(\hbar)}{2\pi}\frac{\partial}{\partial t_i}\mathcal{F}_{\rm NS}( \bs{t}(\hbar),\hbar) + \frac{\hbar^2}{2\pi}\frac{\partial}{\partial \hbar}
	\left( \frac{\mathcal{F}_{\rm NS}( \bs{t}(\hbar),\hbar)}{\hbar} \right)\\
	&\quad+    \mathcal{F}_{\rm GV}\left({2\pi\over \hbar} \bs{t}(\hbar), \frac{4\pi^2}{\hbar} \right) \\
&= \sum_{\balpha \in \Delta_+} \sum_{w\ge 1}{1\over 2\pi w^2} \cot\left( {\hbar w \over 2}\right) \re^{-w (\balpha, \bs{t}(\hbar))}\left(1+w (\balpha, \bs{t}(\hbar))\right)\\
+&\sum_{\balpha \in \Delta_+} \sum_{w\ge 1} \csc\left( {\hbar w \over 2}\right)^2 \re^{-w (\balpha, \bs{t}(\hbar))}{\hbar \over 4 \pi w}- \sum_{\balpha \in \Delta_+}\sum_{v \ge 1} {1\over 2v} {\csc^2\left( {2\pi^2 v \over \hbar} \right)} (\re^{- {2\pi\over \hbar }v (\balpha, \bs{t}(\hbar))}).\\
\ea
\ee
It is useful to note that
 $\mathfrak{F}^{\text{1loop}}( \bs{t}(\hbar),\hbar)$ has an integral expression which can be found in \cite[eq.~(3.9)]{Bonelli:2016idi} and follows from \cite{Hatsuda:2015owa}
\be \label{integral} \ba \mathfrak{F}^{\text{1loop}}( \bs{t}(\hbar),\hbar)=& \sum_{\balpha \in \Delta_+}\bigg(-\frac{\hbar^2}{8\pi^4}\text{Li}_3(\re^{- 2\pi(\balpha, \bs{t}(\hbar))\over \hbar})\\
&+2\text{Re}\int_0^{\infty\E^{\ri 0}}\rd x\frac{x}{\E^{2\pi x}-1}\log(1+\re^{- 4\pi(\balpha, \bs{t}(\hbar))\over \hbar}-2\re^{- 2\pi(\balpha, \bs{t}(\hbar))\over \hbar}\cosh\frac{4\pi^2 x}{\hbar}\bigg)~.\ea\ee
Using the above integral representation, it is easy to see that in the dual 4d limit 
we get 

\be \label{1l4f}\ba\E^{\mathfrak{F}^{\text{1loop}}\left({\bf t}(\hbar) +2 \pi \ri { \bs w}, \hbar\right)} ~ \xrightarrow{\text{\eqref{dual4dcomplex}}} ~&\re^{\frac{1}{2} (N-1) N\left(\frac{\log (\beta )}{6}-\frac{2 \zeta (3)}{\beta ^2}+2\zeta'(-1)\right)}\\
&\prod\limits_{{\bs \alpha }\in \Delta_+}{\bigg(\re^{\frac{\ri \pi^2({\bs \sigma+\bs w},\bs \alpha) }{3 \beta }}{\beta }^{-(\bs \sigma +\bs w  , {\bs \alpha})^2}\bigg)\over G(1+(\bs \sigma+\bs w, \bs \alpha))G(1-(\bs \sigma+\bs w, \bs \alpha))}~.\ea\ee

\subsection{The polynomial part}\label{appeb:poly}
Here we study the 4d limit of the polynomial part \eqref{fpert}, namely 
\be \ba\mathfrak{F}^{\rm p}\left( {t}_N, {\bs t}(\hbar),{\hbar} \right)&=F_{\rm p}\left( \frac{2\pi}{\hbar}{t}_N, \frac{2\pi}{\hbar}{\bs t}(\hbar) , \frac{4\pi^2}{\hbar} \right)\\
 &={1\over  12\pi \hbar} \sum_{\balpha \in \Delta_+} (\bs{t}(\hbar), \balpha)^3 + {t_N \over  4 \pi \hbar N  } \sum_{\balpha \in \Delta_+} (\bs{t}(\hbar), \balpha)^2 + 
{\pi \over 3\hbar}\left( 1-{ \hbar^2 \over 4\pi^2} \right)( \bs{t} (\hbar), {\bs \rho}).\ea\ee
We have

\be\label{polinomial}\ba \mathfrak{F}^{\rm p} ({t_N},{\bs t(\hbar)}+2 \pi \ri \bs {n}, \hbar)
~\xrightarrow{\text{\eqref{dual4dcomplex}}}~& -{ 2\pi^2 \ri  \over 3\beta} ~(\bs{\sigma}+{\bs{n}} , {\bs \rho})+ \left(\log (\beta )+\frac{\log (T)}{2 N}\right)\sum_{\balpha \in \Delta_+}\left({\bs \sigma+\bs w}, \bs \alpha\right)^2 \\\ea\ee

\subsection{Overall normalization}\label{4danl}
We now compute the limit of \eqref{ANg}. 
 We have
 \begin{align}\label{Zconi} Z_{\rm coni}\left(\frac{2\pi}{ N\hbar}t_N,  \hbar N\right) &\to \re^{-N^2 T^{1/N}},\\
 Z_{\rm coni}^{\rm np}\left(\frac{2 \pi}{N \hbar }t_N,{4 \pi^2\over N \hbar}\right)&\to 1,\end{align}
 \be  \re^{{N\over 2}A_c\left(\frac{\hbar }{\pi }\right)-\frac{1}{2} A_c\left(\frac{N \hbar }{\pi }\right)}~\to ~N^{1/12}\re^{(N-1)\zeta'(-1)+{1\over 12}(N-1)\log \beta+\frac{(N-1) N \zeta (3)}{\beta ^2}}\ee
   \be \re^{{ \left(N^2-1\right)}\frac{\pi }{12 N \hbar }(t_N)} \to \re^{-\frac{1}{12} \left(N^2-1\right) \log (\beta )} T^{-\frac{N^2-1}{24 N}}\ee
To derive \eqref{Zconi}, we have used the identity
\be \left(-q^{1/N} z^{1/N},q^{1/N} ,q^{1/N} \right)_{\infty}=\prod_{n=1}^\infty\left(1+q^{n/N} z^{1/N}\right)^n~.  \ee
 Hence
\be\label{An4d}\ba \re^{A_N(t_N, \hbar)}\to & \re^{N^2 T^{1/N}}N^{1/12}T^{-\frac{N^2-1}{24 N}}\re^{(N-1)\zeta'(-1)-{1\over 12}(N(N-1))\log \beta+\frac{(N-1) N \zeta (3)}{\beta ^2}}.\ea \ee
 Note that  $\re^{N^2 T^{1/N}}$ is the well known algebraic solution to non-autonomous Toda equation, see \cite{Bershtein:2018srt} and references therein.
 \subsection{Total}
 By combining \eqref{An4d}, \eqref{1l4f} and  \eqref{polinomial} we obtain
\be 
\label{1lpol}\ba & \re^{N^2 T^{1/N}}T^{-\frac{N^2-1}{24 N}}\re^{(N^2-1)\zeta'(-1)}N^{1/12}  T^{{1\over 2}\left({\bs \sigma+\bs w}\right)^2} \prod\limits_{{\bs \alpha }\in \Delta}{1\over  G(1+(\bs \sigma+\bs w)), \bs \alpha)}~,
\ea\ee
where we used \eqref{eq:73} as well as
 \be \sum\limits_{{\bs \alpha }\in \Delta_+}  ((\bs \sigma +\bs w ) , {\bs \alpha})^2 = N (\bs \sigma +\bs w )^2~.\ee
In particular this means that, if we chose $\bs \sigma$ such that all the $x_i$ in \eqref{mmap} vanish, namely
\be \sigma_{N-k+1}^*=\frac{-2 k+N+1}{2 N}, \quad k=1,\dots N\ee
then we have
 \be\ba
 \sum_{\boldsymbol{w}\in Q_{N-1}}\frac{T^{\frac12(\boldsymbol{\sigma^*}+\boldsymbol{w})^2}}{\prod\limits_{\boldsymbol{\alpha}\in \Delta}G\left(1+(\boldsymbol{\alpha},\boldsymbol{\sigma^*}+\boldsymbol{w})\right)}Z_{\rm inst}^{\rm 4d}(\boldsymbol{\sigma^*}+\boldsymbol{w},T)
=  N^{-1/ 12}\E^{-(N^2-1)\zeta'(-1)}\E^{-N^2 T^{1\over N}}T^{\frac{N^2-1}{24N}}.\ea
\ee
This equality can also be tested explicitly, e.g.~numerically.

\section{Independent numerical tests}
\label{sec:num}

In this section we provide complementary  numerical evidences for \eqref{mresult}.
We work on the codim-$(N-2)$ slices of the parameter space such that $x_1=\cdots=x_{i\neq J}=\cdots=x_{N-1}=0$. 
On such slices, \eqref{mresult} reads
 \be\label{mresult1p}\sum_{\boldsymbol{w}\in Q_{N-1}}\frac{T^{\frac12(\boldsymbol{\sigma}+\boldsymbol{w})^2}Z_{\rm inst}^{\rm 4d}(\boldsymbol{\sigma}+\boldsymbol{w},T)}{\prod\limits_{\boldsymbol{\alpha}\in \Delta}G\left(1+(\boldsymbol{\alpha},\boldsymbol{\sigma}+\boldsymbol{w})\right)}\Big|_{\bs \sigma=\boldsymbol{\sigma}^{(J)}}
=  \frac{T^{\frac{N^2-1}{24N}}}{N^{1/ 12}\E^{(N^2-1)\zeta'(-1)}\E^{N^2 T^{1\over N}}}\det \left(1+ x_{J} A_{J}\right)\ee
  where 
  \be\label{vecalpha} \bs{\sigma}^{(J)}\in \left\{\sum_{i=1}^{N}\sigma_i{\bs e}_i~ \Big |~ x_{i}(\sigma_1, \dots ,\sigma_N) =0,~\forall~ i\neq J, \text{ and } \sum\limits_{i=1}^N  \sigma_i=0\right\}\ee
 and we treat $x_i$'s as functions of $\bs{\sigma}$ given by the mapping \eqref{mmap}.  
Since 
\be \det \left(1+ x_{J} A_{J}\right)=\sum_{N\geq 0}x_J^N\text{Z}_J(N)\ee
where $\text{Z}_J(N)$ is a polynomial of $\Tr A_J^i, i=1,\cdots, N$, 
\be\ba \text{Z}_J(1)&=\Tr A_J,\\
\text{Z}_J(2)&=\frac{1}{2}\left((\Tr A_J)^2-\Tr A_J^2\right)\\
\text{Z}_J(3)&=\frac{1}{6}\left((\Tr A_J)^3-3\Tr A_J\Tr A_J^2+2\Tr A_J^3\right),\\
\vdots&
\ea\ee
 the spectrum of $A_J(x,y)$ only depends on 
\be\Tr A_J^i=\int \prod\limits_{k=1}^{i}\rd x_k \left(\prod\limits_{k=1}^{i-1}A_J(x_k,x_{k+1})\right)A_J(x_i,x_1), \quad i\in\mathbb{N}.\ee
As a result, if two operators $A_J$ and $K_J$ satisfy
 \be\label{AtoK} \Tr A_J^i=\Tr K_J^i,~ \forall ~ i \ee
  we conclude that $K_J$ and $A_J$ have identical spectrum. For convenience, instead of working with $A_J$, we will calculate the spectrum of $K_J$ whose kernel is
\be\label{speck} K_J(x,y)= \frac{\sqrt{V_J(x)}\sqrt{V_J(y)}}{4\pi\cosh\left(\frac{x-y}{2}+\frac{\ri\pi(2J-N)}{2N}\right)},\ee
where 
\be V_J(x)=f\left(x+\frac{\ri\pi(2J-N)}{N}\right)f\left(x\right).\ee
This operator is of trace class with a discrete spectrum which we denote by  \be \left\{\re^{-E_n^{(J)}(T)}\right\}_{n\geq 0}~.\ee
Another useful consequence is that the spectrum of $K_{J}$ and $K_{N-J}$ is identical since $\Tr K_J^i= \Tr K_{N-J}^i$, which can be seen by the change of variable \be x_k\rightarrow x_k+\frac{2J-N}{N}\pi\ri \ee and the redefinitions \be x_k\rightarrow x_{N-k} .\ee 
 Therefore equation \eqref{mresult1p} reads
  \be\label{mresult1p2}\sum_{\boldsymbol{w}\in Q_{N-1}}\frac{T^{\frac12(\boldsymbol{\sigma}+\boldsymbol{w})^2}Z_{\rm inst}^{\rm 4d}(\boldsymbol{\sigma}+\boldsymbol{w},T)}{\prod\limits_{\boldsymbol{\alpha}\in \Delta}G\left(1+(\boldsymbol{\alpha},\boldsymbol{\sigma}+\boldsymbol{w})\right)}\Big|_{\bs \sigma=\boldsymbol{\sigma}^{(J)}}
=  \frac{T^{\frac{N^2-1}{24N}}}{N^{1/ 12}\E^{(N^2-1)\zeta'(-1)}\E^{N^2 T^{1\over N}}}\det \left(1+ x_{J} K_{J}\right)~.\ee
We test \eqref{mresult1p2} by checking  that, for fixed $T$, we have
 \be\label{ndet}\ba  \tau_0(\bs{0}, \bs{\sigma}^{(J)}, T)=0\qquad \text{iff}  \qquad  x_J(\bs{\sigma}^{(J)})=-\re^{E_n^{(J)}(T)} \ea 
\ee
where $\re^{E_n^{(J)}(T)}$ is computed independently by using the numerical methods of \cite[eq.~(2.8),(2.9)]{okuyama2016tba} together with \cite[Appendix C]{ghmabjm}. The numerical value   is denoted by $E_n^{(J)*}(T)$\footnote{The $*$ in the superscript is to stress that this is the result obtained from numerical calculation of the spectrum.}.

\subsection*{Example: $N=3$}
\label{num:su3}
As explained above, we only need to calculate the spectrum of $A_1$ whose kernel is given by
\be A_1(x,y)=  {     f(x-\frac{\ri \pi  }{3})  f(y)  \over 4 \pi \cosh \left({x- y\over 2}-\frac{\ri \pi }{6}\right)},\ee
To explain the technique, we elaborate on the the checking of 
\be\label{SU3num}{
\sum_{\boldsymbol{w}\in Q_{2}}\frac{T^{\frac12(\boldsymbol{\sigma}+\boldsymbol{w})^2}Z_{\rm inst}^{\rm 4d}(\boldsymbol{\sigma}+\boldsymbol{w},T)}{\prod\limits_{\boldsymbol{\alpha}\in \Delta}G\left(1+(\boldsymbol{\alpha},\boldsymbol{\sigma}+\boldsymbol{w})\right)}\Big|_{\bs \sigma=\boldsymbol{\sigma}^{(1)}}=  3^{-1/ 12}\E^{-8\zeta'(-1)}\E^{-9 T^{1\over 3}}T^{\frac{1}{9}}\det \left(1+ x_{1} A_{1}\right)},\ee
corresponding to setting $x_2=0$ while keeping $x_1\neq 0$ in \eqref{mresult1p}. To be more specific, in this example, we have
\be\label{SU3x2} x_2= \E^{2\pi\ri (\sigma_1+\sigma_2)}+\E^{2\pi\ri (\sigma_1+\sigma_3)}+\E^{2\pi\ri (\sigma_2+\sigma_3)}, \qquad \text{ where } \sigma_3=-\sigma_2-\sigma_1 .\ee
Solving  equation $ \eqref{SU3x2} = 0$ we obtain  that $\bs{\sigma}^{(1)}$ defined in \eqref{vecalpha} is given by\footnote{ Equation \eqref{vecaSU3} depends on the branches for the square root functions. We can choose anyone and they all give correct result. Here we are using the principle branch. }
\begin{align}\label{vecaSU3} \nonumber\bs{\sigma}^{(1)}=&\sigma_1{\bs e}_1-\frac{\ri}{2\pi}\log\frac{-\E^{-4 \ri \pi  \sigma _1} \sqrt{1-4 \E^{6 \ri \pi  \sigma _1}}-\E^{-4 \ri \pi 
   \sigma _1}}{\sqrt{2}}{\bs e}_2\\
   &+(-\sigma_1+\frac{\ri}{2\pi}\log\frac{-\E^{-4 \ri \pi  \sigma _1} \sqrt{1-4 \E^{6 \ri \pi  \sigma _1}}-\E^{-4 \ri \pi 
   \sigma _1}}{2}){\bs e }_3.\end{align}
   The corresponding value of $x_1$ according to \eqref{mmap} is
 \be\label{x1sigma}\ba x_1=\E^{2\pi\ri\sigma_1}-\E^{-4\pi\ri\sigma_1}.\ea\ee
Next,
we want to find the values of $\sigma_1$ for which
\be\label{checkNO}   \sum_{\boldsymbol{w}\in Q_{2}}\frac{T^{\frac12(\boldsymbol{\sigma}+\boldsymbol{w})^2}}{\prod_{\boldsymbol{\alpha}\in \Delta}G\left(1+(\boldsymbol{\alpha},\boldsymbol{\sigma}+\boldsymbol{w})\right)}Z_{\rm inst}^{\rm 4d}(\boldsymbol{\sigma}+\boldsymbol{w},T)\big|_{\bs{\sigma}=\bs{\sigma}^{(1)}}  =0\ee
where the relevant Nekrasov partition function  $Z_{\rm inst}^{\rm 4d}$ is the one corresponding to $N=3$ in \eqref{exn3}. %
 Particularly, we fix $T$ and use different instanton counting number $n^{\text{inst}}$'s, to get numerical solution for the only parameter left, namely $\sigma_1$. We then use this value to compute the corresponding value of $x_1$ according to  \eqref{x1sigma}.
   Finally we get the energy using $E_0^{(1)}=\log(-x_1)$. We find that with the increasion of instanton number in \eqref{exn3}, $E_0^{(1)}$ converges to $E_0^{(1)*}$. An example can be found in 
\autoref{tb1}.
   
   Note we can find different solutions for \eqref{checkNO}, which corresponds to $E_n^{(1)}$ for some $n\geq 0$. We need to choose the one giving the minimal $\log(-x_1)$ so that it is the ground state energy $E_0^{(1)}$ we are looking for.

\begin{table}[t] 
\centering
   \begin{tabular}{l l l}
  \\
 $n^{\text{inst}}$ & $E_0^{(1)}((\frac{1}{3})^6)$  \\
\hline
  0&\underline{5.65}272224402310\\
 1 & \underline{5.65649}732964673 \\
 2 & \underline{5.656499622}13448\\
 3 & \underline{5.6564996223763}8\\ 
  4 &                 5.65649962237639 \\ 
 \hline
$E_0^{(1)*}((\frac{1}{3})^6)$ &  5.65649962237639    \\
\end{tabular}
\\
\caption{Comparison of $E_0^{(1)}$ for $T=\frac{1}{3^6}$, with instanton number $n^{\text{inst}}=0,\cdots,4$ and the numerical value  $E_0^{(1)*}$ in the example of $N=3$.
}
 \label{tb1}
 \end{table}
 
 We also performed similar test for $N=4$.

\newpage
\section{The $D^{(N)}_\ell$ coefficients}\label{app:dnres}
As discussed in \autoref{frommm},  by expanding around the Gaussian point we can get the coefficients $D^{(N)}_\ell$ in \eqref{eds} in a systematic way.
Below we report some explicit examples.
 \subsection{The $N=3$ example}

The first few coefficients are
\be\small\ba D_1^{(3)}&=-\frac{2 M_1^3}{3 \sqrt{3}}+\frac{1}{2} \sqrt{3} M_2 M_1^2+\frac{1}{2} \sqrt{3} M_2^2 M_1+\frac{5 M_1}{12 \sqrt{3}}-\frac{2 M_2^3}{3 \sqrt{3}}+\frac{5
   M_2}{12 \sqrt{3}}.\ea\ee

\be\label{d23}\small\ba D_2^{(3)}&=\frac{2 M_1^6}{27}-\frac{1}{3} M_2 M_1^5+\frac{1}{24} M_2^2 M_1^4+\frac{17 M_1^4}{54}+\frac{97}{108} M_2^3 M_1^3-\frac{389}{216} M_2 M_1^3+\frac{1}{24}
   M_2^4 M_1^2\\
  & -\frac{4}{3} M_2^2 M_1^2-\frac{85 M_1^2}{288}-\frac{1}{3} M_2^5 M_1-\frac{389}{216} M_2^3 M_1+\frac{493 M_2 M_1}{432}+\frac{2
   M_2^6}{27}+\frac{17 M_2^4}{54}-\frac{85 M_2^2}{288}.\ea\ee

 \be\small\ba
 D_3^{(3)}&=-\frac{4 M_1^9}{243 \sqrt{3}}-\frac{13 M_1^7}{54 \sqrt{3}}-\frac{439
   M_1^5}{432 \sqrt{3}}+\frac{32021 M_1^3}{31104 \sqrt{3}}+\frac{7 M_1}{144
   \sqrt{3}}+\frac{7 M_2}{144 \sqrt{3}}-\frac{18689 M_2^2
   M_1}{3456 \sqrt{3}}\\
   &+\frac{M_2 M_1^8}{9 \sqrt{3}}+\frac{577 M_2 M_1^6}{324 \sqrt{3}}+\frac{43133
   M_2 M_1^4}{5184 \sqrt{3}}+\frac{13429 M_2^2 M_1^3}{1728
   \sqrt{3}}-\frac{18689 M_2 M_1^2}{3456 \sqrt{3}}\\
   &-\frac{5 M_2^2 M_1^7}{36 \sqrt{3}}-\frac{13}{32} \sqrt{3} M_2^2
   M_1^5-\frac{10633 M_2^3 M_1^4}{2592 \sqrt{3}}+\frac{13429 M_2^3
   M_1^2}{1728 \sqrt{3}}+\frac{32021 M_2^3}{31104 \sqrt{3}}\\
   &-\frac{469 M_2^3 M_1^6}{1296 \sqrt{3}}+\frac{77 M_2^4 M_1^5}{144
   \sqrt{3}}-\frac{10633 M_2^4 M_1^3}{2592 \sqrt{3}}-\frac{13}{32} \sqrt{3}
   M_2^5 M_1^2+\frac{43133 M_2^4 M_1}{5184 \sqrt{3}}-\frac{439 M_2^5}{432
   \sqrt{3}}\\
   &-\frac{4 M_2^9}{243 \sqrt{3}}+\frac{M_1 M_2^8}{9 \sqrt{3}}-\frac{5 M_1^2
   M_2^7}{36 \sqrt{3}}-\frac{13 M_2^7}{54 \sqrt{3}}-\frac{469 M_1^3
   M_2^6}{1296 \sqrt{3}}+\frac{577 M_1 M_2^6}{324 \sqrt{3}}+\frac{77 M_1^4
   M_2^5}{144 \sqrt{3}}.
 \ea\ee
 
 \be\small\ba
 D_4^{(3)}&=\frac{2 M_1^{12}}{2187}+\frac{61 M_1^{10}}{2187}+\frac{3857
   M_1^8}{11664}+\frac{216877 M_1^6}{139968}-\frac{5629919
   M_1^4}{4478976}-\frac{7039 M_1^2}{15552}+\frac{5987 M_2 M_1}{2592}\\
   &-\frac{2}{243} M_2 M_1^{11}-\frac{2315 M_2 M_1^9}{8748}-\frac{25823 M_2
   M_1^7}{7776}-\frac{118985 M_2 M_1^5}{6912}+\frac{10986997 M_2
   M_1^3}{1119744}\\
   &+\frac{19}{972} M_2^2 M_1^{10}+\frac{677 M_2^2 M_1^8}{1296}+\frac{215459
   M_2^2 M_1^6}{62208}-\frac{41389 M_2^2 M_1^4}{3456}+\frac{335209 M_2^2
   M_1^2}{82944}-\frac{7039 M_2^2}{15552}\\
   &+\frac{307 M_2^3 M_1^9}{17496}+\frac{10231 M_2^3 M_1^7}{15552}+\frac{143657
   M_2^3 M_1^5}{15552}-\frac{4874117 M_2^3 M_1^3}{279936}+\frac{10986997
   M_2^3 M_1}{1119744}\\
   &-\frac{1021 M_2^4 M_1^8}{10368}-\frac{25295 M_2^4
   M_1^6}{23328}+\frac{787919 M_2^4 M_1^4}{93312}-\frac{41389 M_2^4
   M_1^2}{3456}-\frac{5629919 M_2^4}{4478976}-\frac{25295 M_1^4
   M_2^6}{23328}\\
   &-\frac{M_2^5 M_1^7}{2592}-\frac{4807 M_2^5 M_1^5}{2592}+\frac{143657 M_2^5
   M_1^3}{15552}+\frac{215459 M_2^6 M_1^2}{62208}-\frac{118985 M_2^5
   M_1}{6912}+\frac{216877 M_2^6}{139968}\\
   &+\frac{677 M_1^2 M_2^8}{1296}+\frac{3857 M_2^8}{11664}-\frac{M_1^5
   M_2^7}{2592}+\frac{10231 M_1^3 M_2^7}{15552}-\frac{25823 M_1
   M_2^7}{7776}+\frac{8113 M_1^6 M_2^6}{46656}\\
   &+\frac{2 M_2^{12}}{2187}-\frac{2}{243} M_1 M_2^{11}+\frac{19}{972} M_1^2
   M_2^{10}+\frac{61 M_2^{10}}{2187}+\frac{307 M_1^3
   M_2^9}{17496}-\frac{2315 M_1 M_2^9}{8748}-\frac{1021 M_1^4 M_2^8}{10368}.
 \ea\ee

\subsection{The $N=4$ example}

\be\small\ba D_1^{(4)}&=-\frac{M_1^3}{\sqrt{2}}+2 M_2 M_1^2+\frac{M_3 M_1^2}{2 \sqrt{2}}+\sqrt{2} M_2^2 M_1+\frac{M_3^2 M_1}{2 \sqrt{2}}+\frac{3 M_1}{4
   \sqrt{2}}-\frac{M_2^3}{4}-\frac{M_3^3}{\sqrt{2}}\\
   &+2 M_2 M_3^2+\frac{M_2}{8}+\sqrt{2} M_2^2 M_3+\frac{3 M_3}{4 \sqrt{2}}.\ea\ee
   
\be\label{d24}\small\ba D_2^{(4)}&=\frac{M_1^6}{4}-\sqrt{2} M_2 M_1^5-\frac{1}{4} M_3 M_1^5+M_2^2 M_1^4-\frac{3}{16} M_3^2 M_1^4+\frac{M_2 M_3 M_1^4}{\sqrt{2}}+\frac{9 M_1^4}{8}+\frac{17
   M_2^3 M_1^3}{4 \sqrt{2}}\\
   &+\frac{5}{8} M_3^3 M_1^3-\frac{M_2 M_3^2 M_1^3}{\sqrt{2}}-\frac{85 M_2 M_1^3}{8 \sqrt{2}}-\frac{1}{2} M_2^2 M_3
   M_1^3-\frac{17}{16} M_3 M_1^3+\frac{1}{2} M_2^4 M_1^2-\frac{3}{16} M_3^4 M_1^2\\
   &-\frac{M_2 M_3^3 M_1^2}{\sqrt{2}}-4 M_2^2 M_1^2+5 M_2^2 M_3^2
   M_1^2-\frac{1}{4} M_3^2 M_1^2+\frac{31 M_2^3 M_3 M_1^2}{8 \sqrt{2}}+\frac{57 M_2 M_3 M_1^2}{16 \sqrt{2}}-\frac{79 M_1^2}{64}\\
  & -\frac{M_2^5 M_1}{2
   \sqrt{2}}-\frac{1}{4} M_3^5 M_1+\frac{M_2 M_3^4 M_1}{\sqrt{2}}-\frac{79 M_2^3 M_1}{16 \sqrt{2}}-\frac{1}{2} M_2^2 M_3^3 M_1-\frac{17}{16} M_3^3
   M_1+\frac{31 M_2^3 M_3^2 M_1}{8 \sqrt{2}}\\
   &+\frac{57 M_2 M_3^2 M_1}{16 \sqrt{2}}+\frac{179 M_2 M_1}{32 \sqrt{2}}+2 M_2^4 M_3 M_1+\frac{11}{2} M_2^2
   M_3 M_1+\frac{29 M_3 M_1}{32}+\frac{M_2^6}{32}+\frac{M_3^6}{4}\\
   &-\sqrt{2} M_2 M_3^5+\frac{M_2^4}{8}+M_2^2 M_3^4+\frac{9 M_3^4}{8}+\frac{17 M_2^3
   M_3^3}{4 \sqrt{2}}-\frac{85 M_2 M_3^3}{8 \sqrt{2}}-\frac{11 M_2^2}{128}+\frac{1}{2} M_2^4 M_3^2\\
   &-4 M_2^2 M_3^2-\frac{79 M_3^2}{64}-\frac{M_2^5 M_3}{2
   \sqrt{2}}-\frac{79 M_2^3 M_3}{16 \sqrt{2}}+\frac{179 M_2 M_3}{32 \sqrt{2}}.\ea\ee

\be\small\ba D_3^{(4)}&=-\frac{M_1^9}{12 \sqrt{2}}+\frac{1}{2} M_2 M_1^8+\frac{M_3 M_1^8}{8 \sqrt{2}}-\frac{3 M_2^2 M_1^7}{2 \sqrt{2}}+\frac{M_3^2 M_1^7}{16
   \sqrt{2}}-\frac{1}{2} M_2 M_3 M_1^7-\frac{21 M_1^7}{16 \sqrt{2}}-\frac{35}{48} M_2^3 M_1^6\\
   &-\frac{35 M_3^3 M_1^6}{96 \sqrt{2}}+\frac{1}{8} M_2 M_3^2
   M_1^6+\frac{265}{32} M_2 M_1^6+\frac{M_2^2 M_3 M_1^6}{\sqrt{2}}+\frac{13 M_3 M_1^6}{8 \sqrt{2}}+\frac{7 M_2^4 M_1^5}{2 \sqrt{2}}+\frac{7 M_3^4
   M_1^5}{32 \sqrt{2}}\\
   &+\frac{3}{4} M_2 M_3^3 M_1^5-\frac{31 M_2^2 M_1^5}{2 \sqrt{2}}-\frac{31 M_2^2 M_3^2 M_1^5}{8 \sqrt{2}}+\frac{39 M_3^2 M_1^5}{64
   \sqrt{2}}-\frac{15}{16} M_2^3 M_3 M_1^5-\frac{197}{32} M_2 M_3 M_1^5\\
   &-\frac{393 M_1^5}{64 \sqrt{2}}+\frac{7}{4} M_2^5 M_1^4+\frac{7 M_3^5 M_1^4}{32
   \sqrt{2}}-\frac{3}{4} M_2 M_3^4 M_1^4-\frac{581}{32} M_2^3 M_1^4+\frac{7 M_2^2 M_3^3 M_1^4}{8 \sqrt{2}}-\frac{67 M_3^3 M_1^4}{64
   \sqrt{2}}\\
   &+\frac{259}{64} M_2^3 M_3^2 M_1^4-\frac{227}{128} M_2 M_3^2 M_1^4+\frac{8017}{192} M_2 M_1^4+\frac{9 M_2^4 M_3 M_1^4}{4 \sqrt{2}}-\frac{3
   M_2^2 M_3 M_1^4}{4 \sqrt{2}}+\frac{1603 M_3 M_1^4}{384 \sqrt{2}}\\
   &-\frac{35 M_2^6 M_1^3}{96 \sqrt{2}}-\frac{35 M_3^6 M_1^3}{96 \sqrt{2}}+\frac{3}{4}
   M_2 M_3^5 M_1^3-\frac{65 M_2^4 M_1^3}{4 \sqrt{2}}+\frac{7 M_2^2 M_3^4 M_1^3}{8 \sqrt{2}}-\frac{67 M_3^4 M_1^3}{64 \sqrt{2}}-\frac{69}{32} M_2^3
   M_3^3 M_1^3\\
   &+\frac{565}{64} M_2 M_3^3 M_1^3+\frac{11861 M_2^2 M_1^3}{384 \sqrt{2}}+\frac{35 M_2^4 M_3^2 M_1^3}{4 \sqrt{2}}-\frac{61 M_2^2 M_3^2
   M_1^3}{4 \sqrt{2}}+\frac{971 M_3^2 M_1^3}{384 \sqrt{2}}+\frac{33}{8} M_2^5 M_3 M_1^3\\
   &+\frac{157}{64} M_2^3 M_3 M_1^3-\frac{2361}{128} M_2 M_3
   M_1^3+\frac{5443 M_1^3}{768 \sqrt{2}}-\frac{3}{16} M_2^7 M_1^2+\frac{M_3^7 M_1^2}{16 \sqrt{2}}+\frac{1}{8} M_2 M_3^6 M_1^2-\frac{57}{16} M_2^5
   M_1^2\\
   &-\frac{31 M_2^2 M_3^5 M_1^2}{8 \sqrt{2}}+\frac{39 M_3^5 M_1^2}{64 \sqrt{2}}+\frac{259}{64} M_2^3 M_3^4 M_1^2-\frac{227}{128} M_2 M_3^4
   M_1^2+\frac{14897}{768} M_2^3 M_1^2+\frac{35 M_2^4 M_3^3 M_1^2}{4 \sqrt{2}}\\
   &-\frac{61 M_2^2 M_3^3 M_1^2}{4 \sqrt{2}}+\frac{971 M_3^3 M_1^2}{384
   \sqrt{2}}+\frac{11}{4} M_2^5 M_3^2 M_1^2-\frac{237}{16} M_2^3 M_3^2 M_1^2-\frac{183}{32} M_2 M_3^2 M_1^2-\frac{41741 M_2 M_1^2}{1536}\\
   &+\frac{65 M_2^6
   M_3 M_1^2}{64 \sqrt{2}}-\frac{161 M_2^4 M_3 M_1^2}{16 \sqrt{2}}-\frac{5379 M_2^2 M_3 M_1^2}{256 \sqrt{2}}-\frac{3311 M_3 M_1^2}{768
   \sqrt{2}}+\frac{M_2^8 M_1}{16 \sqrt{2}}+\frac{M_3^8 M_1}{8 \sqrt{2}}\\
   &-\frac{1}{2} M_2 M_3^7 M_1+\frac{195 M_2^6 M_1}{128 \sqrt{2}}+\frac{M_2^2 M_3^6
   M_1}{\sqrt{2}}+\frac{13 M_3^6 M_1}{8 \sqrt{2}}-\frac{15}{16} M_2^3 M_3^5 M_1-\frac{197}{32} M_2 M_3^5 M_1\\
   &+\frac{2849 M_2^4 M_1}{192
   \sqrt{2}}+\frac{9 M_2^4 M_3^4 M_1}{4 \sqrt{2}}-\frac{3 M_2^2 M_3^4 M_1}{4 \sqrt{2}}+\frac{1603 M_3^4 M_1}{384 \sqrt{2}}+\frac{33}{8} M_2^5 M_3^3
   M_1+\frac{157}{64} M_2^3 M_3^3 M_1\\
   &-\frac{2361}{128} M_2 M_3^3 M_1-\frac{18883 M_2^2 M_1}{1536 \sqrt{2}}+\frac{65 M_2^6 M_3^2 M_1}{64
   \sqrt{2}}-\frac{161 M_2^4 M_3^2 M_1}{16 \sqrt{2}}-\frac{5379 M_2^2 M_3^2 M_1}{256 \sqrt{2}}\\
   &-\frac{1}{2} M_2^7
   M_3 M_1-\frac{89}{8} M_2^5 M_3 M_1-\frac{3621}{128} M_2^3 M_3 M_1+\frac{4925}{256} M_2 M_3 M_1+\frac{5 M_1}{32
   \sqrt{2}}-\frac{M_2^9}{384}-\frac{M_3^9}{12 \sqrt{2}}\\
   &+\frac{1}{2} M_2 M_3^8-\frac{9 M_2^7}{256}-\frac{3 M_2^2 M_3^7}{2 \sqrt{2}}-\frac{21 M_3^7}{16
   \sqrt{2}}-\frac{35}{48} M_2^3 M_3^6+\frac{265}{32} M_2 M_3^6-\frac{67 M_2^5}{512}+\frac{7 M_2^4 M_3^5}{2 \sqrt{2}}-\\
   &\frac{393 M_3^5}{64 \sqrt{2}}+\frac{7}{4} M_2^5 M_3^4-\frac{581}{32} M_2^3 M_3^4+\frac{8017}{192} M_2 M_3^4+\frac{269
   M_2^3}{3072}-\frac{35 M_2^6 M_3^3}{96 \sqrt{2}}-\frac{65 M_2^4 M_3^3}{4 \sqrt{2}}+\\
   &\frac{11861 M_2^2 M_3^3}{384 \sqrt{2}}+\frac{5443 M_3^3}{768
   \sqrt{2}}-\frac{3}{16} M_2^7 M_3^2-\frac{57}{16} M_2^5 M_3^2+\frac{14897}{768} M_2^3 M_3^2-\frac{41741 M_2 M_3^2}{1536}+\frac{M_2}{128}+\\
   &\frac{M_2^8
   M_3}{16 \sqrt{2}}+\frac{195 M_2^6 M_3}{128 \sqrt{2}}+\frac{2849 M_2^4 M_3}{192 \sqrt{2}}-\frac{18883 M_2^2 M_3}{1536 \sqrt{2}}+\frac{5 M_3}{32
   \sqrt{2}}-\frac{31 M_2^2 M_3^5}{2
   \sqrt{2}}-\frac{3311 M_3^2 M_1}{768 \sqrt{2}}.
\ea\ee

 \subsection{The $D^{(N)}_2$ at generic $N$}
 
In principle the method discussed above can be used to obtain $D^{(N)}_m$ for generic $N$ and $m$. However, the expression quickly becomes cumbersome unless we fix $N$ to some value. For $m=1$ this is given in \eqref{Einfty1}. For $m=2$ we get

\be\small\ba\label{Einfty2} 
D_2^{(N)}=&\sum_{l=1}^{N-1}\left(-\frac{ \left(-124 \cos \left(\frac{2 \pi  l}{N}\right)+\cos
   \left(\frac{4 \pi  l}{N}\right)-237\right) \csc ^4\left(\frac{\pi 
   l}{N}\right)}{2304\sin\left(\frac{\pi l}{N}\right)^2}M_l^2 \left(M_l^2-1\right)\right)\\
&+\sum_{l=1}^{N-1}\frac{ \left(1-3 \csc ^2\left(\frac{\pi 
   l}{N}\right)\right)^2}{288\sin\left(\frac{\pi l}{N}\right)^2} M_l^2 \left(M_l^4-1\right)\\
&+\sum_{l=1}^{N-1}\frac{M_l^2 \left(4 \left(M_l^2+8\right) M_l^2+45\right)}{1152\sin\left(\frac{\pi l}{N}\right)^2}\\
&+\sum_{l=1}^{N-1}-\frac{\left(1-3 \csc ^2\left(\frac{\pi 
   l}{N}\right)\right)}{288\sin\left(\frac{\pi l}{N}\right)^2} M_l^2 \left(2 M_l^4+7 M_l^2-9\right) \\
&+\sum\limits_{1\leq l< l'\leq N-1}\frac{ \left(1-3 \csc ^2\left(\frac{\pi 
   l}{N}\right)\right)^2}{288\sin\left(\frac{\pi l}{N}\right)\sin\left(\frac{\pi l'}{N}\right)} M_l(M_l^2-1) M_{l'}(M_{l'}^2-1)\\
&+\sum\limits_{1\leq l< l'\leq N-1}\frac{M_l(1+2M_l^2)M_{l'}(1+2M_{l'}^2)}{576\sin\left(\frac{\pi l}{N}\right)\sin\left(\frac{\pi l'}{N}\right)}\\
&+\sum_{l\neq l'=1}^{N-1}-\frac{\left(1-3 \csc ^2\left(\frac{\pi 
   l'}{N}\right)\right)}{288\sin(\frac{\pi l}{N})\sin\left(\frac{\pi l'}{N}\right)}M_l(2M_l^2+1) M_{l'}(M_{l'}^2-1)\\
&-\sum\limits_{1\leq l< l'\leq N-1}\frac{1}{96} M_l M_{l'} \sin \left(\frac{\pi  l}{N}\right) \sin \left(\frac{\pi 
   l'}{N}\right) \left(\cos \left(\frac{\pi  (l+l')}{N}\right)+5\right)
   \csc ^2\left(\frac{\pi  (l-l')}{2 N}\right) \csc ^4\left(\frac{\pi 
   (l+l')}{2 N}\right) \\
  &\quad \times\left(6 M_l M_{l'} \csc \left(\frac{\pi 
   l}{N}\right) \csc \left(\frac{\pi  l'}{N}\right)+\left(2 M_l^2+1\right) \csc
   ^2\left(\frac{\pi  l}{N}\right)+\left(2 M_{l'}^2+1\right) \csc
   ^2\left(\frac{\pi  l'}{N}\right)\right)\\
&+\sum\limits_{1\leq l< l'\leq N-1}\frac{1}{2 \left(\cos \left(\frac{\pi  l}{N}\right)-\cos
   \left(\frac{\pi  l'}{N}\right)\right)^4}
M_l M_{l'} \\
&\quad\times\bigg(M_l^3 M_{l'} \sin ^2\left(\frac{\pi 
   l'}{N}\right)\\
   &\quad\quad+ M_l M_{l'} \left(M_{l'}^2 \sin ^2\left(\frac{\pi 
   l}{N}\right)+2 \left(-3 \sin \left(\frac{\pi  l}{N}\right) \sin \left(\frac{\pi 
   l'}{N}\right)+\sin ^2\left(\frac{\pi  l}{N}\right)+\sin ^2\left(\frac{\pi 
   l'}{N}\right)\right)\right)\\
   &\quad\quad -2 M_{l'}^2 \sin ^2\left(\frac{\pi 
   l}{N}\right)-2 M_l^2 \sin \left(\frac{\pi  l'}{N}\right) \left(\sin
   \left(\frac{\pi  l'}{N}\right)-M_{l'}^2 \sin \left(\frac{\pi 
   l}{N}\right)\right)+4 \sin \left(\frac{\pi  l}{N}\right) \sin \left(\frac{\pi 
   l'}{N}\right)\\
   &\quad\quad -\sin ^2\left(\frac{\pi  l}{N}\right)-\sin ^2\left(\frac{\pi 
   l'}{N}\right)\bigg)   \\
&-\sum_{l\neq l'=1}^{N-1}\frac{M_l \left(M_l^2-1\right) M_{l'} \left(\cos \left(\frac{2 \pi 
   l}{N}\right)+5\right) \csc ^3\left(\frac{\pi  l}{N}\right) \left(\left(M_l^2+2\right)
   \sin \left(\frac{\pi  l'}{N}\right)+M_l M_{l'} \sin \left(\frac{\pi 
   l}{N}\right)\right)}{24 \left(\cos \left(\frac{\pi  l}{N}\right)-\cos \left(\frac{\pi 
   l'}{N}\right)\right)^2}\\
&+\sum_{l\neq l'=1}^{N-1}\frac{M_l \left(2 M_l^2+1\right) M_{l'} \left(\left(M_l^2+4\right) \left(-\csc
   \left(\frac{\pi  l}{N}\right)\right) \sin \left(\frac{\pi  l'}{N}\right)-M_l
   M_{l'}\right)}{24 \left(\cos \left(\frac{\pi  l}{N}\right)-\cos \left(\frac{\pi
    l'}{N}\right)\right)^2}\\
&\text{[continued on next page]}
\ea\ee

\be\small\ba
&\text{[continued from last page]}\\
&+\sum\limits_{1\leq l< l'<l''\leq N-1}\bigg(\frac{\sin \left(\frac{\pi  l}{N}\right) \sin ^2\left(\frac{\pi  l'}{N}\right) \sin \left(\frac{\pi  l''}{N}\right)}{\left(\cos \left(\frac{\pi  l}{N}\right)-\cos
   \left(\frac{\pi  l'}{N}\right)\right)^2 \left(\cos \left(\frac{\pi  l'}{N}\right)-\cos \left(\frac{\pi  l''}{N}\right)\right)^2}\\
&\quad\quad\quad\quad\quad\quad\quad\quad\quad\times\bigg( M_l \left(M_{l'}^2+2\right) M_{l'}^2 M_{l''} \csc ^2\left(\frac{\pi  l'}{N}\right)+M_l M_{l'}^3 M_{l''}^2 \csc \left(\frac{\pi 
   l'}{N}\right) \csc \left(\frac{\pi  l''}{N}\right)\\
   &\quad\quad\quad\quad\quad\quad\quad\quad\quad\quad\quad+M_l^2 M_{l'}^3 M_{l''} \csc \left(\frac{\pi  l}{N}\right) \csc \left(\frac{\pi 
   l'}{N}\right)+M_l^2 M_{l'}^2 M_{l''}^2 \csc \left(\frac{\pi  l}{N}\right) \csc \left(\frac{\pi  l''}{N}\right)\bigg)\\
   &\quad\quad\quad\quad\quad\quad\quad\quad+(l\leftrightarrow l')+(l\leftrightarrow l'')\bigg)\\
&+\sum\limits_{1\leq l< l'<l''\leq N-1}\bigg(\frac{\left(1-3 \csc ^2\left(\frac{\pi  l}{N}\right)\right) \sin \left(\frac{\pi  l'}{N}\right) \sin \left(\frac{\pi  l''}{N}\right)}{12 \left(\cos \left(\frac{\pi 
   l'}{N}\right)-\cos \left(\frac{\pi  l''}{N}\right)\right)^2}\\
   &\quad\quad\quad\quad\quad\quad\quad\quad\quad\times M_l \left(M_l^2-1\right) \csc \left(\frac{\pi  l}{N}\right) \left(M_{l'}^2 M_{l''} \csc \left(\frac{\pi  l'}{N}\right)+M_{l'} M_{l''}^2 \csc
   \left(\frac{\pi  l''}{N}\right)\right)\\
   &\quad\quad\quad\quad\quad\quad\quad\quad+(l\leftrightarrow l')+(l\leftrightarrow l'')\bigg)\\
&+\sum\limits_{1\leq l< l'<l''\leq N-1}\bigg(-\frac{\sin \left(\frac{\pi  l'}{N}\right) \sin \left(\frac{\pi  l''}{N}\right)}{24 \left(\cos \left(\frac{\pi  l'}{N}\right)-\cos \left(\frac{\pi 
   l''}{N}\right)\right)^2}\\
    &\quad\quad\quad\quad\quad\quad\quad\quad\quad\times M_l \left(2 M_l^2+1\right) \csc \left(\frac{\pi  l}{N}\right) \left(M_{l'}^2 M_{l''} \csc \left(\frac{\pi  l'}{N}\right)+M_{l'} M_{l''}^2 \csc
   \left(\frac{\pi  l''}{N}\right)\right)\\
    &\quad\quad\quad\quad\quad\quad\quad\quad+(l\leftrightarrow l')+(l\leftrightarrow l'')\bigg)\\
&+\sum\limits_{1\leq l< l'<l''<l'''\leq N-1}\bigg(\frac{\sin \left(\frac{\pi  l}{N}\right) \sin \left(\frac{\pi  l'}{N}\right) \sin \left(\frac{\pi  l''}{N}\right) \sin \left(\frac{\pi 
   l'''}{N}\right)}{\left(\cos \left(\frac{\pi  l}{N}\right)-\cos \left(\frac{\pi  l'}{N}\right)\right)^2 \left(\cos \left(\frac{\pi  l''}{N}\right)-\cos
   \left(\frac{\pi  l'''}{N}\right)\right)^2}\\
   &\quad\quad\quad\quad\quad\quad\quad\quad\quad\quad\times\bigg(M_l M_{l'}^2 M_{l''}^2 M_{l'''} \csc \left(\frac{\pi  l'}{N}\right) \csc \left(\frac{\pi  l''}{N}\right)+M_l^2 M_{l'} M_{l''}^2
   M_{l'''} \csc \left(\frac{\pi  l}{N}\right) \csc \left(\frac{\pi  l''}{N}\right)\\
   &\quad\quad\quad\quad\quad\quad\quad\quad\quad\quad\quad\quad+M_l M_{l'}^2 M_{l''} M_{l'''}^2 \csc \left(\frac{\pi 
   l'}{N}\right) \csc \left(\frac{\pi  l'''}{N}\right)+M_l^2 M_{l'} M_{l''} M_{l'''}^2 \csc \left(\frac{\pi  l}{N}\right) \csc \left(\frac{\pi 
   l'''}{N}\right)\bigg)\\
   &\quad\quad\quad\quad\quad\quad\quad\quad\quad\quad+(l'\leftrightarrow l'')+(l'\leftrightarrow l''')\bigg)
\ea\ee
If we fix $N=3$, this reduces to \eqref{d23}, while for $N=4$, we recover \eqref{d24}. It should be possible to reorganize  the expression above in a simpler combinatorial expression. 

\section{Bilinear relations around infinity}\label{appbilin}

\subsection{Structure constants}
\label{appen:stconst}
In this subsection, we want to determine the form of $C(\mathbf{M})$ in our Ansatz \eqref{eq:ansatz1}. 

Substitute \eqref{eq:ansatz1} into \eqref{eq:bilinearInfinity1}. Note that we get a series of equations for different powers of~$\mathfrak{r}$. Besides, we can choose arbitrary \(\boldsymbol{\varepsilon}\) in those equations, which goes back to the choice of $\mathbf{M}$ in \eqref{eq:45}. However, to get the explicit expression of $C(\mathbf{M})$, we don't need to solve all of such equations. It is enough to use just several of them. We use the first few of them with simpler expression. Precisely, we choose the following 3 equations: 
\begin{itemize}
\item 
As usual, we first look at the equation coming from the coefficient of \(\mathfrak{r}^0\).   The first non-trivial coefficient for \(\mathfrak{r}^0\) appears when  \(\boldsymbol{\varepsilon}=\mathbf{e}_j+\mathbf{e}_k\)\footnote{We warn the reader that $\mathbf{e}_j$ in this section should not be confused with weights of the fundamental representation. In this section,   $\mathbf{e}_j$  denotes the $N-1$  dimensional vector $(0,\cdots, 0,1,0,\cdots 0)$ where the only non-zero component is the $j$th component.}.  In this particular case, only the first line of \eqref{eq:bilinearInfinity1} contributes and the \(\boldsymbol{\Delta}\) is among \(0, -\mathbf{e}_j, -\mathbf{e}_k \text{ and } -\mathbf{e}_j-\mathbf{e}_k\). To be more precise, by requiring the coefficient to vanish, we have the following equation
\begin{equation}\begin{gathered}
\label{eq:53}
C(\mathbf{M}+\frac12 \mathbf{e}_j+\frac12 \mathbf{e}_k)C(\mathbf{M}-\frac12 \mathbf{e}_j-\frac12 \mathbf{e}_k)\left( (\sin \frac{\pi j}{N}+\sin \frac{\pi k}{N})^2-\sin^2 \frac{\pi(j+k)}{N} \right)+
\\
+
C(\mathbf{M}+\frac12 \mathbf{e}_j-\frac12 \mathbf{e}_k)C( \mathbf{M}-\frac12 \mathbf{e}_j+\frac12 \mathbf{e}_k)\left( (\sin \frac{\pi j}{N}-\sin \frac{\pi k}{N})^2-\sin^2 \frac{\pi(j-k)}{N} \right)=0,
\end{gathered}
\end{equation}
which can be rewritten as
\begin{equation}
\label{eq:54}
\frac{C(\mathbf{M}+\frac12 \mathbf{e}_j+\frac12 \mathbf{e}_k)C( \mathbf{M}-\frac12 \mathbf{e}_j-\frac12 \mathbf{e}_k)}{C( \mathbf{M}+\frac12 \mathbf{e}_j-\frac12 \mathbf{e}_k)C( \mathbf{M}-\frac12 \mathbf{e}_j+\frac12 \mathbf{e}_k)}=
\left(\frac{\sin\frac{(j-k)\pi}{2N}}{\sin\frac{(j+k)\pi}{2N}}\right)^2.
\end{equation}
We introduce
\begin{equation}
\label{eq:55}
f_k(\mathbf{K})=\frac{C(\mathbf{K}+\frac12 \mathbf{e}_k)}{C(\mathbf{K}-\frac12 \mathbf{e}_k)},
\end{equation}
then
\begin{equation}
\label{eq:56}
\frac{f_k(\mathbf{M}+\frac12 \mathbf{e}_j)}{f_k( \mathbf{M}-\frac12 \mathbf{e}_j)}=\left(\frac{\sin\frac{(j-k)\pi}{2N}}{\sin\frac{(j+k)\pi}{2N}}\right)^2
\end{equation}
is solved by
\begin{equation}
\label{eq:57}
f_k(\mathbf{M})=\tilde{\phi}_k(M_k)\prod_{j\neq k}\left(\frac{\sin\frac{(j-k)\pi}{2N}}{\sin\frac{(j+k)\pi}{2N}}\right)^{2M_j}.
\end{equation}
We can further solve for $C(\mathbf{M})$ and obtain
\begin{equation}
\label{eq:58}
C(\mathbf{M})=\phi(M_1,\ldots, \widehat{M_k},\ldots , M_{N-1} )\phi_k(M_k)
\prod_{j<k}\left(\frac{\sin\frac{(j-k)\pi}{2N}}{\sin\frac{(j+k)\pi}{2N}}\right)^{2M_jM_k}.
\end{equation}
Using the permutation symmetry of the set of all variables, we conclude
\begin{equation}
\label{eq:cAnsatz}
C(\mathbf{M})=\prod_{k=1}^{N-1}\phi_k(M_k)
\prod_{j<k}\left(\frac{\sin\frac{(j-k)\pi}{2N}}{\sin\frac{(j+k)\pi}{2N}}\right)^{2M_jM_k}.
\end{equation}
\item In order to get \(\phi_k(M_k)\), we use the coefficient of \(\mathfrak{r}^{-1}\) when \(\boldsymbol{\varepsilon}=0\). The coefficient consists of three parts, corresponding to the first line of \eqref{eq:bilinearInfinity1} with \(\boldsymbol{\Delta}=\pm \mathbf{e}_k\)  and the second line with \(\boldsymbol{\Delta}=0\), respectively:
\begin{equation}
\label{eq:48}
\sum_k 8\sin^4 \frac{\pi k}{N} C(\mathbf{M}+\mathbf{e}_k) C(\mathbf{M}-\mathbf{e}_k)
-2C(\mathbf{M})^2\sum_kM_k\sin \frac{\pi k}{N}=0.
\end{equation}
Plugging in \eqref{eq:cAnsatz}, we find that $\phi_k(M_k)$ satisfies
\begin{equation}
\label{eq:equationM0}
\sum_k 4\sin^4 \frac{\pi k}{N} \frac{\phi_k(M_k+1)\phi_k(M_k-1)}{\phi(M_k)^2}
=\sum_kM_k\sin \frac{\pi k}{N}.
\end{equation}
This equation itself is hard to use, but combined with the equation obtained in the next item, it will enable us to solve for \(\phi_k(M_k)\).
\item We further use the coefficient of \(\mathfrak{r}^{-1}\) for \(\boldsymbol{\varepsilon}=\mathbf{e}_j\). The coefficient involves the first line of \eqref{eq:bilinearInfinity1} when \(\boldsymbol{\Delta}=\pm\mathbf{e}_k\)  and \(\boldsymbol{\Delta}=-\mathbf{e}_j\pm\mathbf{e}_k\), and the second line when \(\boldsymbol{\Delta}=0\) and \(\boldsymbol{\Delta}=-\mathbf{e}_j\):
\begin{equation}\begin{gathered}
\label{eq:62}
0=
4 C(\mathbf{M}+\frac12 \mathbf{e}_j)C(\mathbf{M}-\frac12 \mathbf{e}_j) \left( M_j \sin \frac{\pi j}{N}- \left( \sin \frac{\pi \mathbf{k}}{N},\mathbf{M} \right) \right)+
\\
+
2\sum_{k\neq j}C(\mathbf{M}+\frac12\mathbf{e}_j+\mathbf{e}_k)C(\mathbf{M}-\frac12\mathbf{e}_j-\mathbf{e}_k)\left( (\sin \frac{\pi j}{N}+2\sin \frac{\pi k}{N})^2-\sin^2 \frac{\pi(2k+j)}{N} \right)+
\\
+
2\sum_{k\neq j}C(\mathbf{M}+\frac12\mathbf{e}_j-\mathbf{e}_k)C(\mathbf{M}-\frac12\mathbf{e}_j+\mathbf{e}_k)\left( (\sin \frac{\pi j}{N}-2\sin \frac{\pi k}{N})^2-\sin^2 \frac{\pi(-2k+j)}{N} \right),
\end{gathered}
\end{equation}
and after some rewriting:
\begin{equation}\begin{gathered}
\label{eq:relation1e}
0=
-2 \sum_{k\neq j} M_k\sin \frac{\pi k}{N}+
\\
+
\sum_{k\neq j}\frac{C(\mathbf{M}+\frac12\mathbf{e}_j+\mathbf{e}_k)C(\mathbf{M}-\frac12\mathbf{e}_j-\mathbf{e}_k)}{C(\mathbf{M}+\frac12\mathbf{e}_j)C(\mathbf{M}-\frac12\mathbf{e}_j)}
\left( (\sin \frac{\pi j}{N}+2\sin \frac{\pi k}{N})^2-\sin^2 \frac{\pi(2k+j)}{N} \right)+
\\
+
\sum_{k\neq j}\frac{C(\mathbf{M}+\frac12\mathbf{e}_j-\mathbf{e}_k)C(\mathbf{M}-\frac12\mathbf{e}_j+\mathbf{e}_k)}{C(\mathbf{M}+\frac12\mathbf{e}_j)C(\mathbf{M}-\frac12\mathbf{e}_j)}
\left( (\sin \frac{\pi j}{N}-2\sin \frac{\pi k}{N})^2-\sin^2 \frac{\pi(-2k+j)}{N} \right).
\end{gathered}\end{equation}
Plug in \eqref{eq:cAnsatz} into the combination
\begin{equation}\begin{gathered}
\label{eq:64}
\frac{C(\mathbf{M}+\frac12\mathbf{e}_j+\mathbf{e}_k)C(\mathbf{M}-\frac12\mathbf{e}_j-\mathbf{e}_k)}{C(\mathbf{M}+\frac12\mathbf{e}_j)C(\mathbf{M}-\frac12\mathbf{e}_j)}=\frac{\phi_k(M_k+1)\phi_k(M_k-1)}{\phi_k(M)^2}\times
\\
\times
\prod_{l\neq k} \left( \frac{\sin \frac{(l-k)\pi}{2N}}{\sin \frac{(l+k)\pi}{2N}} \right)^{2(M_l+\frac12 \delta_{jl})}
\left( \frac{\sin \frac{(l-k)\pi}{2N}}{\sin \frac{(l+k)\pi}{2N}} \right)^{-2(M_l-\frac12 \delta_{jl})}=
\\
=
\frac{\phi_k(M_k+1)\phi_k(M_k-1)}{\phi_k(M)^2}
\left( \frac{\sin \frac{(j-k)\pi}{2N}}{\sin \frac{(j+k)\pi}{2N}} \right)^2.
\end{gathered}
\end{equation}
Using \eqref{eq:64}, \eqref{eq:relation1e} becomes
\begin{equation}
\label{eq:65}
\sum_{k\neq j}M_k \sin\frac{\pi k}{N}=\sum_{k\neq j} 4\sin^4 \left(\frac{\pi k}{N}\right)\frac{\phi_k(M_k+1)\phi_k(M_k+1)}{\phi_k(M_k)^2}.
\end{equation}
Subtracting this equation from \eqref{eq:equationM0}, we get
\begin{equation}
\label{eq:59}
\frac{\phi_k(M_k+1)\phi_k(M_k-1)}{\phi_k(M_k)^2}=\frac{M_k}{4\sin^3 \frac{\pi k}{N}},
\end{equation}
which is solved by
\begin{equation}
\label{eq:60}
\phi_k(M_k)=G(M_k+1) \left(\sin \frac{\pi k}{N} \right)^{-\frac{3}{2}M_k^2}2^{-M_k^2} a_k b_k^{M_k}.
\end{equation}
Therefore,
\begin{equation}
\label{eq:52}
C(\mathbf{M})=\prod_{k=1}^{N-1}G(M_k+1)\left(\sin \frac{\pi k}{N} \right)^{-\frac{3}{2}M_k^2}2^{-M_k^2}\prod_{j<k}\left(\frac{\sin\frac{(j-k)\pi}{2N}}{\sin\frac{(j+k)\pi}{2N}}\right)^{2M_jM_k} A \prod b_k^{M_k},
\end{equation}
which is in agreement with the matrix model prediction \eqref{myC}.
\end{itemize}

\subsection{Relations for conformal blocks}
In this subsection, we reproduce $D_k(\mathbf{M})$ in our ansatz \eqref{eq:ansatz1}. To be exact, we work out the recurrence relation determining $D_k(\mathbf{M})$ for arbitrary $k$.

In order to simplify the equation for $D_k(\mathbf{M})$, we define
\be \ell_{\boldsymbol{\varepsilon},\boldsymbol{\Delta}}=\frac{C(\mathbf{M}+\frac12 \boldsymbol{\varepsilon}+\boldsymbol{\Delta})C(\mathbf{M}-\frac12 \boldsymbol{\varepsilon}-\boldsymbol{\Delta})}{C(\mathbf{M}+\frac12 \boldsymbol{\varepsilon})C(\mathbf{M}-\frac12 \boldsymbol{\varepsilon})}~.\ee

By using \eqref{eq:52}, we get
\begin{equation}\begin{gathered}
\label{eq:61}
\ell_{\boldsymbol{\varepsilon},\boldsymbol{\Delta}}=\prod_{k=1-\frac12\varepsilon_l}^{|\Delta_l+\frac12 \varepsilon_l|}\prod_{j=1+\frac12\varepsilon_l}^{|\Delta_l+\frac12\varepsilon_l|}\left(M_l+j-k-\frac12\varepsilon_l\right)
\times
\\
\times
\prod_{k=1}^{N-1}\left( \sin \frac{\pi k}{N} \right)^{-3\Delta_k(\Delta_k+\varepsilon_k)}
2^{-2\Delta_k(\Delta_k+\varepsilon_k)}
\prod_{j\neq k}\left(\frac{\sin\frac{(j-k)\pi}{2N}}{\sin\frac{(j+k)\pi}{2N}}\right)^{2(\Delta_j+\varepsilon_j)\Delta_k}
\end{gathered}
\end{equation}
Now divide \eqref{eq:61} by \(C(\mathbf{M}+\frac12 \boldsymbol{\varepsilon})C(\mathbf{M}-\frac12 \boldsymbol{\varepsilon})\) and substitute \eqref{eq:61} into it. We extract the coefficient of \(\mathfrak{r}^{-l}\):
\begin{equation}\begin{gathered}
\label{eq:bilinearInfinity2}
\sum_{(\boldsymbol{\Delta},\boldsymbol{\Delta}+\boldsymbol{\varepsilon})+n+m=l}
\ell_{\boldsymbol{\varepsilon},\boldsymbol{\Delta}}
D_n(\mathbf{M}')D_m(\mathbf{M}'')
\left(\left(\boldsymbol{\varepsilon}+2\boldsymbol{\Delta},\sin \frac{\pi \mathbf{k}}{N} \right)^2-\sin^2 \left(\boldsymbol{\varepsilon}+2\boldsymbol{\Delta},\frac{\pi \mathbf{k}}{N}\right)\right)+
\\
+2\sum_{(\boldsymbol{\Delta},\boldsymbol{\Delta}+\boldsymbol{\varepsilon})+n+m+1=l}
\ell_{\boldsymbol{\varepsilon},\boldsymbol{\Delta}}
D_n(\mathbf{M}')D_m(\mathbf{M}'')
 \left(\left(\boldsymbol{\varepsilon}+2\boldsymbol{\Delta},\mathbf{M}\right)\left(\boldsymbol{\varepsilon}+2\boldsymbol{\Delta},\sin \frac{\pi \mathbf{k}}{N} \right)-\left(\sin\frac{\pi \mathbf{k}}{N},\mathbf{M} \right) \right)+
\\
+2\sum_{(\boldsymbol{\Delta},\boldsymbol{\Delta}+\boldsymbol{\varepsilon})+n+m+1=l}
\ell_{\boldsymbol{\varepsilon},\boldsymbol{\Delta}}
D_n(\mathbf{M}')D_m(\mathbf{M}'')(n-m)
\left(\boldsymbol{\varepsilon}+2\boldsymbol{\Delta},\sin\frac{\pi \mathbf{k}}{N}\right)+
\\
+
\sum_{(\boldsymbol{\Delta},\boldsymbol{\Delta}+\boldsymbol{\varepsilon})+n+m+2=l}
\ell_{\boldsymbol{\varepsilon},\boldsymbol{\Delta}}
D_n(\mathbf{M}')D_m(\mathbf{M}'')
\left(\boldsymbol{\varepsilon}+2\boldsymbol{\Delta},\mathbf{M}\right)^2
+
\\
+
2\sum_{(\boldsymbol{\Delta},\boldsymbol{\Delta}+\boldsymbol{\varepsilon})+n+m+2=l}
\ell_{\boldsymbol{\varepsilon},\boldsymbol{\Delta}}
D_n(\mathbf{M}')D_m(\mathbf{M}'')(n-m)
\left(\boldsymbol{\varepsilon}+2\boldsymbol{\Delta},\mathbf{M}\right)
+
\\
+
\sum_{(\boldsymbol{\Delta},\boldsymbol{\Delta}+\boldsymbol{\varepsilon})+n+m+2=l}
\ell_{\boldsymbol{\varepsilon},\boldsymbol{\Delta}}
D_n(\mathbf{M}')D_m(\mathbf{M}'')(n-m)^2
=0~.
\end{gathered}
\end{equation}
We denote the l.h.s.~of \eqref{eq:bilinearInfinity2}  by $\widehat{F}_l^{\boldsymbol{\varepsilon}}(\mathbf{M})$.
\eqref{eq:bilinearInfinity2} can be turned into a recurrence relation for \(D_k(\mathbf{M})\). For this purpose, we need to single out the terms involving  $D_l(\mathbf{M})$ and use \(F_l^{\boldsymbol{\varepsilon}}(\mathbf{M})\) to name the remaining terms consisting of all other $D_{n<l}$'s.
Parallel to \autoref{appen:stconst}, we use the equations for the same \(\boldsymbol{\varepsilon}\)'s, i.e. \(\boldsymbol{\varepsilon}=\mathbf{e}_i+\mathbf{e}_j\), \(\boldsymbol{\varepsilon}=0\) and \(\boldsymbol{\varepsilon}=\mathbf{e}_j\). So the corresponding $\boldsymbol{\Delta}$'s are also the same as the ones in \autoref{appen:stconst}.

\begin{itemize}
\item For $\boldsymbol{\varepsilon}=\mathbf{e}_i+\mathbf{e}_j$ and \(\boldsymbol{\Delta}=0, -\mathbf{e}_j, -\mathbf{e}_k \text{ and } -\mathbf{e}_j-\mathbf{e}_k\), just as in \autoref{appen:stconst}, the prefactors for the terms with \(n=0,m=l\) and \(n=l,m=0\)  are
\begin{equation}
\label{eq:51}
\ell_{\mathbf{e}_j+\mathbf{e}_k,0}\left( (\sin \frac{\pi j}{N}+\sin \frac{\pi k}{N})^2-\sin^2 \frac{\pi(j+k)}{N} \right)=4\sin\frac{\pi j}{N}\sin\frac{\pi k}{N}\sin^2 \frac{\pi(j+k)}{2n},
\end{equation}
\begin{equation}
\label{eq:63}
\ell_{\mathbf{e}_j+\mathbf{e}_k,-\mathbf{e}_k}\left( (\sin \frac{\pi j}{N}-\sin \frac{\pi k}{N})^2-\sin^2 \frac{\pi(j-k)}{N} \right)=
-4\sin\frac{\pi j}{N}\sin\frac{\pi k}{N}\sin^2 \frac{\pi(j+k)}{2N},
\end{equation}
where we used that
\begin{equation}
\label{eq:84}
\ell_{\mathbf{e}_j+\mathbf{e}_k,\mathbf{0}}=\ell_{\mathbf{e}_j+\mathbf{e}_k,-\mathbf{e}_j-\mathbf{e}_k}=1,\qquad
\ell_{\mathbf{e}_j+\mathbf{e}_k,-\mathbf{e}_j}=\ell_{\mathbf{e}_j+\mathbf{e}_k,-\mathbf{e}_k}=\left(\frac{\sin \frac{(j+k)\pi}{2N}}{\sin \frac{(j-k)\pi}{2N}}\right)^2.
\end{equation}

From \eqref{eq:bilinearInfinity2}, we obtain
\begin{multline}
\label{eq:recursion2}
\widehat{F}_n^{\mathbf{e}_j+\mathbf{e}_k}(\mathbf{M})=8\sin\frac{\pi j}{N}\sin\frac{\pi k}{N}\sin^2 \frac{\pi(j+k)}{2N}
\left( \re^{\frac12 \partial_j}-\re^{-\frac12 \partial_j} \right)
\left( \re^{\frac12 \partial_k}-\re^{-\frac12 \partial_k} \right)
D_n(\mathbf{M})+\\
+F_n^{\mathbf{e}_j+\mathbf{e}_k}(\mathbf{M})=0,
\end{multline}
where \(\re^{\partial_j}\equiv\re^{\partial_{M_j}}\) is a shift operator in \(M_j\).

\item Now we consider the case \(\boldsymbol{\varepsilon}=\mathbf{e}_j\) parallel to what we just did for $\boldsymbol{\varepsilon}=\mathbf{e}_i+\mathbf{e}_j$. We have
\begin{equation}
\label{eq:66}
\ell_{\mathbf{e}_j,\pm \mathbf{e}_k}=\ell_{\mathbf{e}_j,-\mathbf{e}_j\mp\mathbf{e}_k}=\frac1{4}M_k \left(\sin \frac{\pi k}{N} \right)^{-3}\left( \frac{\sin \frac{\pi(j-k)}{2N}}{\sin \frac{\pi(j+k)}{2N}} \right)^{\pm 2}.
\end{equation}
The relation \eqref{eq:bilinearInfinity2} in this case becomes
\begin{equation}\begin{gathered}
\label{eq:recursion1}
\widehat{F}_{n+1}^{\mathbf{e}_j}(\mathbf{M})=2\sum_{k\neq j}\left(\sin \frac{\pi k}{N} + \left(\cos \frac{2\pi k}{N}-\cos \frac{\pi j}{N}\cos \frac{\pi k}{N}\right)\frac{\sin \frac{\pi j}{N}}{\sin^2 \frac{\pi k}{N}} \right)M_k \left( \re^{\frac12\partial_j+\partial_k}+\re^{-\frac12\partial_j-\partial_k} \right)D_n(\mathbf{M})+
\\
+
2\sum_{k\neq j}\left(\sin \frac{\pi k}{N} - \left(\cos \frac{2\pi k}{N}-\cos \frac{\pi j}{N}\cos \frac{\pi k}{N}\right)\frac{\sin \frac{\pi j}{N}}{\sin^2 \frac{\pi k}{N}} \right)M_k \left( \re^{\frac12\partial_j-\partial_k}+\re^{-\frac12\partial_j+\partial_k} \right)D_n(\mathbf{M})-
\\
-4\sum_{k\neq j} \sin \frac{\pi k}{N}\left( \re^{\frac12\partial_j}+\re^{-\frac12\partial_j} \right)D_n(\mathbf{M})+F_{n+1}^{\mathbf{e}_j}(\mathbf{M})=0.
\end{gathered}
\end{equation}
This relation can be rewritten in the following way:
\begin{equation}\begin{gathered}
\label{eq:70}
2\sum_{k\neq j}\left(\cos \frac{2\pi k}{N}-\cos \frac{\pi j}{N}\cos \frac{\pi k}{N}\right)\frac{\sin \frac{\pi j}{N}}{\sin^2 \frac{\pi k}{N}}
M_k \left( \re^{\frac12\partial_j}-\re^{-\frac12\partial_j} \right)\left( \re^{\partial_k}-\re^{-\partial_k} \right)D_n(\mathbf{M})
+
\\
+
2\sum_{k\neq j}\sin \frac{\pi k}{N}M_k\left( \re^{\frac14 \partial_j}-\re^{-\frac14\partial_j} \right)^2 \left( \re^{\frac12\partial_k}-\re^{-\frac12\partial_k} \right)^2D_n(\mathbf{M})+F_n^{\mathbf{e}_j}(\mathbf{M})=0.
\end{gathered}
\end{equation}
We see that this equation is more complicated than \eqref{eq:recursion2}, it will be good to have a substitution of it. Indeed, in the next item, we find a simpler recurrence relation and its combination with \eqref{eq:recursion2} is enough to phase out \eqref{eq:70}.

\item The last case to consider is \(\boldsymbol{\varepsilon}=0\), here
\begin{equation}
\label{eq:71}
\ell_{0,\mp \mathbf{e}_k}=M_k \frac{1}{4 \sin^3 \frac{\pi k}{N}}.
\end{equation}
The relation \eqref{eq:bilinearInfinity2}  becomes
\begin{multline}
\label{eq:72}
\widehat{F}^{\mathbf{0}}_{n+1}(\mathbf{M})=2\sum_k\sin \left(\frac{\pi k}{N}\right)M_k\left( \re^{\partial_k}+\re^{-\partial_k} \right)D_n(\mathbf{M})-4\sum_k \sin \left(\frac{\pi k}{N}\right)M_kD_n(\mathbf{M})
+F_{n+1}^{\mathbf{0}}(\mathbf{M})
=0,
\end{multline}
or equivalently,
\begin{equation}
\label{eq:recursion0}
2\sum_k\sin \left(\frac{\pi k}{N}\right)M_k\left( \re^{\frac12\partial_k}-\re^{-\frac12\partial_k} \right)^2D_n(\mathbf{M})+F_{n+1}^{\mathbf{0}}(\mathbf{M})=0.
\end{equation}
\end{itemize}

Obviously, constant and linear functions of $M_i$'s consist the kernel of the combination of shift operators in \eqref{eq:recursion0} and \eqref{eq:recursion2}, i.e. \(\partial_j\partial_k+\text{higher derivatives}\) and \(\sum_{k=1}^{N-1}\sin \left( \frac{\pi k}{N} \right)M_k\partial_k^2\). So these recurrence relations can only determine $D_n(\mathbf{M})$ up to a function in this kernel.  However, this problem can be resolved using other information about the tau function that we already have. The constant terms are determined by requiring
 \be D_{n>0}(0)=0,\ee
and coefficients of $M_i$'s for each $D_n(\mathbf{M})$ can be obtained from the expansion of the Bessel function in \eqref{eq:Bessels}, for example,
\begin{equation}
\label{eq:75}
D_1(\mathbf{e}_j)=-\frac1{8\sin \frac{\pi j}{N}}.
\end{equation}

\subsection{First non-trivial term from the recurrence relations}

Now let us see how this recursion works.
In order to find \(D_1(\mathbf{M})\) we first write two difference equations \eqref{eq:recursion2}, \eqref{eq:recursion0}.
Equation \eqref{eq:recursion2} reads
\begin{equation}\begin{gathered}
\label{eq:32}
8\sin\frac{\pi j}{N}\sin\frac{\pi k}{N}\sin^2 \frac{\pi(j+k)}{2N}
\left( \re^{\frac12 \partial_j}-\re^{-\frac12 \partial_j} \right)
\left( \re^{\frac12 \partial_k}-\re^{-\frac12 \partial_k} \right)
D_1(\mathbf{M})+\\+
2\sum\limits_{\boldsymbol{\Delta}}\ell_{\mathbf{e}_j+\mathbf{e}_k,\boldsymbol{\Delta}}
 \left(\left(\boldsymbol{\varepsilon}+2\boldsymbol{\Delta},\mathbf{M}\right)\left(\boldsymbol{\varepsilon}+2\boldsymbol{\Delta},\sin \frac{\pi \mathbf{k}}{N} \right)-\left(\sin\frac{\pi \mathbf{k}}{N},\mathbf{M} \right) \right)+\\
+\sum_{\boldsymbol{\Delta}'}\ell_{\mathbf{e}_i+\mathbf{e}_j,\mathbf{\Delta}'}\left(\left(\boldsymbol{\varepsilon}+2\boldsymbol{\Delta},\sin \frac{\pi \mathbf{k}}{N} \right)^2-\sin^2 \left(\boldsymbol{\varepsilon}+2\boldsymbol{\Delta},\frac{\pi \mathbf{k}}{N}\right)\right)=0,
\end{gathered}
\end{equation}
where \(\boldsymbol{\Delta}\in \{0,-\mathbf{e}_j,-\mathbf{e}_k,-\mathbf{e}_j-\mathbf{e}_k\}\), \(\boldsymbol{\Delta}'\in \{\pm \mathbf{e}_m,\pm\mathbf{e}_m-\mathbf{e}_j,\pm \mathbf{e}_m-\mathbf{e}_k,\pm \mathbf{e}_m-\mathbf{e}_j-\mathbf{e}_k\}\), where \(m\neq i\), \(m\neq j\), \(i\neq j\).

Equation \eqref{eq:recursion0} reads
\begin{multline}
\label{eq:36}
2\sum_k\sin \frac{\pi k}{N} M_k\left(\re^{\frac12 \partial_k}-\re^{-\frac12 \partial_k}\right)^2 D_1(\mathbf{M})+
2\sum_{\boldsymbol{\Delta}=\pm \mathbf{e}_j} \ell_{\mathbf{0},\boldsymbol{\Delta}}\left( 
(2\boldsymbol{\Delta},\mathbf{M})\left(2\boldsymbol{\Delta},\sin \frac{\pi \mathbf{k}}{N}\right)-\left( \sin \frac{\pi \mathbf{k}}{N},\mathbf{M} \right) \right)+\\+
\sum_{\boldsymbol{\Delta}\in \{\pm\mathbf{e}_j\pm \mathbf{e}_k\}}
\ell_{\mathbf{0},\boldsymbol{\Delta}}
\left(\left(\boldsymbol{\varepsilon}+2\boldsymbol{\Delta},\sin \frac{\pi \mathbf{k}}{N} \right)^2-\sin^2 \left(\boldsymbol{\varepsilon}+2\boldsymbol{\Delta},\frac{\pi \mathbf{k}}{N}\right)\right)=0.
\end{multline}
Summation in \eqref{eq:32} goes over \(\boldsymbol{\Delta}=0,-\mathbf{e}_j,-\mathbf{e}_k,-\mathbf{e}_j-\mathbf{e}_k\).
Using \eqref{eq:84} we can write down
\begin{equation}\begin{gathered}
\label{eq:85}
8\sin\frac{\pi j}{N}\sin\frac{\pi k}{N}\sin^2 \frac{\pi(j+k)}{2N}
\left( \re^{\frac12 \partial_j}-\re^{-\frac12 \partial_j} \right)
\left( \re^{\frac12 \partial_k}-\re^{-\frac12 \partial_k} \right)
D_1(\mathbf{M})+\\+
4 \left( (M_k+M_j)\left( \sin \frac{\pi k}{N}+\sin \frac{\pi j}{N} \right)-
\left( \sin \frac{\pi \mathbf{k}}{N},\mathbf{M} \right) \right)+\\+
4 \left(\frac{\sin\frac{(j+k)\pi}{2N} }{\sin\frac{(j-k)\pi}{2N}} \right)^2 \left( (M_k-M_j)\left( \sin \frac{\pi k}{N}-\sin \frac{\pi j}{N} \right)-
\left( \sin \frac{\pi \mathbf{k}}{N},\mathbf{M} \right) \right)+\\+
\sum_{m\neq j,k}4\left(1+ \left(\frac{\sin\frac{(j+k)\pi}{2N} }{\sin\frac{(j-k)\pi}{2N}} \right)^2 \right) M_m\sin \frac{\pi m}{N}=0.
\end{gathered}
\end{equation}
This equation can further be simplified
\begin{equation}
\label{eq:8}
\left( \re^{\frac12 \partial_j}-\re^{-\frac12 \partial_j} \right)
\left( \re^{\frac12 \partial_k}-\re^{-\frac12 \partial_k} \right)
D_1(\mathbf{M})=
\frac{M_k \sin \frac{\pi j}{N}+M_j\sin \frac{\pi k}{N}}{2\sin^2 \frac{\pi(j-k)}{2N}\sin^2 \frac{\pi(j+k)}{2N}}.
\end{equation}
It defines a part of the coefficients of \(D_1(\mathbf{M})\):
\begin{equation}
\label{eq:9}
D_1(\mathbf{M})=\sum_{j<k}\frac{M_k^2M_j\sin \frac{\pi j}{N}+M_j^2M_k\sin \frac{\pi k}{N}}{\left(\cos \frac{\pi k}{N}-\cos \frac{\pi j}{N}\right)^2}+\sum_{j=1}^{N-1} f_j(M_j)=\sum_{j\neq k}\frac{M_k^2M_j\sin \frac{\pi j}{N}}{\left(\cos \frac{\pi k}{N}-\cos \frac{\pi j}{N}\right)^2}+\tilde{D}_1(\mathbf{M}).
\end{equation}
Now we come back to equation \eqref{eq:36} to find remaining coefficients.
We rewrite it explicitly
\begin{multline}
\label{eq:10}
2\sum_k\sin \frac{\pi k}{N} M_k\left(\re^{\frac12 \partial_k}-\re^{-\frac12 \partial_k}\right)^2 D_1(\mathbf{M})+
4\sum_j \frac{M_j}{4\sin^3 \frac{\pi j}{N}}\left(4 M_j \sin \frac{\pi j}{N}-\left( \sin \frac{\pi \mathbf{k}}{N},\mathbf{M} \right) \right)
+\\+
\sum_{\boldsymbol{\Delta}\in \{\pm\mathbf{e}_j\pm \mathbf{e}_k\}}
\ell_{\mathbf{0},\boldsymbol{\Delta}}
\left(\left(\boldsymbol{\varepsilon}+2\boldsymbol{\Delta},\sin \frac{\pi \mathbf{k}}{N} \right)^2-\sin^2 \left(\boldsymbol{\varepsilon}+2\boldsymbol{\Delta},\frac{\pi \mathbf{k}}{N}\right)\right)=0,
\end{multline}
or more explicitly
\begin{multline}
\label{eq:11}
2\sum_k\sin \frac{\pi k}{N}M_k\left(\re^{\frac12 \partial_k}-\re^{-\frac12 \partial_k}\right)^2 D_1(\mathbf{M})+
\sum_k \frac{3M_k^2}{\sin^2 \frac{\pi k}{N}}-
\sum_{j<k} M_jM_k\left(\frac{\sin \frac{\pi k}{N}}{\sin^3 \frac{\pi j}{N}}+\frac{\sin \frac{\pi j}{N}}{\sin^3 \frac{\pi k}{N}}\right)
+\\+
\sum_{\boldsymbol{\Delta}\in \{\pm\mathbf{e}_j\pm \mathbf{e}_k\}}
\ell_{\mathbf{0},\boldsymbol{\Delta}}
\left(\left(\boldsymbol{\varepsilon}+2\boldsymbol{\Delta},\sin \frac{\pi \mathbf{k}}{N} \right)^2-\sin^2 \left(\boldsymbol{\varepsilon}+2\boldsymbol{\Delta},\frac{\pi \mathbf{k}}{N}\right)\right)=0.
\end{multline}
We can check that all \(M_jM_k\) terms cancel, and we are left with simpler equation
\begin{equation}
\label{eq:12}
2\sum_k\sin \frac{\pi k }{N}M_k \left( \re^{\frac12\partial_k}-\re^{-\frac12\partial_k} \right)^2\tilde{D}_1(\mathbf{M})+\sum_k \frac{3M_k^2}{\sin^2 \frac{\pi k}{N}}=0,
\end{equation}
which tells us that
\begin{equation}
\label{eq:25}
\tilde{D}_1(\mathbf{M})=-\sum_k \frac{M_k^3}{4\sin^3 \frac{\pi k}{N}}+\sum_k a_kM_k.
\end{equation}
Combining together \eqref{eq:9}, \eqref{eq:25}, and \eqref{eq:75} we get the final answer
\begin{equation}
\label{eq:27}
D_1(\mathbf{M})=
\sum_{j\neq k}\frac{M_k^2M_j\sin \frac{\pi j}{N}}{\left(\cos \frac{\pi k}{N}-\cos \frac{\pi j}{N}\right)^2}-
\sum_k \frac{M_k^3-M_k}{4\sin^3 \frac{\pi k}{N}}-\sum_k \frac{M_k}{8\sin \frac{\pi k}{N}},
\end{equation}
which coincides with \eqref{Einfty1}.

\subsection{Estimation of the degree}

We already know that \(D_1(\mathbf{M})\) is a polynomial in \(M_k\) of degree not greater than 3.
Now we want to prove that \(\deg D_l(\mathbf{M})\le 3l\).
Since all these coefficients can be found recursively, it is enough to show that this property is preserved by recurrence, namely, to show that

\begin{equation}
\label{eq:29}
\deg(F_l^{\mathbf{e}_j+\mathbf{e}_k}(\mathbf{M}))\le 3l-2
\end{equation}
and
\begin{equation}
\label{eq:30}
\deg(F_{l+1}^{\mathbf{0}}(\mathbf{M}))\le 3l-1,
\end{equation}
or
\begin{equation}
\deg(F_l^{\mathbf{0}}(\mathbf{M}))\le 3l-4.
\end{equation}
First we can easily check using \eqref{eq:61} that
\begin{equation}
\label{eq:38}
\deg(\ell_{\boldsymbol{\epsilon},\boldsymbol{\Delta}})=(\boldsymbol{\Delta},\boldsymbol{\Delta}+\boldsymbol{\varepsilon}).
\end{equation}
Now consider \(F_l^{\boldsymbol{\varepsilon}}(\mathbf{M})\) for \(\boldsymbol{\varepsilon}=\mathbf{e}_j+\mathbf{e}_k\):
\begin{equation}
\begin{gathered}
\label{eq:42}
F^{\boldsymbol{\varepsilon}}_l(\mathbf{M})=\sideset{}{'}\sum_{\substack{(\boldsymbol{\Delta},\boldsymbol{\Delta}+\boldsymbol{\varepsilon})+n+m=l\\(\boldsymbol{\Delta},\boldsymbol{\Delta}+\boldsymbol{\varepsilon})=0}}
\ell_{\boldsymbol{\varepsilon},\boldsymbol{\Delta}}
D_n(\mathbf{M}')D_m(\mathbf{M}'')
\left(\left(\boldsymbol{\varepsilon}+2\boldsymbol{\Delta},\sin \frac{\pi \mathbf{k}}{N} \right)^2-\sin^2 \left(\boldsymbol{\varepsilon}+2\boldsymbol{\Delta},\frac{\pi \mathbf{k}}{N}\right)\right)+
\\
+\sum_{\substack{(\boldsymbol{\Delta},\boldsymbol{\Delta}+\boldsymbol{\varepsilon})+n+m=l\\(\boldsymbol{\Delta},\boldsymbol{\Delta}+\boldsymbol{\varepsilon})\ge 1}}
\ell_{\boldsymbol{\varepsilon},\boldsymbol{\Delta}}
D_n(\mathbf{M}')D_m(\mathbf{M}'')
\left(\left(\boldsymbol{\varepsilon}+2\boldsymbol{\Delta},\sin \frac{\pi \mathbf{k}}{N} \right)^2-\sin^2 \left(\boldsymbol{\varepsilon}+2\boldsymbol{\Delta},\frac{\pi \mathbf{k}}{N}\right)\right)+
\\
+2\sum_{(\boldsymbol{\Delta},\boldsymbol{\Delta}+\boldsymbol{\varepsilon})+n+m+1=l}
\ell_{\boldsymbol{\varepsilon},\boldsymbol{\Delta}}
D_n(\mathbf{M}')D_m(\mathbf{M}'')
 \left(\left(\boldsymbol{\varepsilon}+2\boldsymbol{\Delta},\mathbf{M}\right)\left(\boldsymbol{\varepsilon}+2\boldsymbol{\Delta},\sin \frac{\pi \mathbf{k}}{N} \right)-\left(\sin\frac{\pi \mathbf{k}}{N},\mathbf{M} \right) \right)+
\\
+2\sum_{(\boldsymbol{\Delta},\boldsymbol{\Delta}+\boldsymbol{\varepsilon})+n+m+1=l}
\ell_{\boldsymbol{\varepsilon},\boldsymbol{\Delta}}
D_n(\mathbf{M}')D_m(\mathbf{M}'')(n-m)
\left(\boldsymbol{\varepsilon}+2\boldsymbol{\Delta},\sin\frac{\pi \mathbf{k}}{N}\right)+
\\
+
\sum_{(\boldsymbol{\Delta},\boldsymbol{\Delta}+\boldsymbol{\varepsilon})+n+m+2=l}
\ell_{\boldsymbol{\varepsilon},\boldsymbol{\Delta}}
D_n(\mathbf{M}')D_m(\mathbf{M}'')
\left(\boldsymbol{\varepsilon}+2\boldsymbol{\Delta},\mathbf{M}\right)^2
+
\\
+
2\sum_{(\boldsymbol{\Delta},\boldsymbol{\Delta}+\boldsymbol{\varepsilon})+n+m+2=l}
\ell_{\boldsymbol{\varepsilon},\boldsymbol{\Delta}}
D_n(\mathbf{M}')D_m(\mathbf{M}'')(n-m)
\left(\boldsymbol{\varepsilon}+2\boldsymbol{\Delta},\mathbf{M}\right)
+
\\
+
\sum_{(\boldsymbol{\Delta},\boldsymbol{\Delta}+\boldsymbol{\varepsilon})+n+m+2=l}
\ell_{\boldsymbol{\varepsilon},\boldsymbol{\Delta}}
D_n(\mathbf{M}')D_m(\mathbf{M}'')(n-m)^2
=0,
\end{gathered}
\end{equation}
where \(\sum'\) means summation excluding terms with \(n=l, m=0\) and \(n=0,m=l\).
The first term of \eqref{eq:38} needs special treatment, but degrees of the other terms can be easily computed:
\begin{multline}
\label{eq:68}
\deg \left(\ell_{\boldsymbol{\varepsilon},\boldsymbol{\Delta}}
D_n(\mathbf{M}')D_m(\mathbf{M}'')
\left(\left(\boldsymbol{\varepsilon}+2\boldsymbol{\Delta},\sin \frac{\pi \mathbf{k}}{N} \right)^2-\sin^2 \left(\boldsymbol{\varepsilon}+2\boldsymbol{\Delta},\frac{\pi \mathbf{k}}{N}\right)\right)\right)\le
\\
\le 3n+3m+(\boldsymbol{\Delta},\boldsymbol{\Delta}+\boldsymbol{\varepsilon})
\le 3l-2(\boldsymbol{\Delta},\boldsymbol{\Delta}+\boldsymbol{\varepsilon})
\le 3l-2,
\end{multline}
\begin{multline}
\label{eq:74}
\deg \left(\ell_{\boldsymbol{\varepsilon},\boldsymbol{\Delta}}
D_n(\mathbf{M}')D_m(\mathbf{M}'')
 \left(\left(\boldsymbol{\varepsilon}+2\boldsymbol{\Delta},\mathbf{M}\right)\left(\boldsymbol{\varepsilon}+2\boldsymbol{\Delta},\sin \frac{\pi \mathbf{k}}{N} \right)-\left(\sin\frac{\pi \mathbf{k}}{N},\mathbf{M} \right) \right)+
\right)\le
\\
\le 3n+3m+(\boldsymbol{\Delta},\boldsymbol{\Delta}+\boldsymbol{\varepsilon})+1
\le 3l-2(\boldsymbol{\Delta},\boldsymbol{\Delta}+\boldsymbol{\varepsilon})-3+1
\le 3l-2,
\end{multline}
\begin{multline}
\label{eq:77}
\deg \left( \ell_{\boldsymbol{\varepsilon},\boldsymbol{\Delta}}
D_n(\mathbf{M}')D_m(\mathbf{M}'')(n-m)
\left(\boldsymbol{\varepsilon}+2\boldsymbol{\Delta},\sin\frac{\pi \mathbf{k}}{N}\right)
 \right)
\le 3m+3n+(\boldsymbol{\Delta},\boldsymbol{\Delta}+\boldsymbol{\varepsilon})\le
\\
\le 3l-2(\boldsymbol{\Delta},\boldsymbol{\Delta}+\boldsymbol{\varepsilon})-3
\le 3l-3,
\end{multline}
\begin{multline}
\label{eq:86}
\deg \left(\ell_{\boldsymbol{\varepsilon},\boldsymbol{\Delta}}
D_n(\mathbf{M}')D_m(\mathbf{M}'')
\left(\boldsymbol{\varepsilon}+2\boldsymbol{\Delta},\mathbf{M}\right)^2
\right)
\le 3m+3n+(\boldsymbol{\Delta},\boldsymbol{\Delta}+\boldsymbol{\varepsilon})+2 \le
\\
\le 3l-2(\boldsymbol{\Delta},\boldsymbol{\Delta}+\boldsymbol{\varepsilon})-6+2 \le 3l-4,
\end{multline}
\begin{multline}
\label{eq:87}
\deg \left( \ell_{\boldsymbol{\varepsilon},\boldsymbol{\Delta}}
D_n(\mathbf{M}')D_m(\mathbf{M}'')(n-m)
\left(\boldsymbol{\varepsilon}+2\boldsymbol{\Delta},\mathbf{M}\right)
 \right)
\le 3m+3n+(\boldsymbol{\Delta},\boldsymbol{\Delta}+\boldsymbol{\varepsilon})+1 \le
\\
\le 3l-2(\boldsymbol{\Delta},\boldsymbol{\Delta}+\boldsymbol{\varepsilon})-6+1 \le 3l-5,
\end{multline}
\begin{multline}
\label{eq:88}
\deg \left(\ell_{\boldsymbol{\varepsilon},\boldsymbol{\Delta}}
D_n(\mathbf{M}')D_m(\mathbf{M}'')(n-m)^2
\right)
\le 3m+3n+(\boldsymbol{\Delta},\boldsymbol{\Delta}+\boldsymbol{\varepsilon}) \le
\\
\le 3l-2(\boldsymbol{\Delta},\boldsymbol{\Delta}+\boldsymbol{\varepsilon})-6+\le 3l-6.
\end{multline}
Going to the first term, we can rewrite it using \eqref{eq:51}, \eqref{eq:63}:
\begin{multline}
\label{eq:89}
\sideset{}{'}\sum_{\substack{(\boldsymbol{\Delta},\boldsymbol{\Delta}+\boldsymbol{\varepsilon})+n+m=l\\(\boldsymbol{\Delta},\boldsymbol{\Delta}+\boldsymbol{\varepsilon})=0}}
\ell_{\boldsymbol{\varepsilon},\boldsymbol{\Delta}}
D_n(\mathbf{M}')D_m(\mathbf{M}'')
\left(\left(\boldsymbol{\varepsilon}+2\boldsymbol{\Delta},\sin \frac{\pi \mathbf{k}}{N} \right)^2-\sin^2 \left(\boldsymbol{\varepsilon}+2\boldsymbol{\Delta},\frac{\pi \mathbf{k}}{N}\right)\right)=
\\
=\sum_{\substack{j<k\\m,n>0}}
4\sin\frac{\pi j}{N}\sin\frac{\pi k}{N}\sin^2 \frac{\pi(j+k)}{2n}
\left( \re^{\frac12 \tilde{\partial}_j}-\re^{-\frac12 \tilde{\partial}_j} \right)
\left( \re^{\frac12 \tilde{\partial}_k}-\re^{-\frac12 \tilde{\partial}_k} \right)
D_n(\mathbf{M})\cdot D_m(\mathbf{M}),
\end{multline}
where \(\tilde{\partial}_j\) stands for Hirota derivative, which is the difference of derivatives acting on the first and on the second function.
It's precise definition is the following:
\begin{equation}
\label{eq:90}
A(\mathbf{M}+\boldsymbol{\delta})B(\mathbf{M}-\boldsymbol{\delta})=
\sum_{n_1,\ldots, n_{N-1}=0}^{\infty} \prod_{i=1}^{N-1}\frac{\delta_i^{n_i}}{n_i!} \prod \tilde{\partial}_i^{n_i} A(\boldsymbol{M})\cdot B(\mathbf{M}).
\end{equation}
It is clear that Hirota derivatives also decrease power by 1, and since the lowest derivative is \(\tilde{\partial}_j\tilde{\partial}_k\), the total power of this expression is \(3m+3n-2=3l-2\), as it should be:
\begin{equation}
\label{eq:91}
\deg \left( \sideset{}{'}\sum_{\substack{(\boldsymbol{\Delta},\boldsymbol{\Delta}+\boldsymbol{\varepsilon})+n+m=l\\(\boldsymbol{\Delta},\boldsymbol{\Delta}+\boldsymbol{\varepsilon})=0}}
\ell_{\boldsymbol{\varepsilon},\boldsymbol{\Delta}}
D_n(\mathbf{M}')D_m(\mathbf{M}'')
\left(\left(\boldsymbol{\varepsilon}+2\boldsymbol{\Delta},\sin \frac{\pi \mathbf{k}}{N} \right)^2-\sin^2 \left(\boldsymbol{\varepsilon}+2\boldsymbol{\Delta},\frac{\pi \mathbf{k}}{N}\right)\right) \right)
=3l-2.
\end{equation}

Now we try to do the same computation for \(F^{\mathbf{0}}_l(\mathbf{M})\):
\begin{equation}\begin{gathered}
\label{eq:92}
F^{\boldsymbol{0}}_l(\mathbf{M})=\sideset{}{'}\sum_{\substack{(\boldsymbol{\Delta},\boldsymbol{\Delta})+n+m=l\\(\boldsymbol{\Delta},\boldsymbol{\Delta})=1}}
\ell_{\boldsymbol{0},\boldsymbol{\Delta}}
D_n(\mathbf{M}')D_m(\mathbf{M}'')
\left(\left(2\boldsymbol{\Delta},\sin
\frac{\pi \mathbf{k}}{N} \right)^2-\sin^2
\left(2\boldsymbol{\Delta},\frac{\pi
\mathbf{k}}{N}\right)\right)- \\
-2\sideset{}{'}\sum_{\substack{(\boldsymbol{\Delta},\boldsymbol{\Delta})+n+m+1=l\\(\boldsymbol{\Delta},\boldsymbol{\Delta})=0}}
\ell_{\boldsymbol{0},\boldsymbol{\Delta}}
D_n(\mathbf{M}')D_m(\mathbf{M}'')
\left(\sin\frac{\pi
\mathbf{k}}{N},\mathbf{M} \right)+ \\
+\sum_{\substack{(\boldsymbol{\Delta},\boldsymbol{\Delta})+n+m=l\\(\boldsymbol{\Delta},\boldsymbol{\Delta})\ge 2}}
\ell_{\boldsymbol{0},\boldsymbol{\Delta}}
D_n(\mathbf{M}')D_m(\mathbf{M}'')
\left(\left(2\boldsymbol{\Delta},\sin
\frac{\pi \mathbf{k}}{N} \right)^2-\sin^2
\left(2\boldsymbol{\Delta},\frac{\pi
\mathbf{k}}{N}\right)\right)+ \\
+2\sum_{\substack{(\boldsymbol{\Delta},\boldsymbol{\Delta})+n+m+1=l\\(\boldsymbol{\Delta},\boldsymbol{\Delta})\ge 1}}
\ell_{\boldsymbol{0},\boldsymbol{\Delta}}
D_n(\mathbf{M}')D_m(\mathbf{M}'')
\left(\left(2\boldsymbol{\Delta},\mathbf{M}\right)\left(2\boldsymbol{\Delta},\sin
\frac{\pi \mathbf{k}}{N} \right)-\left(\sin\frac{\pi
\mathbf{k}}{N},\mathbf{M} \right) \right)+ \\
+2\sum_{(\boldsymbol{\Delta},\boldsymbol{\Delta})+n+m+1=l}
\ell_{\boldsymbol{0},\boldsymbol{\Delta}}
D_n(\mathbf{M}')D_m(\mathbf{M}'')(n-m)
\left(2\boldsymbol{\Delta},\sin\frac{\pi
\mathbf{k}}{N}\right)+ \\ +
\sum_{(\boldsymbol{\Delta},\boldsymbol{\Delta})+n+m+2=l}
\ell_{\boldsymbol{0},\boldsymbol{\Delta}}
D_n(\mathbf{M}')D_m(\mathbf{M}'')
\left(2\boldsymbol{\Delta},\mathbf{M}\right)^2
+ \\ +
2\sum_{(\boldsymbol{\Delta},\boldsymbol{\Delta})+n+m+2=l}
\ell_{\boldsymbol{0},\boldsymbol{\Delta}}
D_n(\mathbf{M}')D_m(\mathbf{M}'')(n-m)
\left(2\boldsymbol{\Delta},\mathbf{M}\right)
+ \\ +
\sum_{(\boldsymbol{\Delta},\boldsymbol{\Delta})+n+m+2=l}
\ell_{\boldsymbol{0},\boldsymbol{\Delta}}
D_n(\mathbf{M}')D_m(\mathbf{M}'')(n-m)^2
=0,
\end{gathered}\end{equation}
where notation \(\sum'\) means summation that excludes \(n=l-1, m=0\) and \(n=0, m=l-1\).
Estimates for the degrees of these terms except the first two:
\begin{multline}
\label{eq:93}
\deg \left(\ell_{\boldsymbol{0},\boldsymbol{\Delta}}
D_n(\mathbf{M}')D_m(\mathbf{M}'')
\left(\left(2\boldsymbol{\Delta},\sin
\frac{\pi \mathbf{k}}{N} \right)^2-\sin^2
\left(2\boldsymbol{\Delta},\frac{\pi
\mathbf{k}}{N}\right)\right) \right)\le
\\
\le 3n+3m+(\boldsymbol{\Delta},\boldsymbol{\Delta})
\le 3l-2(\boldsymbol{\Delta},\boldsymbol{\Delta})\le 3l-4,
\end{multline}
\begin{multline}
\label{eq:94}
\deg \left( \ell_{\boldsymbol{0},\boldsymbol{\Delta}}
D_n(\mathbf{M}')D_m(\mathbf{M}'')
\left(\left(2\boldsymbol{\Delta},\mathbf{M}\right)\left(2\boldsymbol{\Delta},\sin
\frac{\pi \mathbf{k}}{N} \right)-\left(\sin\frac{\pi
\mathbf{k}}{N},\mathbf{M} \right) \right) \right)\le
\\
\le 3m+3m+(\boldsymbol{\Delta},\boldsymbol{\Delta})+1
\le 3l-2(\boldsymbol{\Delta},\boldsymbol{\Delta})-3+1
\le 3l-4,
\end{multline}
\begin{multline}
\label{eq:95}
\deg \left( \ell_{\boldsymbol{0},\boldsymbol{\Delta}}
D_n(\mathbf{M}')D_m(\mathbf{M}'')(n-m)
\left(2\boldsymbol{\Delta},\sin\frac{\pi
\mathbf{k}}{N}\right) \right)\le
\\
\le 3m+3m+(\boldsymbol{\Delta},\boldsymbol{\Delta})
\le 3l-2(\boldsymbol{\Delta},\boldsymbol{\Delta})-3
\le 3l-5,
\end{multline}
\begin{multline}
\label{eq:96}
\deg \left( \ell_{\boldsymbol{0},\boldsymbol{\Delta}}
D_n(\mathbf{M}')D_m(\mathbf{M}'')
\left(2\boldsymbol{\Delta},\mathbf{M}\right)^2
 \right)\le
\\
\le 3m+3m+(\boldsymbol{\Delta},\boldsymbol{\Delta})+2
\le 3l-2(\boldsymbol{\Delta},\boldsymbol{\Delta})-6+2
\le 3l-6,
\end{multline}
\begin{multline}
\label{eq:97}
\deg \left( \ell_{\boldsymbol{0},\boldsymbol{\Delta}}
D_n(\mathbf{M}')D_m(\mathbf{M}'')(n-m)
\left(2\boldsymbol{\Delta},\mathbf{M}\right)
 \right)\le
\\
\le 3m+3m+(\boldsymbol{\Delta},\boldsymbol{\Delta})+1
\le 3l-2(\boldsymbol{\Delta},\boldsymbol{\Delta})-6+1
\le 3l-7,
\end{multline}
\begin{multline}
\label{eq:98}
\deg \left( \ell_{\boldsymbol{0},\boldsymbol{\Delta}}
D_n(\mathbf{M}')D_m(\mathbf{M}'')(n-m)^2
 \right)\le
\\
\le 3m+3m+(\boldsymbol{\Delta},\boldsymbol{\Delta})
\le 3l-2(\boldsymbol{\Delta},\boldsymbol{\Delta})-6
\le 3l-6,
\end{multline}
where we used extensively the fact that if some expression so proportional to \(\boldsymbol{\Delta}\), it will vanish when \((\boldsymbol{\Delta},\boldsymbol{\Delta})=0\), so one should have at least \((\boldsymbol{\Delta},\boldsymbol{\Delta})=1\).

Let us analyze the last term:
\begin{equation}\begin{gathered}
\label{eq:99}
\sideset{}{'}\sum_{\substack{(\boldsymbol{\Delta},\boldsymbol{\Delta})+n+m=l\\(\boldsymbol{\Delta},\boldsymbol{\Delta})=1}}
\ell_{\boldsymbol{0},\boldsymbol{\Delta}}
D_n(\mathbf{M}')D_m(\mathbf{M}'')
\left(\left(2\boldsymbol{\Delta},\sin
\frac{\pi \mathbf{k}}{N} \right)^2-\sin^2
\left(2\boldsymbol{\Delta},\frac{\pi
\mathbf{k}}{N}\right)\right)- \\
-2\sideset{}{'}\sum_{\substack{(\boldsymbol{\Delta},\boldsymbol{\Delta})+n+m+1=l\\(\boldsymbol{\Delta},\boldsymbol{\Delta})=0}}
\ell_{\boldsymbol{0},\boldsymbol{\Delta}}
D_n(\mathbf{M}')D_m(\mathbf{M}'')
\left(\sin\frac{\pi
\mathbf{k}}{N},\mathbf{M} \right)=
\\
=\sum_k\sum_{m+n+1=l}\sin \frac{\pi k}{N}M_k\left( D_n(\mathbf{M}+\mathbf{e}_k)D_m(\mathbf{M}-\mathbf{e}_k)+ D_n(\mathbf{M}-\mathbf{e}_k)D_m(\mathbf{M}+\mathbf{e}_k) - 2D_n(\mathbf{M})D_m(\mathbf{M}) \right)=
\\
=\sum_k\sum_{m+n+1=l} \sin \frac{\pi k}{N}M_k \left( e^{\frac12 \tilde{\partial}_k}-e^{-\frac12 \tilde{\partial}_k} \right)^2D_n(\mathbf{M})\cdot D_m(\mathbf{M}),
\end{gathered}\end{equation}
so its degree
\begin{equation}
\label{eq:100}
\deg \left( M_k \left( e^{\frac12 \tilde{\partial}_k}-e^{-\frac12 \tilde{\partial}_k} \right)^2D_n(\mathbf{M})\cdot D_m(\mathbf{M}) \right)
\le 3m+3n-1
\le 3l-3-1\le 3l-4.
\end{equation}
This way we completed the proof of \eqref{eq:29} and \eqref{eq:30}.

\bibliography{painleve}
\bibliographystyle{utphys}
\end{document}